\documentclass{ieeeaccess}
\usepackage{cite}
\usepackage{amsmath,amssymb,amsfonts}
\DeclareMathOperator*{\argmax}{arg\!\,max}
\usepackage{booktabs}
\usepackage{algorithmic}
\usepackage{graphicx}
\usepackage{hyperref}
\usepackage{textcomp}
\usepackage{mathtools}
\usepackage{makecell}
\usepackage{multirow}
\usepackage{array}
\usepackage[outdir=./]{epstopdf}
\usepackage[font={sf,scriptsize},labelfont={bf,color=accessblue},caption=false]{subfig}

\newcommand{\rulesep}{\unskip\ \vrule\ }

\def\BibTeX{{\rm B\kern-.05em{\sc i\kern-.025em b}\kern-.08em
    T\kern-.1667em\lower.7ex\hbox{E}\kern-.125emX}}
\begin{document}
\history{Date of publication xxxx 00, 0000, date of current version xxxx 00, 0000.}
\doi{10.1109/ACCESS.2017.DOI}

\DeclarePairedDelimiter\ceil{\lceil}{\rceil}
\DeclarePairedDelimiter\floor{\lfloor}{\rfloor}

\title{Deep Spoken Keyword Spotting: An Overview}
\author{\uppercase{Iv\'an L\'opez-Espejo}\authorrefmark{1},
\uppercase{Zheng-Hua Tan}\authorrefmark{1},     \IEEEmembership{Senior Member, IEEE}, \uppercase{John Hansen}\authorrefmark{2}, \IEEEmembership{Fellow, IEEE}, \uppercase{and Jesper Jensen}\authorrefmark{1,3}}
\address[1]{Department of Electronic Systems, Aalborg University, 9220 Aalborg, Denmark (e-mail: \{ivl,zt,jje\}@es.aau.dk)}
\address[2]{Erik Jonsson School of Engineering and Computer Science, The University of Texas at Dallas, TX 75080 Richardson, USA (e-mail: john.hansen@utdallas.edu)}
\address[3]{Oticon A/S, 2765 Smørum, Denmark (e-mail: jesj@oticon.com)}
\tfootnote{This work was supported, in part, by Demant Foundation.}

\markboth
{I. L\'opez-Espejo \headeretal: Deep Spoken KWS: An Overview}
{I. L\'opez-Espejo \headeretal: Deep Spoken KWS: An Overview}

\corresp{Corresponding author: Iv\'an L\'opez-Espejo (e-mail: ivl@es.aau.dk).}

\begin{abstract}
Spoken keyword spotting (KWS) deals with the identification of keywords in audio streams and has become a fast-growing technology thanks to the paradigm shift introduced by deep learning a few years ago. This has allowed the rapid embedding of deep KWS in a myriad of small electronic devices with different purposes like the activation of voice assistants. Prospects suggest a sustained growth in terms of social use of this technology. Thus, it is not surprising that deep KWS has become a hot research topic among speech scientists, who constantly look for KWS performance improvement and computational complexity reduction. This context motivates this paper, in which we conduct a literature review into deep spoken KWS to assist practitioners and researchers who are interested in this technology. Specifically, this overview has a comprehensive nature by covering a thorough analysis of deep KWS systems (which includes speech features, acoustic modeling and posterior handling), robustness methods, applications, datasets, evaluation metrics, performance of deep KWS systems and audio-visual KWS. The analysis performed in this paper allows us to identify a number of directions for future research, including directions adopted from automatic speech recognition research and directions that are unique to the problem of spoken KWS.
\end{abstract}


\titlepgskip=-15pt

\maketitle

\section{Introduction}
\label{sec:intro}

\PARstart{I}{nteracting} with machines via voice is not science fiction anymore. Quite the opposite, speech technologies have become ubiquitous in nowadays society. The proliferation of voice assistants like Amazon's Alexa, Apple's Siri, Google's Assistant and Microsoft's Cortana is good proof of this \cite{Hoy18}. A distinctive feature of voice assistants is that, in order to be used, they first have to be activated by means of a spoken wake-up word or keyword, thereby avoiding running far more computationally expensive automatic speech recognition (ASR) when it is not required \cite{Assaf17}. More specifically, voice assistants deploy a technology called spoken keyword spotting ---or simply keyword spotting---, which can be understood as a subproblem of ASR \cite{Oriol14}. Particularly, keyword spotting (KWS) can be defined as the task of identifying keywords in audio streams comprising speech. And, apart from activating voice assistants, KWS has plenty of applications such as speech data mining, audio indexing, phone call routing, etc. \cite{Yimeng16}.

Over the years, different techniques have been explored for KWS. One of the earliest approaches is based on the use of large-vocabulary continuous speech recognition (LVCSR) systems \cite{Weintraub93, Miller07, Guoguo13}. These systems are employed to decode the speech signal, and then, the keyword is searched in the generated lattices (i.e., in the representations of the different sequences of phonetic units that, given the speech signal, are likely enough). One of the advantages of this approach is the flexibility to deal with changing/non-predefined keywords \cite{Yiyan18, Shan18, Rajeev21} (although there is often a drop in performance when keywords are out of vocabulary \cite{Guoguo15}). The main disadvantage of LVCSR-based KWS systems might reside in the computational complexity dimension: these systems need to generate rich lattices, which requires high computational resources \cite{Shan18, Chai19} and also introduces latency \cite{Sun17}. While this should not be an issue for some applications like \emph{offline} audio search \cite{Can11, Shan18}, LVCSR systems are not suitable for the lately-popular KWS applications\footnote{By lately-popular KWS applications we mean activation of voice assistants, voice control, etc.} intended for small electronic devices (e.g., smartphones, smart speakers and wearables) characterized by notable memory, computation and power constraints \cite{Parada15, Bai19, Chai19, Jingyong19}.

\begin{figure}
    \centering
    \includegraphics[width=0.9\linewidth]{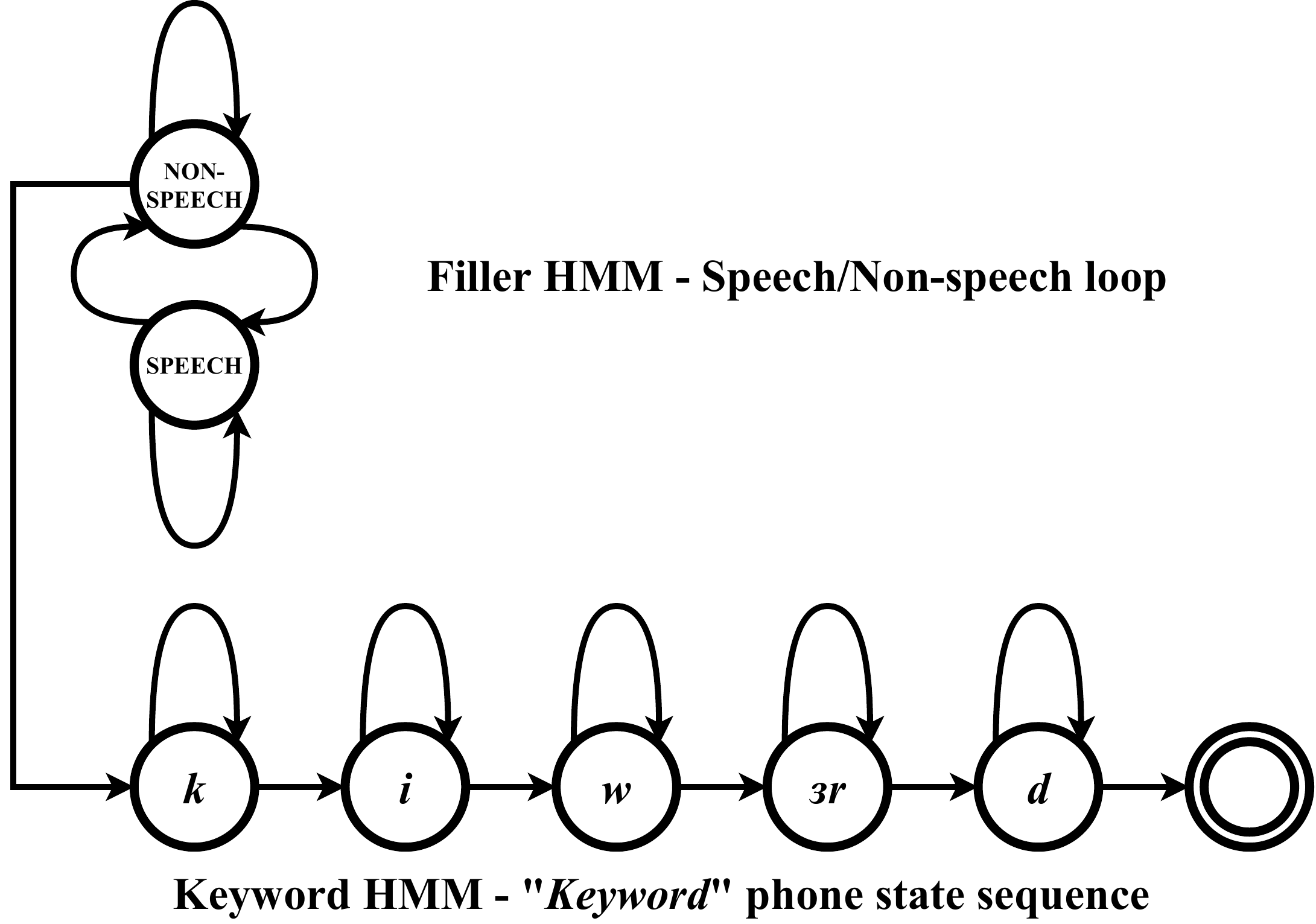}
    \caption{Scheme of a keyword/filler HMM-based KWS system \cite{Sun17} when the system keyword is ``\emph{keyword}''. While typically the keyword is modeled by a context-dependent triphone-based HMM, a monophone-based HMM is depicted instead for illustrative purposes. The filler HMM is often a speech/non-speech monophone loop.}
    \label{fig:hmm}
\end{figure}

A still attractive and lighter alternative to LVCSR is the keyword/filler hidden Markov model (HMM) approach, which was proposed around three decades ago \cite{HMM89, Rose90, Wilpon91}. By this, a keyword HMM and a filler HMM are trained to model keyword and non-keyword audio segments, respectively, as illustrated by Figure \ref{fig:hmm}. Originally, the acoustic features were modeled by means of Gaussian mixture models (GMMs) to produce the state emission likelihoods in keyword/filler HMM-based KWS \cite{HMM89, Rose90, Wilpon91}. Nowadays, similarly to the case of ASR, deep neural networks (DNNs) have replaced GMMs with this purpose \cite{Chen13, Parada14, Ming15, Sankaran16} due to the consistent superior performance of the former. Viterbi decoding \cite{Viterbi} is applied at runtime to find the best path in the decoding graph, and, whenever the likelihood ratio of the keyword model \emph{versus} filler model is larger than a predefined threshold, the KWS system is triggered \cite{Sun17}. While this type of KWS systems is rather compact and good performing, it still needs Viterbi decoding, which, depending on the HMM topology, can be computationally demanding \cite{Parada14, Chai19}.

The arrival of 2014 represented a milestone for KWS technology as a result of the publication of the first deep spoken KWS system \cite{Parada14}. In this paradigm (being new at the time), the sequence of word posterior probabilities yielded by a DNN is directly processed to determine the possible existence of keywords without the intervention of any HMM (see Figure \ref{fig:pipeline}). The deep KWS paradigm has recently attracted much attention \cite{Bai19, Wang19} due to a threefold reason:
\begin{enumerate}
    \item It does not require a complicated sequence search algorithm (i.e., Viterbi decoding); instead, a significantly simpler posterior handling suffices;
    \item The complexity of the DNN producing the posteriors (acoustic model) can be easily adjusted \cite{Shan18, Wang19} to fit the computational resource constraints;
    \item It brings consistent significant improvements over the keyword/filler HMM approach in small-footprint (i.e., low memory and low computational complexity) scenarios in both clean and noisy conditions \cite{Parada14, Jingyong19}.
\end{enumerate}
This threefold reason makes it very appealing to deploy the deep KWS paradigm to a variety of consumer electronics with limited resources like earphones and headphones \cite{Benjamin16}, smartphones, smart speakers and so on. Thus, much research on deep KWS has been conducted since 2014 until today, e.g., \cite{Parada14, Parada15, Parada15c, Sun16, Tang18, Wang19, Alvarez19, Oleg20}. And, what is more, we can expect that deep KWS will continue to be a hot topic in the future despite all the progress made.

In this paper, we present an overview of the deep spoken keyword spotting technology. We believe that this is a good time to look back and analyze the development trajectory of deep KWS to elucidate future challenges. It is worth noticing that only a small number of KWS overview articles is presently available in the literature \cite{OVKWS1,OVKWS2,OVKWS3,OVKWS4}; at best, they shallowly encompass state-of-the-art deep KWS approaches, along with the most relevant datasets. Furthermore, while some relatively recent ASR overview articles covering acoustic modeling ---which is a central part of KWS, see Figure \ref{fig:pipeline}--- can also be found \cite{OVASR1,OVASR2}, still (deep) KWS involves inherent issues, which need to be specifically addressed. \emph{Some} of these inherent issues are related to posterior handling (see Figure \ref{fig:pipeline}), the class-imbalance problem \cite{Hou20}, technology applications, datasets and evaluation metrics. To sum up, we can state that \emph{1)} deep spoken KWS is currently a hot topic\footnote{A proof of this is the organization of events like Auto-KWS 2021 Challenge \cite{AutoKWS}.}, \emph{2)} available KWS overview articles are outdated and/or they offer only a limited treatment of the latest progress, and \emph{3)} deep KWS involves unique inherent issues compared to general-purpose ASR. Thus, this article aims at providing practitioners and researchers who are interested in the topic of keyword spotting with an up to date comprehensive overview of this technology.

The rest of this article is organized as follows: in Section \ref{sec:overview}, the general approach to deep spoken KWS is introduced. Then, in Sections \ref{sec:features}, \ref{sec:modeling} and \ref{sec:posterior}, respectively, the three main components constituting a modern KWS system are analyzed, i.e., speech feature extraction, acoustic modeling and posterior handling. In Section \ref{sec:robustness}, we review current methods to strengthen the robustness of KWS systems against different sources of distortion. Applications of KWS are discussed in Section \ref{sec:applications}. Then, in Section \ref{sec:datasets}, we analyze the speech corpora currently employed for experimentally validating the latest KWS developments. The most important evaluation metrics for KWS are examined in Section \ref{sec:metrics}. In Section \ref{sec:performance}, a comparison among some of the latest deep KWS systems in terms of both KWS performance and computational complexity is presented. Section \ref{sec:avkws} comprises a short review of the literature on audio-visual KWS. Finally, concluding remarks and comments about the future directions in the field are given in Section \ref{sec:conclusions}.

\section{Deep Spoken Keyword Spotting Approach}
\label{sec:overview}

\begin{figure*}
\begin{picture}(100,100)
\put(0,0){\includegraphics[width=\linewidth]{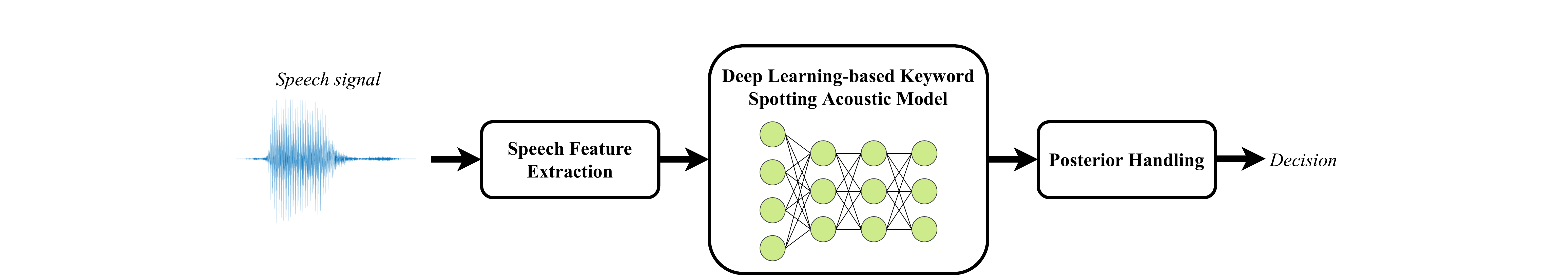}}
\put(120,52){\small $x(m)$}
\put(209,52){\small $\mathbf{X}_{\{i\}}$}
\put(322,52){\small $\mathbf{y}^{\{i\}}$}
\put(262,82){\small $\mathbf{f}(\cdot|\theta)$}
\end{picture}
\caption{General pipeline of a modern deep spoken keyword spotting system: \emph{1)} features are extracted from the speech signal, \emph{2)} a DNN acoustic model uses these features to produce posteriors over the different keyword and filler (non-keyword) classes, and \emph{3)} the temporal sequence of these posteriors is processed (Posterior Handling) to determine the possible existence of keywords.}
\label{fig:pipeline}
\end{figure*}
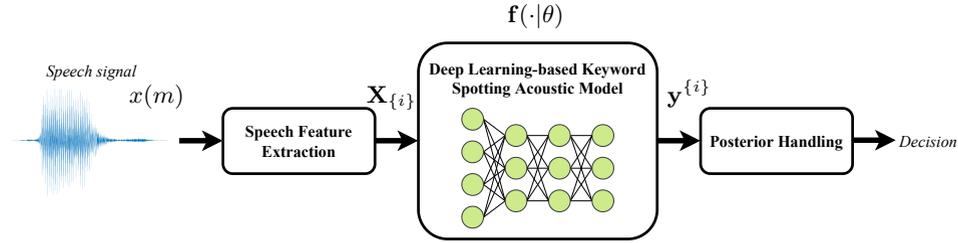

\begin{figure}
\begin{picture}(100,100)
\put(0,0){\includegraphics[width=\linewidth]{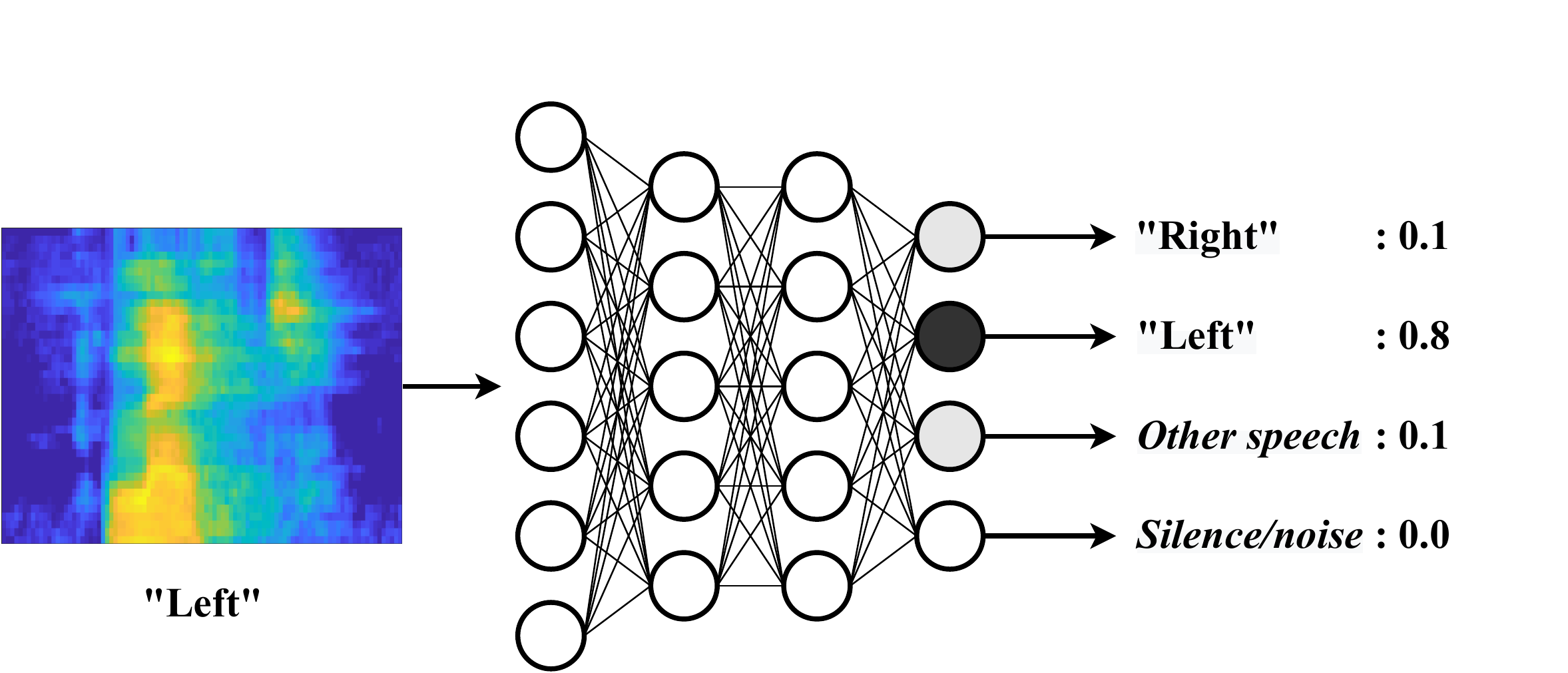}}
\put(25,75){\small $\mathbf{X}_{\{i\}}$}
\put(105,85){\small $\mathbf{f}(\cdot|\theta)$}
\put(215,80){\small $\mathbf{y}^{\{i\}}$}
\end{picture}
\caption{Illustrative example on how a DNN acoustic model performs. There are $N=4$ different classes representing the keywords ``right'' and ``left'', \emph{other speech} and \emph{silence/noise}. The acoustic model receives a speech segment $\mathbf{X}_{\{i\}}$ (log-Mel spectrogram) comprising the keyword ``left''. The DNN produces a posterior distribution over the $N=4$ different classes. Keyword ``left'' is given the highest posterior probability, 0.8.}
\label{fig:acoustic_model}
\end{figure}
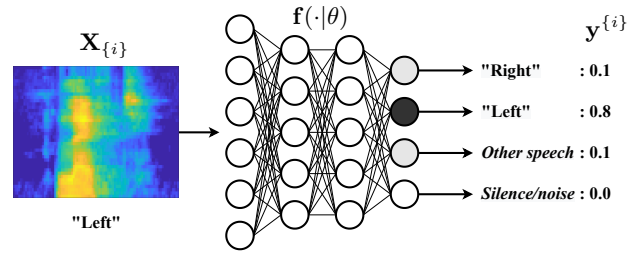

Figure \ref{fig:pipeline} depicts the general pipeline of a modern deep spoken keyword spotting system \cite{Parada14, Parada15, Parada15c, Bruno18, Liu19, Peter20}, which is composed of three main blocks: \emph{1)} the \emph{speech feature extractor} converting the input signal to a compact speech representation, \emph{2)} the \emph{deep learning-based acoustic model} producing posteriors over the different keyword and filler (non-keyword) classes from the speech features (see the example of Figure \ref{fig:acoustic_model}), and \emph{3)} the \emph{posterior handler} processing the temporal sequence of posteriors to determine the possible existence of keywords in the input signal.

Let $x(m)$ be a finite acoustic time signal comprising speech. In the first place, the speech feature extractor computes an alternative representation of $x(m)$, namely, $\mathbf{X}$. It is desirable $\mathbf{X}$ to be \emph{compact} (i.e., lower-dimensional, to limit the computational complexity of the task), \emph{discriminative} in terms of the phonetic content and \emph{robust} to acoustic variations \cite{Lopez17}. Speech features $\mathbf{X}$ are traditionally represented by a two-dimensional matrix composed of a time sequence of $K$-dimensional feature vectors $\mathbf{x}_t$ ($t=0,...,T-1$) as in
\begin{equation}
    \mathbf{X} = \left(\mathbf{x}_0,...,\mathbf{x}_t,...,\mathbf{x}_{T-1}\right)\in\mathbb{R}^{K\times T},
\end{equation}
where $T$, the total number of feature vectors, depends on the length of the signal $x(m)$. Speech features $\mathbf{X}$ can be based on a diversity of representation types, such as, e.g., spectral \cite{Parada14, Parada15c, Xiong19}, cepstral \cite{Fernandez07, Bai19} and time-domain ones \cite{Emad19}. Further details about the different types of speech features used for KWS are provided in Section \ref{sec:features}.

The DNN acoustic model receives $\mathbf{X}$ as input and outputs a sequence of posterior probabilities over the different keyword and non-keyword classes. Particularly, the acoustic model sequentially consumes time segments
\begin{equation}
    \mathbf{X}_{\{i\}}=\left(\mathbf{x}_{is-P},...,\mathbf{x}_{is},...,\mathbf{x}_{is+F}\right)
    \label{eq:feat_chunks}
\end{equation}
of $\mathbf{X}$ until the whole feature sequence $\mathbf{X}$ is processed. In Eq. (\ref{eq:feat_chunks}), $i=\ceil{\frac{P}{s}},...,\floor{\frac{T-1-F}{s}}$ is an integer segment index and $s$ represents the time frame shift. Moreover, $P$ and $F$ denote, respectively, the number of past and future frames (temporal context) in each segment $\mathbf{X}_{\{i\}}\in\mathbb{R}^{K\times(P+F+1)}$. While $s$ is typically designed to have some degree of overlap between consecutive segments $\mathbf{X}_{\{i\}}$ and $\mathbf{X}_{\{i+1\}}$, many works consider acoustic models classifying non-overlapping segments that are sufficiently long (e.g., one second) to cover an entire keyword \cite{Tang18, Bai19, Zeng19, Chen19, Choi19, Menglong20, Ximin20, Emre20}. With regard to $P$ and $F$, a number of approaches considers $F<P$ to reduce latency without significantly sacrificing performance \cite{Parada14, Parada15c, Bruno18, Chai19}. In addition, voice activity detection \cite{Tan20} is sometimes used to reduce power consumption by only inputting to the acoustic model segments $\mathbf{X}_{\{i\}}$ in which voice is present \cite{Parada14, Guoguo15, Loren18, Yuan19, Yang20}.

Then, let us suppose that the DNN acoustic model $\mathbf{f}(\cdot|\theta):\mathbb{R}^{K\times (P+F+1)}\rightarrow I^N$ has $N$ output nodes meaning $N$ different classes, where $\theta$ and $I=[0,\;1]$ denote the parameters of the acoustic model and the unit interval, respectively. Normally, the output nodes represent either words \cite{Parada14, Parada15c, Tang18, Bruno18, Samuel18, Bai19, Zeng19, Chen19, Choi19, Chai19, Peter20, Emre20, Menglong20, Ximin20, Higuchi20, Yang20} or subword units like context-independent phonemes \cite{He17, Alvarez19, Xuan19, Sharma20}, the latter especially in the context of sequence-to-sequence models \cite{Graves06, Graves12, Graves13} (see Subsection \ref{ssec:rnns} for further details). Let subscript $n$ refer to the $n$-th element of a vector. For every input segment $\mathbf{X}_{\{i\}}$, the acoustic model yields
\begin{equation}
    \mathbf{y}_n^{\{i\}}=\mathbf{f}_n\left.\left(\mathbf{X}_{\{i\}}\right|\theta\right),\;\;\;n=1,...,N,
\end{equation}
where $\mathbf{y}_n^{\{i\}}=P\left(\left.C_n\right|\mathbf{X}_{\{i\}},\theta\right)$ is the posterior of the $n$-th class $C_n$ given the input feature segment $\mathbf{X}_{\{i\}}$. To ensure that $\sum_{n=1}^N\mathbf{y}_n^{\{i\}}=1\;\;\forall i$, deep KWS systems commonly employ a fully-connected layer with softmax activation \cite{BishopPR} as an output layer, e.g., \cite{He17, Du18, Emad19, Lee19, Riviello19, Bai19, Peter20, Ximin20, Simon20, Park20, Wu20}. The parameters of the model, $\theta$, are usually estimated by discriminatively training $\mathbf{f}(\cdot|\theta)$ by backpropagation from annotated speech data characterizing the different $N$ classes. The most popular loss function that is employed to this end is cross-entropy loss \cite{Bridle90, GoodfellowDL}.

Figure \ref{fig:acoustic_model} shows an example, illustrating the above paragraph, in which there are $N=4$ different classes. Two of these classes represent the keywords ``right'' ($C_1$) and ``left'' ($C_2$). The other two classes are the filler classes \emph{other speech} ($C_3$) and \emph{silence/noise} ($C_4$). A segment $\mathbf{X}_{\{i\}}$ consisting of a log-Mel spectrogram comprising the keyword ``left'' is input to the DNN acoustic model. Then, this generates a posterior distribution $\mathbf{y}^{\{i\}}$ over the $N=4$ classes. Keyword ``left'' is given the highest posterior probability, namely, $\mathbf{y}_2^{\{i\}}=P\left(\left.C_2\right|\mathbf{X}_{\{i\}},\theta\right)=0.8$.

Most of the research that has been conducted on deep KWS has focused on its key part, which is the design of increasingly accurate and decreasingly computationally complex acoustic models $\mathbf{f}(\cdot|\theta)$ \cite{Yundong18, Oleg20}.

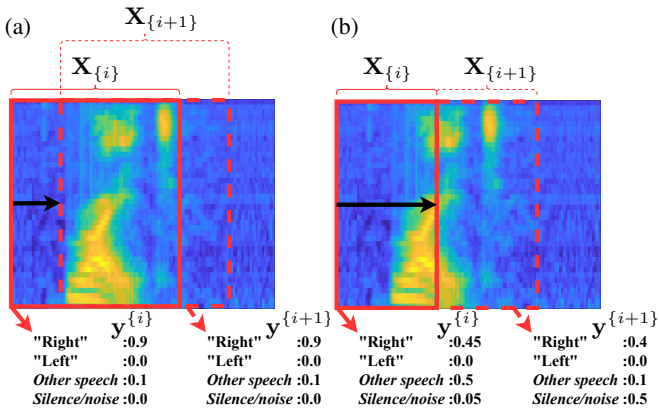
\begin{figure}
\begin{picture}(100,160)
\put(0,0){\includegraphics[width=0.5\linewidth]{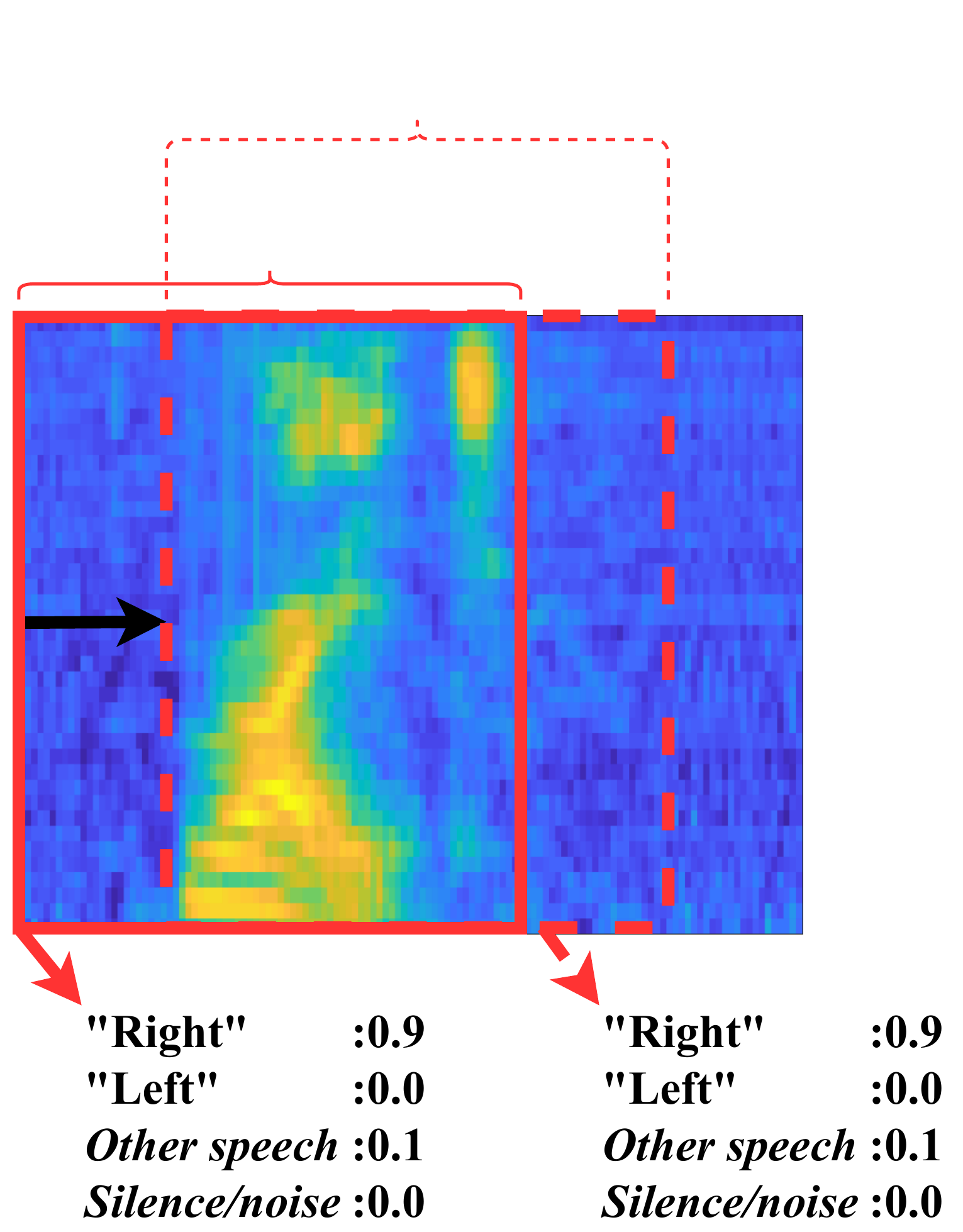}}
\put(0,140){\small (a)}
\put(25,125){\small $\mathbf{X}_{\{i\}}$}
\put(45,144){\small $\mathbf{X}_{\{i+1\}}$}
\put(40,28){\small $\mathbf{y}^{\{i\}}$}
\put(98,28){\small $\mathbf{y}^{\{i+1\}}$}
\end{picture}\hspace{0.7cm}
\begin{picture}(100,160)
\put(0,0){\includegraphics[width=0.5\linewidth]{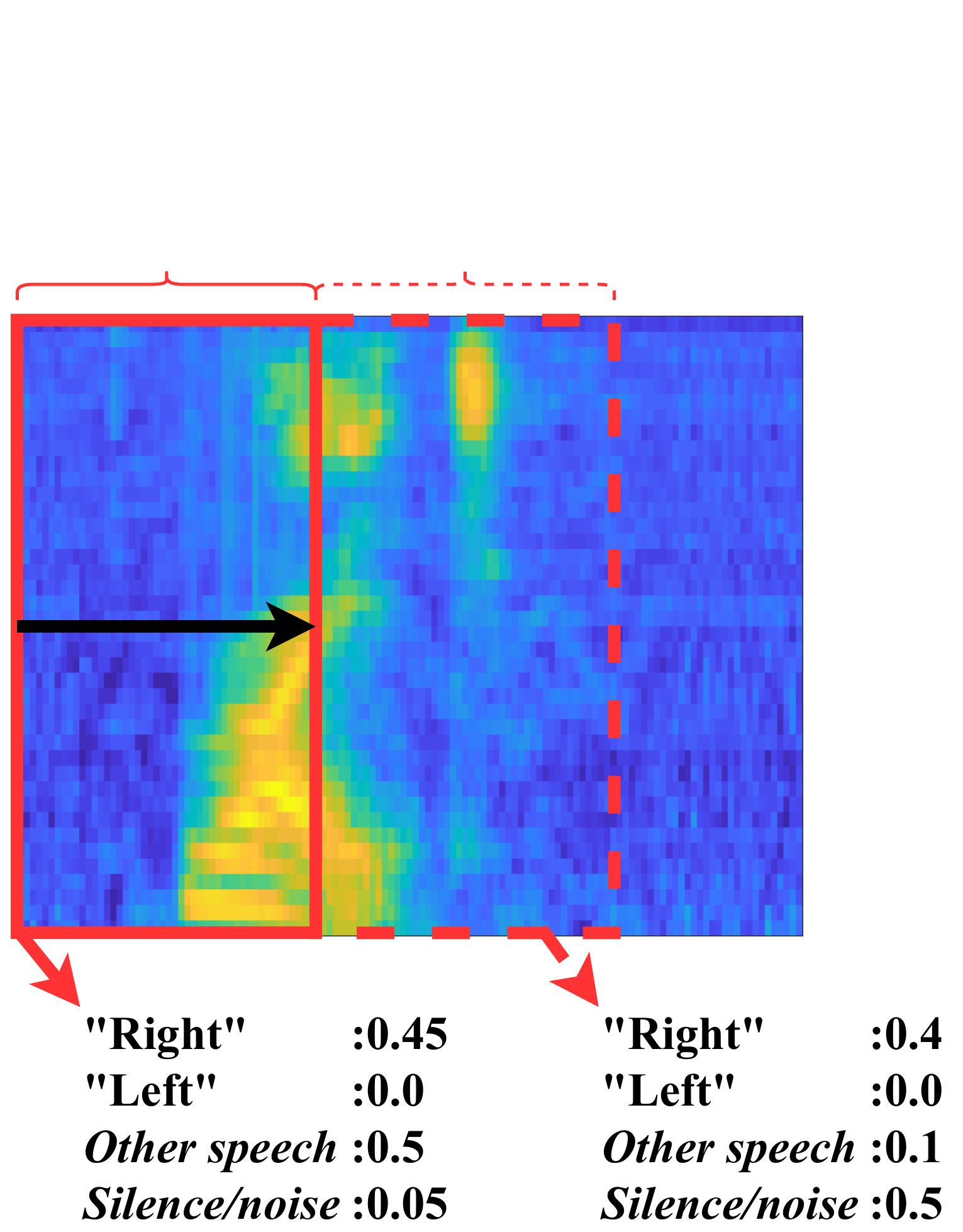}}
\put(0,140){\small (b)}
\put(12,125){\small $\mathbf{X}_{\{i\}}$}
\put(50,125){\small $\mathbf{X}_{\{i+1\}}$}
\put(40,28){\small $\mathbf{y}^{\{i\}}$}
\put(98,28){\small $\mathbf{y}^{\{i+1\}}$}
\end{picture}\\
\caption{Example of the processing of two consecutive feature segments $\mathbf{X}_{\{i\}}$ and $\mathbf{X}_{\{i+1\}}$, from $\mathbf{X}$ comprising the keyword ``right'', by a DNN acoustic model: (a) when using an overlapping segmentation window, and (b) when using a smaller, non-overlapping one.}
\label{fig:posteriors}
\end{figure}

Finally, KWS is not a static task but a dynamic one in which the KWS system has to continuously listen to the input signal $x(m)$ to yield the sequence of posteriors $\mathbf{y}^{\{i\}}$, $i=\ceil{\frac{P}{s}},...,\floor{\frac{T-1-F}{s}}$, in order to detect keywords in real-time. In the example in Figure \ref{fig:acoustic_model}, a straightforward way to do this could just be choosing the class $\hat{C}^{\{i\}}$ with the highest posterior, that is,
\begin{equation}
    \hat{C}^{\{i\}}=\argmax_{C_n}\mathbf{y}_n^{\{i\}}=\argmax_{C_n}P\left(\left.C_n\right|\mathbf{X}_{\{i\}},\theta\right).
    \label{eq:argmax}
\end{equation}
Nevertheless, this approach is not robust, as discussed in what follows. Continuing with the illustration of Figure \ref{fig:acoustic_model}, Figure \ref{fig:posteriors} exemplifies the processing by the acoustic model of two consecutive feature segments $\mathbf{X}_{\{i\}}$ and $\mathbf{X}_{\{i+1\}}$ from $\mathbf{X}$ comprising the keyword ``right''. Figure \ref{fig:posteriors}a shows the typical case of using an overlapping segmentation window. As we can see, following the approach of Eq. (\ref{eq:argmax}) might lead to detecting the same keyword realization twice, yielding a false alarm. In addition, Figure \ref{fig:posteriors}b depicts the case in which a non-overlapping segmentation window is employed. In this situation, the energy of the keyword realization leaks into two different segments in such a manner that neither the posterior $P\left(\left.C_1\right|\mathbf{X}_{\{i\}},\theta\right)$ nor $P\left(\left.C_1\right|\mathbf{X}_{\{i+1\}},\theta\right)$ is sufficiently strong for the keyword to be detected, thereby yielding a miss detection. Hence, a proper handling of the sequence of posteriors $\mathbf{y}^{\{i\}}$ ($i=\ceil{\frac{P}{s}},...,\floor{\frac{T-1-F}{s}}$) is a very important component for effective keyword detection \cite{Fernandez07, Parada14, Parada15, Sun16, Yimeng16, Assaf17, Kumar18, Bruno18, Yue19, Alice19, Liu19, Yuan19, Xiong19, Peter20, Yixin20}. Posterior handling is examined in Section \ref{sec:posterior}.

\section{Speech Feature Extraction}
\label{sec:features}

In the following subsections, we walk through the most relevant speech features revolving around deep KWS: Mel-scale-related features, recurrent neural network features, low-precision features, learnable filterbank features and other features.

\subsection{Mel-Scale-Related Features}
\label{ssec:mel_feats}

\begin{figure}
\includegraphics[width=\linewidth]{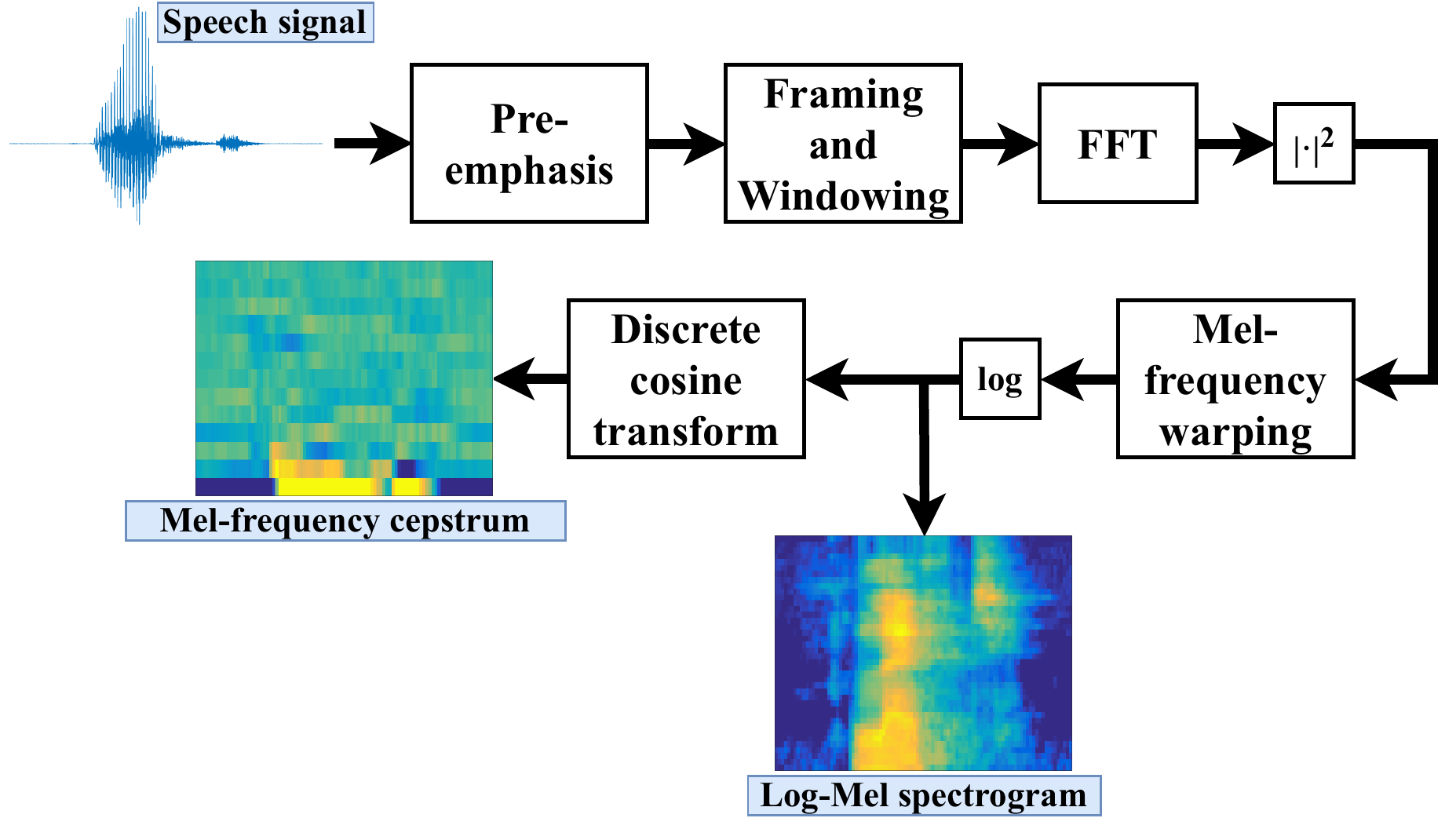}
\caption{Classical pipeline for extracting log-Mel spectral and Mel-frequency cepstral speech features using the fast Fourier transform (FFT).}
\label{fig:mfcc_pipeline}
\end{figure}

Speech features based on the perceptually-motivated Mel-scale filterbank \cite{Stevens37}, like the log-Mel spectral coefficients and Mel-frequency cepstral coefficients (MFCCs) \cite{Mermelstein80}, have been widely used over decades in the fields of ASR and, indeed, KWS. Despite the multiple attempts to learn optimal, alternative representations from the speech signal (see Subsection \ref{ssec:raw_feats} for more details), Mel-scale-related features are still nowadays a solid, competitive and safe choice \cite{Lopez21}. Figure \ref{fig:mfcc_pipeline} depicts the well-known classical pipeline for extracting log-Mel spectral and MFCC features. In deep KWS, both types of speech features are commonly normalized to have zero mean and unit standard deviation before being input to the acoustic model, thereby stabilizing and speeding up training as well as improving model generalization \cite{LeCun12}.

Mel-scale-related features are, by far, the most widely used speech features in deep KWS. For example, MFCCs with temporal context and, sometimes, their first- and second-order derivatives are used in \cite{Fernandez07, Martin13, Fuchs17, Raphael18, Tang18, Bai19, Verdegay19, Pattanayak19, Menglong20, Chen20, Emre20, Ximin20, Axel21, Wang21}. As can be seen from Figure \ref{fig:mfcc_pipeline}, MFCCs are obtained from the application of the discrete cosine transform to the log-Mel spectrogram. This transform produces approximately decorrelated features, which are well-suited to, e.g., acoustic models based on GMMs that, for computational efficiency reasons, use diagonal covariance matrices. However, deep learning models are able to exploit spectro-temporal correlations, yielding the use of the log-Mel spectrogram instead of MFCCs equivalent or better ASR and KWS performance \cite{Watanabe17}. As a result, a good number of deep KWS works considers log-Mel or Mel filterbank speech features with temporal context, e.g., \cite{Parada14, Parada15, Parada15c, Sun16, He17, Arik17, Shan18, Loren18, Samuel18, Yiyan18, Xiong19, Arden19, Zeng19, Wang19, Alvarez19, Hanna19, Lee19, Alice19, Wu20, Meng20, Sharma20, Peter20, Park20, Xuan20, Yan20, Peng20, Kim21, Yao21}. In addition, \cite{Yixin20} proposes instead the use of the first derivative of the log-Mel spectrogram to improve robustness against signal gain changes. The number of filterbank channels in the above works ranges from 20 to 128. In spite of this wide channel range, experience suggests that (deep) KWS performance is not significantly sensitive to the value of this parameter as long as the Mel-frequency resolution is not very poor \cite{Lopez21}. This fact could promote the use of a lower number of filterbank channels in order to limit computational complexity.

\subsection{Recurrent Neural Network Features}
\label{ssec:rnn_feats}

Recurrent neural networks (RNNs) are helpful to summarize variable-length data sequences into fixed-length, compact feature vectors, also known as \emph{embeddings}. Due to this fact, RNNs are very suitable for template matching problems like query-by-example (QbE) KWS, which involves keyword detection by determining the similarity between feature vectors (successively computed from the input audio stream) and keyword templates. In, e.g., \cite{Guoguo15, Hou16, Sacchi19, Yuan19, Huang21}, long short-term memory (LSTM) and gated recurrent unit (GRU) neural networks are employed to extract word embeddings. Generally, these are compared, by means of any distance function like cosine similarity \cite{Amit01} and particularly for QbE KWS, with keyword embeddings obtained during an enrollment phase.

While QbE KWS based on RNN feature extraction follows a different approach from that outlined in Section \ref{sec:overview}, which is the main scope of this manuscript, we have considered it pertinent to allude to it for the following twofold reason. First, there is little difference between the general pipeline of Figure \ref{fig:pipeline} and QbE KWS based on RNN feature extraction, since acoustic modeling is implicitly carried out by the RNN\footnote{Actually, in \cite{Guoguo15, Hou16}, the LSTM networks used there are pure acoustic models, and the word embeddings correspond to the activations prior to the output softmax layer.}. Second, QbE KWS based on RNN feature extraction is especially useful for personalized, open-vocabulary KWS, by which a user is allowed to define her/his own keywords by just recording a few keyword samples during an enrollment phase. Alternatively, in \cite{Sacchi19}, a clever RNN mechanism to generate keyword templates from text instead of speech inputs is proposed. Notice that incorporating new keywords in the context of the deep spoken KWS approach introduced in Section \ref{sec:overview} might require system re-training, which is not always feasible.

QbE KWS based on RNN feature extraction has shown to be more efficient and better performing than classical QbE KWS approaches based on LVCSR \cite{Parada09} and dynamic time warping (DTW) \cite{Levin13}. Therefore, the RNN feature approach is a good choice for on-device KWS applications providing keyword personalization.

\subsection{Low-Precision Features}
\label{ssec:low_feats}

A way to diminish the energy consumption and memory footprint of deep KWS systems to be run on resource-constrained devices consists, e.g., of quantization ---i.e., precision reduction--- of the acoustic model parameters. Research like \cite{Yuriy19, Bo19} has demonstrated that it is possible to (closely) achieve the accuracy provided by full-precision acoustic models while drastically decreasing memory footprint by means of 4-bit quantization of model's weights.

The same philosophy can be applied to speech features. Emerging research \cite{Riviello19} studies two kinds of low-precision speech representations: linearly-quantized log-Mel spectrogram and power variation over time, derived from log-Mel spectrogram, represented by only 2 bits. Experimental results show that using 8-bit log-Mel spectra yields same KWS accuracy as employing full-precision MFCCs. Furthermore, KWS performance degradation is insignificant when exploiting 2-bit precision speech features. As the authors of \cite{Riviello19} state, this fact might indicate that much of the spectral information is superfluous when attempting to spot a set of keywords. In \cite{Lopez21}, we independently arrived at the same finding. In conclusion, there appears to be a large room for future work on the design of new extremely-light and compact (from a computational point of view) speech features for small-footprint KWS (see also the next subsection).

\subsection{Learnable Filterbank Features}
\label{ssec:raw_feats}

The development of end-to-end deep learning systems in which feature extraction is optimal in line with the task and training criterion is a recent trend (e.g., \cite{Jung18, Hannah18}). This approach aspires to become an alternative to the use of well-established handcrafted features like log-Mel features and MFCCs, which are preferred for many speech-related tasks, including deep KWS (see Subsection \ref{ssec:mel_feats}).

Optimal filterbank learning is part of such an end-to-end training strategy, and it has been explored for deep KWS in \cite{Simon20, Lopez21}. In this context, filterbank parameters are tuned towards optimizing word posterior generation. Particularly, in \cite{Simon20}, the acoustic model parameters are optimized jointly with the cut-off frequencies of a filterbank based on sinc-convolutions (SincConv) \cite{Ravanelli18}. Similarly, in \cite{Lopez21}, we studied two filterbank learning approaches: one consisting of filterbank matrix learning in the power spectral domain and another based on parameter learning of a psychoacoustically-motivated gammachirp filterbank \cite{Irino99}. While the use of SincConv is not compared with using handcrafted speech features in \cite{Simon20}, in \cite{Lopez21}, we found no statistically significant KWS accuracy differences between employing a learned filterbank and log-Mel features. This finding is in line with research on filterbank learning for ASR, e.g., \cite{Sainath15, Seki17, Neil18}. In \cite{Lopez21}, it is hypothesized that such a finding might be an indication of information redundancy\footnote{With a sufficiently powerful DNN acoustic model, the actual input feature representation is of less importance (as long as it represents the relevant information about the input signal).}. As suggested in Subsection \ref{ssec:low_feats}, this should encourage research on extremely-light and compact speech features for small-footprint KWS. In conclusion, handcrafted speech features currently provide state-of-the-art KWS performance at the same time that optimal feature learning requires further research to become the preferred alternative.

\subsection{Other Speech Features}
\label{ssec:other_feats}

A small number of works has explored the use of alternative speech features with a relatively low computational impact. For example, \cite{Emad19} introduced the so-called multi-frame shifted time similarity (MFSTS). MFSTS are time-domain features consisting of a two-dimensional speech representation comprised of constrained-lag autocorrelation values. Despite their computational simplicity, which can make them attractive for low-power KWS applications, features like MFCCs provide much better KWS accuracy \cite{Emad19}.

A more interesting approach is that examined by \cite{Ravi18, Erika19}, which fuses two different KWS paradigms: DTW and deep KWS. First, a DTW warping matrix measuring the similarity between an input speech utterance and the keyword template is calculated. From the deep KWS perspective, this matrix can be understood as speech features that are input to a deep learning binary (i.e., keyword/non-keyword) classifier playing the role of an ``acoustic model''. This hybrid approach brings the best of both worlds: \emph{1)} the powerful modeling capabilities of deep KWS, and \emph{2)} the flexibility of DTW KWS to deal with both open-vocabulary and language-independent scenarios. In spite of its potentials, further research on this methodology is needed, since, e.g., it is prone to overfitting \cite{Erika19}.

\section{Acoustic Modeling}
\label{sec:modeling}

This section is devoted to review the core of deep spoken KWS systems: the acoustic model. The natural trend is the design of increasingly accurate models while decreasing computational complexity. In an approximate chronological order, Subsections \ref{ssec:ffnns}, \ref{ssec:cnns} and \ref{ssec:rnns} review advances in acoustic modeling based on fully-connected feedforward networks, convolutional networks, and recurrent and time-delay neural networks, respectively. Finally, Subsection \ref{ssec:training} is dedicated to how these acoustic models are trained.

\subsection{Fully-Connected Feedforward Neural Networks}
\label{ssec:ffnns}

Deep spoken KWS made its debut in 2014 \cite{Parada14} employing acoustic modeling based on the most widespread type of neural architecture at the time: the fully-connected feedforward neural network (FFNN). A simple stack of three fully-connected hidden layers with 128 neurons each and rectified linear unit (ReLU) activations, followed by a softmax output layer, greatly outperformed, with fewer parameters, a (at that time) state-of-the-art keyword/filler HMM system in both clean and noisy acoustic conditions. However, since the constant goal is the design of more accurate/robust and computationally lighter acoustic models, the use of fully-connected FFNNs was quickly relegated to a secondary level. Nowadays, state-of-the-art acoustic models use convolutional and recurrent neural networks (see Subsections \ref{ssec:cnns} and \ref{ssec:rnns}), since they can provide better performance with fewer parameters, e.g., \cite{Parada15c, Shan18}. Even so, standard FFNN acoustic models and variants of them\footnote{For example, in \cite{Oleg20}, it is evaluated an FFNN acoustic model integrating an intermediate pooling layer, which yields improved KWS accuracy in comparison with a standard FFNN using a similar number of parameters.} are considered in recent literature for either comparison purposes or studying different aspects of KWS such as training loss functions, e.g., \cite{Shan18, Liu19, Yuan19, Jingyong19}.

Closely related and computationally cheaper alternatives to fully-connected FFNNs are single value decomposition filter (SVDF) \cite{Nakkiran15, Alvarez19, Park20} and spiking neural networks \cite{Mostafa18, Bruno18, Emre20}. Proposed in \cite{Nakkiran15} to approximate fully-connected layers by low-rank approximations, SVDF achieved to reduce by 75\% the FFNN acoustic model size of the first deep KWS system \cite{Parada14} with no drop in performance. A similar idea was explored in \cite{George16}, where a high degree of acoustic model compression is accomplished by means of low-rank weight matrices. The other side of the same coin is that modeling power can be enhanced by increasing the number of neurons while keeping the original number of multiplications fixed \cite{George16}. In this way, the performance of the first deep KWS system \cite{Parada14} was improved without substantially altering the computational resource usage of the algorithm. Higuchi \emph{et al.} \cite{Higuchi20} have shown that an SVDF neural network is a special case of a stacked one-dimensional convolutional neural network (CNN), so the former can be easily implemented as the latter.

On the other hand, spiking neural networks (SNNs) are human brain-inspired neural networks that, in contrast to artificial neural networks (ANNs), process the information in an event-driven manner, which greatly alleviates the computational load when such information is sparse as in KWS \cite{Mostafa18, Bruno18, Emre20}. To make them work, in the first place, real-valued input data like speech features have to be transformed to a sequence of spikes encoding real values in either its frequency (spike rate) or the relative time between spikes. Then, spikes propagate throughout the SNN to eventually fire the corresponding output neurons, which represent word classes in KWS \cite{Bruno18}. SNNs can yield a similar KWS performance to that of equivalent ANNs while providing a computational cost reduction and energy saving above 80\% \cite{Bruno18} and of dozens of times \cite{Emre20}, respectively. Apart from having been applied to fully-connected FFNNs for KWS \cite{Bruno18, Emre20}, the SNN paradigm has also been recently applied to CNN acoustic modeling \cite{Emre20}, which is reviewed in the next subsection.

\subsection{Convolutional Neural Networks}
\label{ssec:cnns}

Moving from fully-connected FFNN to CNN acoustic modeling was a natural step taken back in 2015 \cite{Parada15c}. Thanks to exploiting local speech time-frequency correlations, CNNs are able to outperform, with fewer parameters, fully-connected FFNNs for acoustic modeling in deep KWS \cite{Parada15c, Ravi18, Raphael18, Yiteng18, Menon18, Liu20, Meng20, Mo20, Oleg20, Wu20}. One of the attractive features of CNNs is that the number of multiplications of the model can be easily limited to meet the computational constraints by adjusting different hyperparameters like, e.g., filter striding, and kernel and pooling sizes. Moreover, this may be done without necessarily sacrificing much performance \cite{Parada15c}.

\begin{figure}
    \centering
    \includegraphics[width=0.7\linewidth]{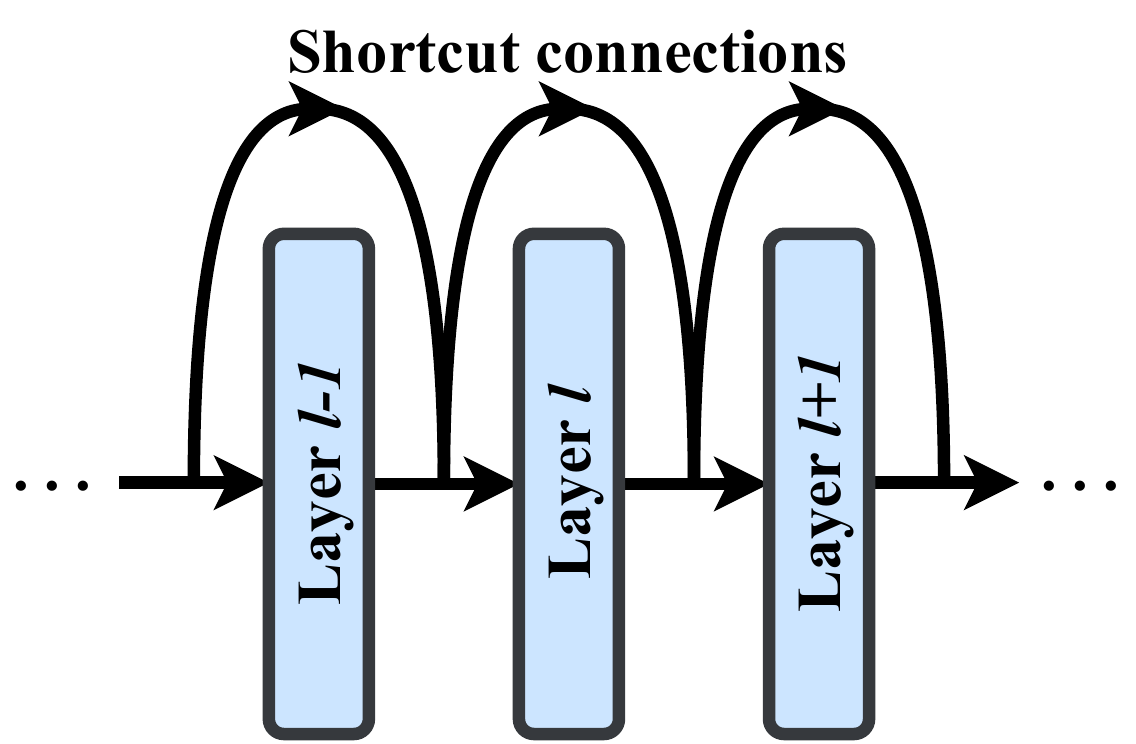}
    \caption{Example of shortcut connections linking non-consecutive layers in residual learning models.}
    \label{fig:resnet}
\end{figure}

Residual learning, proposed by He \emph{et al.} \cite{He16} for image recognition, is widely considered to implement state-of-the-art acoustic models for deep KWS \cite{Tang18, Du18, Alice19, Riviello19, Choi19, Oleg20, Ximin20, Yang20, Menglong20}. In short, residual learning models are constructed by introducing a series of shortcut connections linking non-consecutive layers (as exemplified by Figure \ref{fig:resnet}), which helps to better train very deep CNN models. To the best of our knowledge, Tang and Lin \cite{Tang18} were the first authors exploring deep residual learning for deep KWS. They also integrated dilated convolutions increasing the network's receptive field in order to capture longer time-frequency patterns\footnote{In \cite{Chen19}, the authors achieve this same effect by means of graph convolutional networks \cite{Kipf17}.} without increasing the number of parameters, as also done by a number of subsequent deep KWS systems, e.g., \cite{Emad19, Alice19, Menglong20}. In this way, Tang and Lin greatly outperformed, with less parameters, standard CNNs \cite{Parada15c} in terms of KWS performance, establishing a new state-of-the-art back in 2018. Their powerful deep residual architecture so-called $\texttt{res15}$ has been employed to carry out different KWS studies in areas like robustness for hearing assistive devices \cite{Lopez19, Lopez20}, filterbank learning \cite{Lopez21}, and robustness to acoustic noise \cite{Lopez21b}, among others.

Largely motivated by this success, later work further explored the use of deep residual learning. For example, \cite{Du18} uses a variant of DenseNet \cite{DenseNet}, which can be interpreted as an extreme case of residual network comprising a hive of skip connections and requiring fewer parameters. The use of an acoustic model inspired by WaveNet \cite{Oord16}, involving both skip connections and gated activation units, is evaluated in \cite{Alice19}. Choi \emph{et al.} \cite{Choi19} proposed utilizing one-dimensional convolutions along the time axis (\emph{temporal convolutions}) while treating the (MFCC) features as input channels within a deep residual learning framework (TC-ResNet). This approach could help to overcome the challenge of simultaneously capturing both high and low frequency features by means of not very deep networks ---although we think that this can also be accomplished, to a great extent, by two-dimensional dilated convolutions increasing the network's receptive field---. The proposed temporal convolution yields a significant reduction of the computational burden with respect to a two-dimensional convolution with the same number of parameters. As a result, TC-ResNet matches Tang and Lin's \cite{Tang18} KWS performance while dramatically decreasing both latency and the amount of floating-point operations per second on a mobile device \cite{Choi19}. In \cite{Oleg20}, where an interesting deep KWS system comparison is presented, TC-ResNet, exhibiting one of the least latency and model sizes, is top-ranked in terms of KWS performance, outperforming competitive acoustic models based on standard CNNs, convolutional recurrent neural networks (CRNNs) \cite{Yundong18}, and RNNs with an attention mechanism \cite{Coimbra18} (see also the next subsection), among others. Furthermore, very recently, Zhou \emph{et al.} \cite{Hang21} adopted a technique so-called AdderNet \cite{Hanting20} to replace multiplications by additions in TC-ResNet, thereby drastically reducing its power consumption while maintaining a competitive accuracy.

Another appealing way to reduce the computation and size of standard CNNs is by depthwise separable convolutions \cite{Andrew17}. They work by factorizing a standard convolution into a depthwise one and a pointwise ($1\times 1$) convolution combining the outputs from the depthwise one to generate new feature maps \cite{Andrew17}. Depthwise separable CNNs (DS-CNNs) are a good choice to implement well-performing acoustic models in embedded systems \cite{Xiong19, Peter20}. For example, the authors of \cite{Simon20} are able to reproduce the outstanding performance of TC-ResNet \cite{Choi19} using less parameters thanks to exploiting depthwise separable convolutions. Furthermore, the combination of depthwise separable convolutions with residual learning has been recently explored for deep KWS acoustic modeling \cite{Ximin20, Menglong20, Yang20, Kim21}, generally outperforming all standard residual networks \cite{Tang18}, plain DS-CNNs and TC-ResNet with less computational complexity.

Upon this review, we believe that a modern CNN-based acoustic model should ideally encompass the following three aspects:
\begin{enumerate}
\item A mechanism to exploit long time-frequency dependencies like, e.g., the use of temporal convolutions \cite{Choi19} or dilated convolutions.
\item Depthwise separable convolutions \cite{Andrew17} to substantially reduce both the memory footprint and computation of the model without sacrificing the performance.
\item Residual connections \cite{He16} to fast and effectively train deeper models providing enhanced KWS performance.
\end{enumerate}

\subsection{Recurrent and Time-Delay Neural Networks}
\label{ssec:rnns}

Speech is a temporal sequence with strong time dependencies. Therefore, the utilization of RNNs for acoustic modeling ---and also time-delay neural networks (TDNNs), which are shaped by a set of layers performing on different time scales--- naturally arises. For example, LSTM neural networks \cite{Sepp97}, which overcome the exploding and vanishing gradient problems suffered by standard RNNs, are used for KWS acoustic modeling in, e.g., \cite{Martin13,Yimeng16,Sun16,Kumar18,Alice19}, clearly outperforming FFNNs \cite{Sun16}. When latency is not a strong constraint, bidirectional LSTMs (BiLSTMs) can be used instead to capture both causal and anticausal dependencies for improved KWS performance \cite{Sundar15,Kumar18}. Alternatively, bidirectional GRUs are explored in \cite{Oleg20} for KWS acoustic modeling. When there is no need to model very long time dependencies, as it is the case in KWS, GRUs might be preferred over LSTMs since the former demand less memory and are faster to train while performing similarly or even better \cite{Arik17}.

Besides, \cite{Samuel18} studies a two-stage TDNN consisting of an LVCSR acoustic model followed by a keyword classifier. The authors of \cite{Samuel18} also investigate the integration of frame skipping and caching to decrease computation, thereby outperforming classical CNN acoustic modeling \cite{Parada15c} while halving the number of multiplications.

As we already suggested in Subsection \ref{ssec:cnns}, CNNs might have difficulties to model long time dependencies. To overcome this point, they can be combined with RNNs to build the so-called CRNNs. Thus, it may be stated that CRNNs bring the best of two worlds: first, convolutional layers model local spectro-temporal correlations of speech and, then, recurrent layers follow suit by modeling long-term time dependencies in the speech signal. Some works explore the use of CRNNs for acoustic modeling in deep spoken KWS using either unidirectional or bidirectional LSTMs or GRUs \cite{Arik17,Kumar18,Bo19,Erika19,Zeng19,Oleg20}. Generally, the use of CRNNs allows us for outperforming standalone CNNs and RNNs \cite{Zeng19}.

\subsubsection{Connectionist Temporal Classification}
\label{sssec:ctc}

As for the majority of acoustic models, the above-reviewed RNN acoustic models are typically trained to produce frame-level posterior probabilities. At training time, in case of employing, e.g., cross-entropy loss, frame-level annotated data are required, which may be cumbersome to get. In the context of RNN acoustic modeling, connectionist temporal classification (CTC) \cite{Graves06} is an attractive alternative letting the model unsupervisedly locate and align the phonetic unit labels at training time \cite{Yimeng16}. In other words, frame-level alignments of the target label sequences are not required for training.

Mathematically speaking, let $\mathbf{C}=\left(c_0,...,c_{m-1}\right)$ be the sequence of phonetic units or, e.g., characters corresponding to the sequence of feature vectors $\mathbf{X}=\left(\mathbf{x}_0,...,\mathbf{x}_{T-1}\right)$, where $m<T$ and the accurate alignment between $\mathbf{C}$ and $\mathbf{X}$ is unknown. CTC is an \emph{alignment-free} algorithm whose goal is to maximize \cite{Graves06}
\begin{equation}
    P\left(\mathbf{C}|\mathbf{X}\right)=\sum_{A\in\mathcal{A}_{X,C}}\prod_{t=0}^{T-1} P_t\left(\mathbf{c}|\mathbf{x}_0,...,\mathbf{x}_t\right),
    \label{eq:ctc}
\end{equation}
where $\mathbf{c}$ is the whole set of recognizable phonetic units or characters plus a blank symbol (modeling confusion information of the speech signal \cite{Yimeng16}), and the summation is performed over the set of all valid alignments $\mathcal{A}_{X,C}$. From Eq. (\ref{eq:ctc}), the acoustic model outputs can be understood as the probability distribution over all the possible label sequences given the sequence of input features $\mathbf{X}$ \cite{Fernandez07}.

The very first attempt to apply CTC to KWS was carried out by Fern\'andez \emph{et al.} \cite{Fernandez07} using a BiLSTM for acoustic modeling. At training time, this system just needs, along with the training speech signals, the list of training words in order of occurrence. After this first attempt, several works have explored variants of this approach using different RNN architectures like LSTMs \cite{Bai16,Yimeng16,He17,Xuan19}, BiLSTMs \cite{Martin13,Yan20} and GRUs \cite{Enea19,Xuan19}, as well as considering different phonetic units such as phonemes \cite{Martin13,He17} and Mandarin syllables \cite{Bai16,Yiyan18}. In general, these systems are shown to be superior to both LVCSR- and keyword/filler HMM-based KWS systems with less or no additional computational cost \cite{Bai16,Yimeng16,Yiyan18}. Notice that since CTC requires searching for the keyword phonetic unit sequence on a lattice, this approach is also suitable for open-vocabulary KWS.

\subsubsection{Sequence-to-Sequence Models}
\label{sssec:seq2seq}

\begin{figure}
    \centering
    \includegraphics[width=\linewidth]{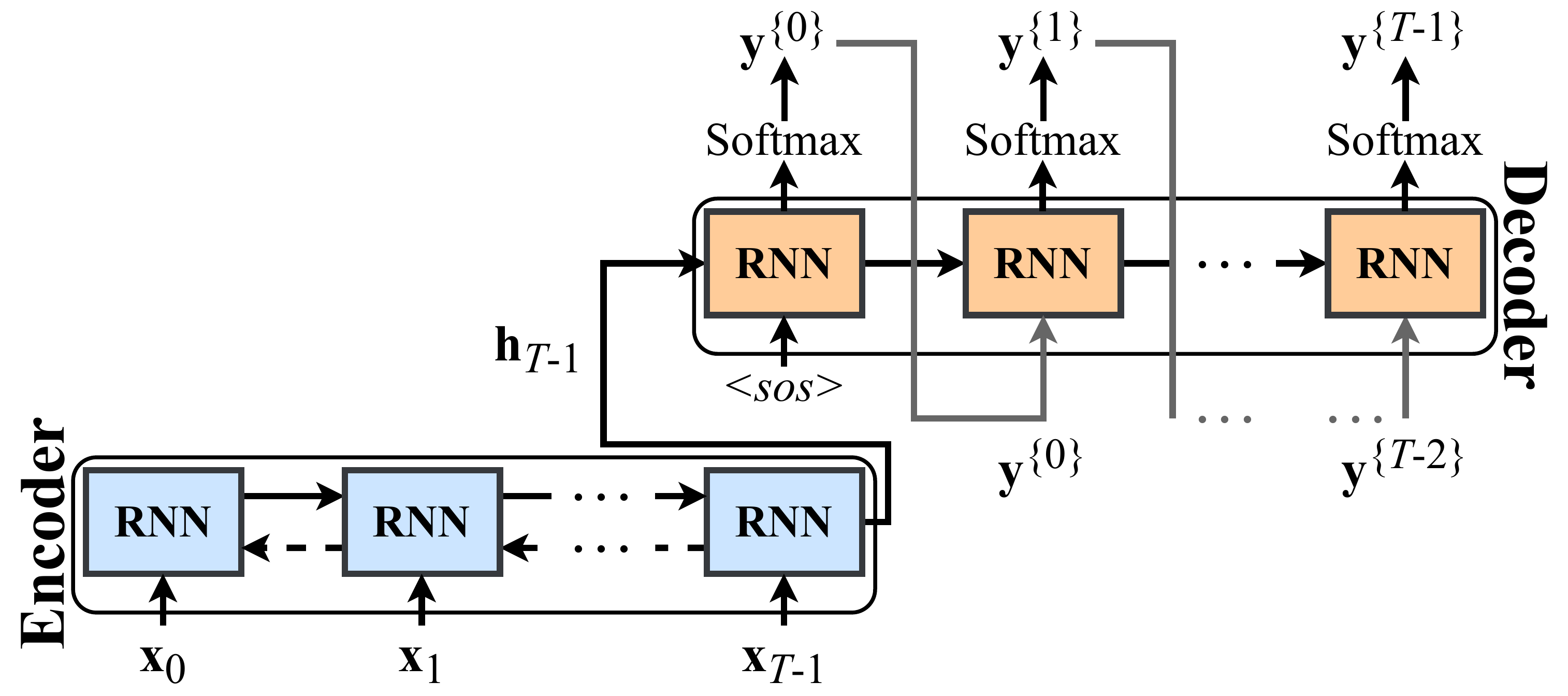}
    \caption{Example of sequence-to-sequence (Seq2Seq) model. Here, ``\emph{<sos>}'' stands for ``start of sequence''. See the text for further details.}
    \label{fig:seq2seq}
\end{figure}

CTC assumes conditional label independence, i.e., past model outputs do not influence current predictions (see Eq. (\ref{eq:ctc})). Hence, in the context of KWS and ASR in general, CTC may need an external language model to perform well. Therefore, a more convenient approach for KWS acoustic modeling might be the use of sequence-to-sequence (Seq2Seq) models, first proposed in \cite{Ilya14} for language translation. Figure \ref{fig:seq2seq} illustrates an example of Seq2Seq model. In short, Seq2Seq models are comprised of an RNN encoder\footnote{In \cite{Shan18}, Shan \emph{et al.} show, for KWS, the superiority of CRNN encoders with respect to GRU ones, which, in turn, are better than LSTM encoders.} summarizing the variable-length input sequence into a fixed-dimensional vector followed by an RNN decoder generating a variable-length output sequence conditioned on both the encoder output and past decoder predictions.

Besides for related tasks like QbE KWS \cite{Chung16}, Seq2Seq models such as an RNN-Transducer (RNN-T) have also been studied for deep spoken KWS \cite{He17,Sharma20,Yao21,Liu21}. RNN-T, integrating both acoustic and language models (and predicting phonemes), is able to outperform a CTC KWS system even when the latter exploits an external phoneme N-gram language model \cite{He17}.

\subsubsection{The Attention Mechanism}
\label{sssec:attention}

As aforementioned, in Seq2Seq models, the encoder has to condense all the needed information into a fixed-dimensional vector regardless the (variable) length of the input sequence, which might be challenging. The attention mechanism \cite{Vaswani17}, similarly to human listening attention, might assist in this context by focusing on the speech sections that are more likely to comprise a keyword \cite{Shan18}.

Let $\mathbf{h}_t$ be the hidden state of the RNN encoder of a Seq2Seq model at time step $t$:
\begin{equation}
    \mathbf{h}_{t}=\mbox{Encoder}\left(\mathbf{x}_{t}, \mathbf{h}_{t-1}\right).
\end{equation}
Before decoding it, the whole input sequence $\mathbf{X}=\left(\mathbf{x}_0,...,\mathbf{x}_{T-1}\right)$ has to be read, since $\mathbf{h}_{T-1}$ is the fixed-dimensional vector summarizing the whole input sequence that is finally input to the decoder (see Figure \ref{fig:seq2seq}). To assist the decoder, a context-relevant subset of $\left\{\mathbf{h}_{0},...,\mathbf{h}_{T-1}\right\}$ can be \emph{attended} to yield $\mathbf{A}$, which is to be used instead of $\mathbf{h}_{T-1}$:
\begin{equation}
    \mathbf{A}=\sum_{t=0}^{T-1}\alpha_{t}\mathbf{h}_{t},
\end{equation}
where $\alpha_{t}=\mbox{Attend}\left(\mathbf{h}_{t}\right)$, being $\mbox{Attend}\left(\cdot\right)$ an attention function \cite{Vaswani17} and $\sum_{t}\alpha_t=1$.

The integration of an attention mechanism (including a variant called multi-head attention \cite{Vaswani17}) in (primarily) Seq2Seq acoustic models in order to focus on the keyword(s) of interest has successfully been accomplished by a number of works, e.g., \cite{He17,Coimbra18,Wang19,Lee19,Zeyu20,Oleg20,Liu21}. These works find that incorporating attention provides KWS performance gains with respect to counterpart Seq2Seq models without attention.

Lastly, let us notice that attention has also been studied in conjunction with TDNNs for KWS \cite{Bai19,Chai19}. Particularly, in \cite{Bai19}, thanks to exploiting shared weight self-attention, Bai \emph{et al.} reproduce the performance of the deep residual learning model \texttt{res15} of Tang and Lin \cite{Tang18} by using 20 times less parameters, i.e., around 12k parameters only.

\subsection{Acoustic Model Training}
\label{ssec:training}

Once the acoustic model architecture has been designed (see the previous subsections) or optimally ``searched'' \cite{Hanna19, Bo21}, it is time to discriminatively estimate its parameters according to an optimization criterion ---defined by a loss function--- by means of backpropagation \cite{Rumelhart86} and using labeled/annotated speech data (see Section \ref{sec:datasets} in the latter respect).

\subsubsection{Loss Functions}

Apart from CTC \cite{Graves06}, which has been examined in the previous subsection, cross-entropy loss \cite{Bridle90, GoodfellowDL} is, by far, the most popular loss function for training deep spoken KWS acoustic models. For example, cross-entropy loss $\mathcal{L}_{\mbox{\scriptsize CE}}$ is considered by \cite{Parada14, George16, Sun16, Arik17, Menon18, Tang18, Kumar18, Liu19, Alvarez19, Bai19, Chai19, Oleg20, Liu20, Peter20}, and, retaking the notation of Section \ref{sec:overview}, can be expressed as
\begin{equation}
    \mathcal{L}_{\mbox{\scriptsize CE}}=-\sum_{i}\sum_{n=1}^N l_n^{\{i\}}\log\left(\mathbf{y}_n^{\{i\}}\right),
    \label{eq:celoss}
\end{equation}
where $l_n^{\{i\}}$ is the binary true (training) label corresponding to the input feature segment $\mathbf{X}_{\{i\}}$. Notice that when the acoustic model is intended to produce subword-level posteriors, commonly, training labels are generated by force alignment using an LVCSR system \cite{Parada14,Liu19,Alvarez19}, which will condition the subsequent KWS system performance.

First proposed in \cite{Dominik10}, max-pooling loss is an alternative to cross-entropy loss that has also been studied for KWS purposes \cite{Sun16,Hou20,Park20}. In the context of KWS, the goal of max-pooling loss is to teach the acoustic model to only trigger at the highest confidence time near the end of the keyword \cite{Sun16}. Let $\hat{\mathbf{L}}$ be the set of all the indices of the input feature segments in a minibatch belonging to any non-keyword class. In addition, let $y_p^{\star}$ be the largest target posterior corresponding to the $p$-th keyword sample in the minibatch, where $p=1,...,P$ and $P$ is the total number of keyword samples in the minibatch. Then, max-pooling loss can be expressed as
\begin{equation}
    \mathcal{L}_{\mbox{\scriptsize MP}}=-\sum_{i\in\hat{\mathbf{L}}}\sum_{n=1}^N l_n^{\{i\}}\log\left(\mathbf{y}_n^{\{i\}}\right)-\sum_{p=1}^P\log\left(y_p^{\star}\right).
    \label{eq:maxpool}
\end{equation}
From (\ref{eq:maxpool}), we can see that max-pooling loss is cross-entropy loss for any non-keyword class (left summand) while, for each keyword sample, the error is backpropagated for a single input feature segment only (right summand). Max-pooling loss has proven to outperform cross-entropy loss in terms of KWS performance, especially when the acoustic model is initialized by cross-entropy loss training \cite{Sun16}. Weakly-constrained and smoothed max-pooling loss variants are proposed in \cite{Hou20} and \cite{Park20}, respectively, which benefit from lowering the dependence on the accuracy of LVCSR force alignment.

\subsubsection{Optimization Paradigms}

In deep KWS, the most frequently used optimizers are stochastic gradient descent (SGD) \cite{SGD52} (normally with momentum), e.g., see \cite{Sundar15, George16, Tang18, Kumar18, Choi19, Chen19, Alvarez19, Emre20, Menglong20, Bo21, Kim21, Liu21, Wang21}, and Adam \cite{Adam}, e.g., see \cite{Shan18, Liu19, Emad19, Alice19, Zeng19, Wang19, Lee19, Bai19, Chai19, Oleg20, Peter20, Simon20, Yan20, Wei21, Axel21}. It is also a common practice to implement a mechanism shrinking the learning rate over epochs \cite{Parada14, George16, Sun16, Kumar18, Shan18, An19, Chen19, Zeng19, Lee19, Bai19, Chai19, Peter20, Menglong20, Simon20, Emre20}. Furthermore, many deep KWS works, e.g., \cite{Shan18, Choi19, Chen19, Menglong20, Kim21, Axel21, Liu21}, deploy a form of parameter regularization like weight decay and dropout. While random acoustic model parameter initialization is the normal approach, initialization based on transfer learning from LVCSR acoustic models has proven to lead to better KWS models by, e.g., alleviating overfitting \cite{Parada14, Samuel18, Yao21}.

\section{Posterior Handling}
\label{sec:posterior}

In order to come up with a final decision about the presence or not of a keyword in an audio stream, the sequence of posteriors yielded by the acoustic model, $\mathbf{y}^{\{i\}}$, needs to be processed. We differentiate between two main posterior handling modes: non-streaming (static) and streaming (dynamic) modes.

\subsection{Non-Streaming Mode}
\label{ssec:nstr_mode}

Non-streaming mode refers to standard multi-class classification of independent input segments comprising a single word each (i.e., isolated word classification). To cover the duration of an entire word, input segments have to be long enough, e.g., around 1 second long \cite{GSCDv1, Warden18}. In this mode, commonly, given an input segment $\mathbf{X}_{\{i\}}$, this is assigned to the class with the highest posterior probability as in Eq. (\ref{eq:argmax}). This approach is preferred over picking classes yielding posteriors above a sensitivity (decision) threshold to be set, since experience tells \cite{Lopez19, Lopez20, Lopez21, Lopez21b} that non-streaming deep KWS systems tend to produce very peaked posterior distributions. This might be attributed to the fact that non-streaming systems do not have to deal with inter-class transition data as in the dynamic case (see the next subsection), but with well-defined, isolated class realizations.

As mentioned in Section \ref{sec:overview}, KWS is not a static task but a dynamic one, which means that a KWS system has to continuously process an input audio stream. Therefore, it is obvious that the non-streaming mode lacks some realism from a practical point of view. Despite this, isolated word classification is considered by a number of deep KWS works, e.g., \cite{Samuel18, Tang18, Bo19, Choi19, Chen19, Zeng19, Riviello19, Bai19, Lopez19, Oleg20, Ximin20, Mo20, Menglong20, Peng20, Chen20, Lopez20, Lopez21, Lopez21b}. We believe that this is because of the simpler experimental framework with respect to that of the dynamic or streaming case. Fortunately, non-streaming performance and streaming performance seem to be highly correlated \cite{Lopez20, Lopez21b}, which makes non-streaming KWS research more relevant than it might look at first sight.

\subsection{Streaming Mode}
\label{ssec:str_mode}

Streaming mode alludes to the continuous processing (normally in real-time) of an input audio stream in which keywords are not isolated/segmented. Hence, in this mode, any given segment may or may not contain (parts of) a keyword. In this case, the acoustic model yields a time sequence of (raw) posteriors $\left\{...,\mathbf{y}^{\{i-1\}},\mathbf{y}^{\{i\}},\mathbf{y}^{\{i+1\}},...\right\}$ with strong local correlations. Due to this, the sequence of raw posteriors, which is inherently noisy, is typically smoothed over time ---e.g., by moving average--- on a class basis \cite{Parada14, Parada15, Sun16, Samuel18, Kumar18, Yuan19, Xiong19, Yue19, Liu19, Wu20, Peter20} before further processing.

Let us denote by $\bar{\mathbf{y}}^{\{i\}}$ the smoothed version of the raw posteriors $\mathbf{y}^{\{i\}}$. Furthermore, let us assume that each of the $N$ classes of a deep KWS system represents a whole word (which is a common case). Then, the smoothed word posteriors $\bar{\mathbf{y}}^{\{i\}}$ are often directly used to determine the presence or not of a keyword either by comparing them with a sensitivity threshold\footnote{This decision threshold might be set by optimizing, on a development set, some kind of figure of merit (see also Section \ref{sec:metrics} on evaluation metrics).} \cite{Sun16, Samuel18, Peter20} or by picking, within a time sliding window, the class with the highest posterior \cite{Kumar18}. Notice that since consecutive input segments $\left\{...,\mathbf{X}_{\{i-1\}},\mathbf{X}_{\{i\}},\mathbf{X}_{\{i+1\}},...\right\}$ may cover fragments of the same keyword realization, false alarms may occur as a result of recognizing the same keyword realization multiple times from the smoothed posterior sequence $\left\{...,\bar{\mathbf{y}}^{\{i-1\}},\bar{\mathbf{y}}^{\{i\}},\bar{\mathbf{y}}^{\{i+1\}},...\right\}$. To prevent this problem, a simple, yet effective mechanism consists of forcing the KWS system not to trigger for a short period of time right after a keyword has been spotted \cite{Sun16, Peter20}.

Differently from the above case, let us now consider the two following scenarios:
\begin{enumerate}
    \item Each of the $N$ classes still represents a whole word but keywords are composed of multiple words (e.g., ``OK Google'').
    \item Each of the $N$ classes represents a subword unit (e.g., a syllable) instead of a whole word.
\end{enumerate}
To tackle such scenarios, the first deep spoken KWS system \cite{Parada14} proposed a simple method processing the smoothed posteriors $\bar{\mathbf{y}}^{\{i\}}$ in order to produce a keyword presence decision. Let us assume that the first class $C_1$ corresponds to the non-keyword class and that the remaining $N-1$ classes represent subunits of a single keyword\footnote{This method can easily be extended to deal with more than one keyword \cite{Parada14}.}. Then, a time sequence of confidence scores $S_c^{\{i\}}$ can be computed as \cite{Parada14}
\begin{equation}
    S_c^{\{i\}}=\sqrt[N-1]{\prod_{n=2}^N \max_{h_{\mbox{\tiny max}}(i) \le k \le i}{\bar{\mathbf{y}}_n^{\{k\}}}},
    \label{eq:sci}
\end{equation}
where $h_{\mbox{\scriptsize max}}(i)$ indicates the onset of the time sliding window. A keyword is detected every time $S_c^{\{i\}}$ exceeds a sensitivity threshold to be tuned. This approach has been widely used in the deep KWS literature, e.g., \cite{Yuan19, Xiong19, Yue19}.

In \cite{Parada15}, Eq. (\ref{eq:sci}) is subject to the constraint that the keyword subunits trigger in the correct order of occurrence within the keyword, which contributes to decreasing false alarms. This improved version of the above posterior handling method is also considered by a number of deep KWS systems, e.g., \cite{Liu19, Wu20}.

When each of the $N$ classes of a deep KWS system represents a subword unit like a syllable or context-independent phoneme, a searchable lattice may be built from the time sequence of posteriors $\mathbf{y}^{\{i\}}$. Actually, this is typically done in the context of CTC \cite{Yimeng16, Yiyan18}. Then, the goal is to find, from the lattice, the most similar subword unit sequence to that of the target keyword. If the score resulting upon the search on the lattice is greater than a predefined score threshold, a keyword is spotted. Notice that this approach, despite its higher complexity, provides a great flexibility by, for example, allowing a user defining her/his own keywords.

\section{Robustness in Keyword Spotting}
\label{sec:robustness}

Normalizing the effect of acoustic variability factors such as background noise and room reverberation is paramount to assure good KWS performance in real-life conditions. This section is intended to review the scarce literature on KWS robust against, primarily but not only, background noise and far-field conditions. The motivation behind primarily dealing with the two latter acoustic variability factors lies in typical use cases of KWS technology\footnote{For example, activation of voice assistants typically takes place at home in far-field conditions and with some TV or music background noise.}.

This section has been arranged according to a taxonomy that segregates front- and back-end methods, which reflects the available literature on KWS robustness. Let us stress that these are normally cross-cutting methods, since they either come from or can be applied to other areas like ASR.

\subsection{Front-End Methods}
\label{ssec:frontend}

Front-end methods refer to those techniques that modify the speech signal before it is fed to the DNN acoustic model. In this subsection, we further differentiate among gain control for far-field conditions, DNN feature enhancement, adaptive noise cancellation and beamforming methods.

\subsubsection{Gain Control for Far-Field Conditions}

Keyword spotting deployment is many times conceived to facilitate real hands-free communication with devices such as smart speakers or in-vehicle systems that are located at a certain distance from the speaker. This means that communication might take place in far-field conditions, and, due to distance attenuation, background noise and reverberation can be particularly harmful.

Prabhavalkar \emph{et al.} \cite{Parada15} were the first to propose the use of automatic gain control (AGC) \cite{JuanPablo11} to provide robustness against background noise and far-field conditions for deep KWS. The philosophy behind AGC is based on selectively amplifying the audio signal depending on whether speech is present or absent. This type of selective amplification is able to yield a significant reduction of miss detections in the far-field scenario \cite{Parada15}.

Later, a more popular \cite{Arik17, Yiteng18, Arden19, Xuan19} and simpler AGC method called PCEN (Per-Channel Energy Normalization) \cite{Yuxuan17} was proposed for KWS. Keeping the original notation of \cite{Yuxuan17}, $E(t,f)$ represents (Mel) filterbank energy at time frame $t$ and frequency bin $f$, and
\begin{equation}
    M(t,f)=(1-s)M(t-1,f)+sE(t,f)
\end{equation}
is a time smoothed version of $E(t,f)$, where $0<s<1$ is a smoothing coefficient. Thus, PCEN is intended to replace the typical log compression of filterbank features as follows:
\begin{equation}
    \mbox{PCEN}(t,f)=\left(\frac{E(t,f)}{\left(\epsilon+M(t,f)\right)^\alpha}+\delta\right)^r-\delta^r,
    \label{eq:PCEN}
\end{equation}
where $\epsilon$ prevents division by zero, $\alpha\in(0,\;1)$ defines the gain normalization strength, and $\delta$ and $r$ determine the root compression. As we can see from Eq. (\ref{eq:PCEN}), the energy contour of $E(t,f)$ is dynamically normalized by $M(t,f)$ on a frequency band basis, which yields significant KWS performance gains under far-field conditions since $M(t,f)$ mirrors the loudness profile of $E(t,f)$ \cite{Yuxuan17}.

An appealing aspect of PCEN is that all its operations are differentiable. As a result, PCEN can be integrated in the DNN acoustic model in order to comfortably tune its set of parameters ---i.e., $s$, $\epsilon$, $\alpha$, $\delta$ and $r$--- towards the optimization of KWS performance during acoustic model training \cite{Yuxuan17}.

\subsubsection{DNN Feature Enhancement}

The powerful modeling capabilities of DNNs can also be exploited to clean the noisy speech features (usually, magnitude spectral features) before these are input to the KWS acoustic model. A variety of approaches can be followed:
\begin{enumerate}
    \item \emph{Enhancement mask estimation}: The aim of this approach is to estimate, from the noisy observation (e.g., noisy Mel spectra \cite{Gu19}) and using a neural network (e.g., a CRNN \cite{Gu19}), a multiplicative de-noising time-frequency mask to be applied to the noisy observation \cite{Gu19, Menne19}. The result is then passed to the acoustic model.
    \item \emph{Noise estimation}: A DNN (e.g., a CNN with dilated convolutions and residual connections \cite{Jung20}) might also be used to provide an estimate of the distortion that contaminates the target speech signal. The estimated distortion can then be subtracted from the noisy observation before feeding the acoustic model with it \cite{Jung20}.
    \item \emph{Clean speech estimation}: In this case, the DNN front-end directly produces an estimate of the clean speech features, from the noisy observation, to be input to the acoustic model. While this approach has been studied for robust ASR \cite{Zixing18}, to the best of our knowledge and surprisingly, this has not been the case for KWS.
    \item \emph{Filter parameter estimation}: The parameters of an enhancement filter (e.g., a Wiener filter \cite{Menne19}) to be applied to the noisy observation before further processing can be estimated by means of a DNN. Similarly to the above case, while this has been studied for robust ASR \cite{Menne19}, this has not been the case for KWS.
\end{enumerate}
Regardless of the chosen approach, the DNN front-end and the KWS acoustic model can be jointly trained following a multi-task learning scheme to account the complementary objectives of the front-end and the acoustic model. By making the DNN front-end aware of the global keyword detection goal \cite{Gu19, Jung20}, superior KWS performance can be achieved in comparison with independent training of the two components.

One conclusion is that, oddly, DNN feature enhancement is a rather unexplored area in the context of KWS. This contrasts with the case of robust ASR, which has widely and successfully studied the application of this type of de-noising front-ends \cite{Zixing18, Menne19}. Immediate future work on robust KWS could address this imbalance, especially by exploring promising time domain solutions that can benefit from phase information \cite{Kai21}.

\subsubsection{Adaptive Noise Cancellation}

\begin{figure}
    \centering
    \includegraphics[width=0.9\linewidth]{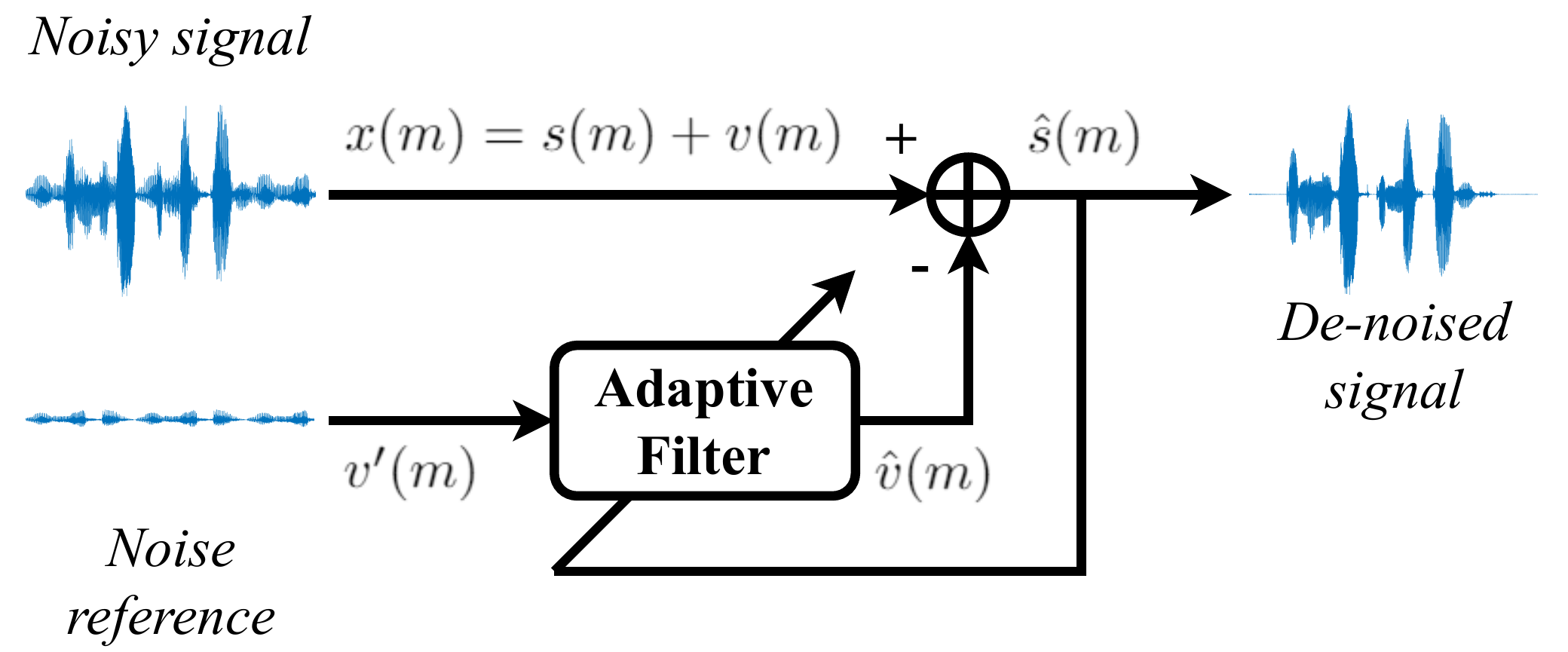}
    \caption{Block diagram of adaptive noise cancellation. A signal of interest $s(m)$ is retrieved from a noisy observation $x(m)=s(m)+v(m)$ by subtracting an estimate of $v(m)$, $\hat{v}(m)$. This estimate is obtained by filtering a noise reference $v'(m)$ that originates from the same noise source as $v(m)$ (i.e., $v(m)$ and $v'(m)$ are highly correlated). The filter weights are continuously adapted to typically minimize the power of the estimated signal of interest $\hat{s}(m)$.}
    \label{fig:anc}
\end{figure}

Presumably thinking of voice assistant use cases, Google developed a series of noise-robust KWS methods based on dual-microphone adaptive noise cancellation (ANC) to particularly deal with speech interference \cite{Yiteng18, Arden19, Yiteng19}. The working principle of ANC is outlined in Figure \ref{fig:anc}. The reason for accounting a dual-microphone scenario is that Google's smart speaker Google Home has two microphones \cite{GoogleSpecs}. It is interesting to point out that the authors of this series of ANC works also tried to apply beamforming and multi-channel Wiener filtering\footnote{Notice that multi-channel Wiener filtering is equivalent to minimum variance distortionless response (MVDR) beamforming followed by single-channel Wiener post-filtering \cite{Ito10, Lefkimmiatis07}.}, but they only found marginal performance gains by doing so \cite{Arden19}.

In \cite{Yiteng18}, Google researchers proposed a de-noising front-end inspired by the human auditory system. In short, the de-noising front-end works by exploiting posterior probability feedback from the KWS acoustic model:
\begin{enumerate}
    \item If the acoustic model finds that voice is absent, the weights of a recursive least squares (RLS) ANC filter working in the short-time Fourier transform (STFT) domain are updated;
    \item If the posterior probabilities computed by the acoustic model are inconclusive (i.e., the presence of a keyword is uncertain), the most recent ANC weights are used to filter/clean the input signal and the presence of a keyword is rechecked.
\end{enumerate}

In a similar vein, a so-called hotword cleaner was reported in \cite{Arden19}, which overcomes one of the shortcomings of the above ANC approach \cite{Yiteng18}: the increased latency and CPU usage derived from having to run the acoustic model twice (one to provide feedback to the de-noising front-end and another for KWS itself). The hotword cleaner \cite{Arden19} leverages the following two characteristics of the KWS scenario to deploy a simple, yet effective de-noising method: \emph{1)} there is typically no speech just before a keyword, and \emph{2)} keywords are of short duration. Bearing these two characteristics in mind, the hotword cleaner \cite{Arden19} simply works by continuously computing fast-RLS ANC weights that are stored and applied to the input signal with a certain delay to clean and not damage the keyword. This methodology was generalized to an arbitrary number of microphones in \cite{Yiteng19}. Overall, all of these ANC-based methods bring significant KWS performance improvements in everyday noisy conditions that include strong TV and music background noise.

\subsubsection{Beamforming}

Spatial filtering, also known as beamforming, enables the exploitation of spatial cues in addition to time and frequency information to boost speech enhancement quality \cite{Kellermann08}. Similarly to the aforementioned case with DNN feature enhancement, KWS lags several steps behind ASR regarding the integration of beamforming as done, e.g., in \cite{Watanabe18}.

To the best of our knowledge, \cite{Xuan20} is the first research studying beamforming for deep KWS. In particular, \cite{Xuan20} applies four \emph{fixed} beamformers that are arranged to uniformly sample the horizontal plane. The acoustic model is then fed with the four resulting beamformed signals plus a reference signal picked from one of the array microphones to avoid degrading performance at higher signal-to-noise ratios (SNRs) \cite{Xuan20}. The acoustic model incorporates an attention mechanism \cite{Vaswani17} to weigh the five input signals, which can be thought as a steering mechanism pointing the effective beam towards the target speaker. Actually, the motivation behind using fixed beamformers lies in the difficulty of estimating the target direction in noisy conditions. However, notice that the attention mechanism implicitly estimates it.

The same authors of \cite{Xuan20} went farther in \cite{Meng20} by replacing the set of fixed beamformers by a set of data-dependent, multiplicative spectral masks playing an equivalent role. The latter masks, which are estimated by a neural network, can be interpreted as \emph{semi-fixed} beamformers. This is because though they are data-dependent, mask look directions (equivalent to look directions of beamformers) are still fixed. This beamforming front-end, which is trained jointly with the acoustic model, outperforms the previous fixed beamforming approach, especially at lower signal-to-interference ratios (SIRs) \cite{Meng20}.

There is still a long way to go regarding the application of beamforming to deep KWS. More specifically, despite the aforementioned steering role of the attention mechanism, we believe that deep beamforming that does not pre-arrange the look direction but estimates it continuously based on microphone signals ---as in, e.g., \cite{Xiao16}--- is worth to explore.

\subsection{Back-End Methods}
\label{ssec:backend}

Back-end methods refer to techniques applied within the acoustic model to primarily improve its generalization ability to a variety of acoustic conditions. The rest of this subsection is devoted to discuss the following matters: multi-style and adversarial training, robustness to keyword data scarcity, the class-imbalance problem and other back-end methods.

\subsubsection{Multi-Style Training}

One of the most popular and effective back-end methods to, especially, deal with background noise and reverberation is multi-style training of the KWS acoustic model (see, e.g., \cite{Parada15, He17, Erika19, Choi19, Pattanayak19, Peter20, Park20, Bluche20, Oleg20, Hou20, Axel21, Park21}). Multi-style training, which has some regularization effect preventing overfitting \cite{Erika19}, simply consists of training the acoustic model with speech data contaminated by a variety of distortions trying to better reflect what is expected to be found at test time.

Usually, distorted speech data are generated by contaminating ---e.g., by background noise addition at different SNR levels--- clean speech data in an artificial manner (see Section \ref{sec:datasets} for practical details). This artificial distortion procedure is known as \emph{data augmentation} \cite{Malek19}. For instance, a series of data augmentation policies like time and frequency masking is defined by a tool like SpecAugment \cite{SpecAugment}. First proposed for end-to-end ASR, SpecAugment has recently become a popular way for generating distorted speech data, also for KWS training purposes \cite{Hou20, Oleg20, Kim21, Wei21, Axel21, Liu21}.

\subsubsection{Adversarial Training}

Deep neural networks often raise the following issue: networks' outputs might not be smooth with respect to inputs \cite{Szegedy14}, e.g., because of the lack of enough training data. This might involve, for example, that a keyword correctly classified by the acoustic model is misclassified when a very small perturbation is added to such a keyword. This kind of subtly distorted input to the network is what we call an \emph{adversarial example}. Interestingly, adversarial examples can be generated by means of techniques like the fast gradient sign method (FGSM) \cite{FGSM} to re-train with them a well-trained KWS acoustic model. The goal of this is to improve robustness by smoothing the distribution of the acoustic model. This approach, which can be interpreted as a type of data augmentation, has shown to be effective to drastically decrease false alarms and miss detections for an attention-based Seq2Seq acoustic model \cite{Wang19}. Alternatively, \cite{Xiong19} proposes to replace, with the same goal, adversarial example re-training by adversarial regularization in the loss function. Wang \emph{et al.} \cite{Xiong19} demonstrate that the latter outperforms the former under far-field and noisy conditions when using a DS-CNN acoustic model for KWS.

\subsubsection{Robustness to Keyword Data Scarcity}

To effectively train a KWS acoustic model, a sufficient amount of speech data is required. This normally includes a substantial number of examples of the specific keyword(s) to be recognized. However, there is a number of possible reasons for which we might suffer from keyword data scarcity. Certainly, collecting additional keyword samples can help to overcome the problem. Nevertheless, speech data collection can be costly and time-consuming, and is often infeasible. Instead, a smart way to obtain additional keyword samples for model training is by synthetically generating them through text-to-speech technology. This type of data augmentation has proven to be highly effective by significantly improving KWS performance in low-resource keyword settings \cite{Sharma20, Lin20, Andrew21}. In particular, in \cite{Sharma20}, it is found that it is important that synthetic speech reflects a wide variety of tones of voice (i.e., speaker diversity) for good KWS performance.

\subsubsection{The Class-Imbalance Problem}
\label{ssec:imbalance}

The class-imbalance problem refers to the fact that, typically, many more non-keyword than keyword samples are available for KWS acoustic model training. Actually, the class-imbalance problem can be understood as a relative keyword data scarcity problem: for obvious reasons, it is almost always easier to access a plethora of non-keyword than keyword samples. The issue lies in that class imbalance can lead to under-training of the keyword class with respect to the non-keyword one.

To reach class balance for acoustic model training, one can imagine many different things that can be done based on data augmentation:
\begin{enumerate}
    \item Generation of adversarial examples yielding miss detections, e.g., through FGSM \cite{FGSM}, to re-train the acoustic model in a class-balanced way;
    \item Generation of additional synthetic keyword samples by means of text-to-speech technology \cite{Sharma20, Lin20}.
\end{enumerate}
To the best of our knowledge, the above two data augmentation approaches have not been studied for tackling the class-imbalance problem.

Differently, a series of works has proposed to essentially focus on challenging non-keyword samples\footnote{A challenging non-keyword sample can be, e.g., one exhibiting similarities with the keyword in terms of phonetics.} at training time instead of fully exploiting all the non-keyword samples available \cite{Liu19, Hou20, Kun20}. For instance, Liu \emph{et al.} \cite{Liu19} suggested to weigh cross-entropy loss $\mathcal{L}_{\mbox{\scriptsize CE}}$ (see Eq. (\ref{eq:celoss})) by $\left(1-\mathbf{y}_n^{\{i\}}\right)^{\gamma}$ to come up with focal loss $\mathcal{L}_{\mbox{\scriptsize FL}}$:
\begin{equation}
    \mathcal{L}_{\mbox{\scriptsize FL}}=-\sum_{i}\sum_{n=1}^N \left(1-\mathbf{y}_n^{\{i\}}\right)^{\gamma}l_n^{\{i\}}\log\left(\mathbf{y}_n^{\{i\}}\right),
    \label{eq:focal_loss}
\end{equation}
where $\gamma$ is a tunable focusing parameter. As one can easily reason, weighing cross-entropy loss as in Eq. (\ref{eq:focal_loss}) helps to focus training on challenging samples. While this weighting procedure is more effective than regular cross-entropy in class-imbalanced scenarios \cite{Liu19}, notice that it might be able to strengthen the model in a wide sense. Because focal loss $\mathcal{L}_{\mbox{\scriptsize FL}}$ operates on a frame basis, \cite{Kun20} improved it by also considering the time context when computing the weight for cross-entropy loss. Particularly, such an improvement is equivalent to assigning bigger weights to those frames belonging to non-keyword samples yielding false alarms.

An alternative approach ---so-called regional hard-example mining--- for dealing with the class-imbalance problem was described in \cite{Hou20}. Regional hard-example mining subsamples the available non-keyword training data to keep a certain balance between keyword and non-keyword samples. Non-keyword sample mining is based on the selection of the most difficult non-keyword samples in the sense that they yield the highest keyword posteriors.

\subsubsection{Other Back-End Methods}

A few other methods for robustness purposes not falling into any of the above categories can be found in the literature. For instance, \cite{Wu20} extracts embeddings characterizing the acoustic environment that are passed to the acoustic model to carry out KWS which is robust to far-field and noisy conditions. In this way, by making the acoustic model \emph{aware} of the acoustic environment, better keyword prediction can be achieved.

We also recently contributed to noise-robust KWS in \cite{Lopez21b}, where we proposed to interpret every typical KWS acoustic model as the concatenation of a keyword embedding extractor followed by a linear classifier consisting of the typical final fully-connected layer with softmax activation for word classification (see Section \ref{sec:overview}). The goal is to, first, multi-style train the keyword embedding extractor by means of a ($C_{N,2}+1$)-pair loss function extending the idea behind tuple-based losses like $N$-pair \cite{Sohn16} and triplet \cite{Schroff15} losses (the latter used both standalone \cite{Sacchi19} and combined with the reversed triplet and hinge losses \cite{Yuan19} for keyword embedding learning). In comparison with these and similar losses also employed for word embedding learning (e.g., a prototypical loss angular variant \cite{Huh21}), in \cite{Lopez21b}, we demonstrate that the ($C_{N,2}+1$)-pair loss reaches larger inter-class and smaller intra-class embedding variation\footnote{This is because the ($C_{N,2}+1$)-pair loss constrains the way the training samples belonging to different classes relate to each other in terms of embedding distance.}. Secondly, the final fully-connected layer with softmax activation is trained by multi-style keyword embeddings employing cross-entropy loss. This two-stage training strategy is much more effective than standard end-to-end multi-style training when facing unseen noises \cite{Lopez21b}. Moreover, another appealing feature of this two-stage training strategy is that it increases neither the number of parameters nor the number of multiplications of the model.

\section{Applications}
\label{sec:applications}

Keyword spotting technology (including deep KWS) has a number of applications, which range from the more traditional ones like voice-dialing, interaction with a call center and speech retrieval to nowadays flagship application, namely, the activation of voice assistants.

In addition to the above, KWS technology could be useful, e.g., to assist disabled people like vision-impaired pedestrians when it comes to the activation of pedestrian call buttons in crosswalks. For example, \cite{Verdegay19} proposes the use of a CRNN-based KWS system \cite{Arik17} for the activation of pedestrian call buttons via voice, thereby contributing to improve accessibility in public areas to people with the above-mentioned disability.

In-vehicle systems can also benefit from voice control. For example, in \cite{Yue19}, Tan \emph{et al.} explore multi-source fusion exploiting variations of vehicle's speed and direction for \emph{online} sensitivity threshold selection. The authors of \cite{Yue19} demonstrate that this strategy improves KWS accuracy with respect to using a fixed, predetermined sensitivity threshold for the posteriors yielded by the DNN acoustic model.

Moreover, it is worth noticing that KWS is a technology that is sometimes better suited than ASR to the solution of certain problems where the latter is typically employed. This is the case, for instance, of by-topic audio classification and audio sentiment detection \cite{Kaushik15, Kaushik17}, since the accuracy of these tasks rather relies on being able to correctly spot a very focused (i.e., quite limited) vocabulary in the utterances. In other words, lexical evidence is sparse for such tasks.

Some work has explored KWS also for voice control of videogames \cite{Sundar15, An19}. Particularly, \cite{Sundar15} points out how KWS becomes an extremely difficult task when it comes to dealing with children controlling videogames with their voice due to excitement and, generally speaking, the nature of children and children's voice \cite{Lee99}. To partially deal with this, the authors of \cite{Sundar15} propose the detection of overlapping keywords in the context of a multiplayer side-scroller game called Mole Madness. Since BiLSTMs have proven to work well for children's speech \cite{Martin11}, a BiLSTM acoustic model with $2^N$ output classes ---where $N$ is the number of keywords--- is used to represent all possible combinations of overlapping keywords. It is found that, under the videogame conditions, modeling the large variations of children's speech time structure is challenging even for a relatively large BiLSTM.

Other KWS applications include voice control of home automation \cite{Vacher15}, even the navigation of complex procedures in the International Space Station \cite{ISS10}, etc.

\subsection{Personalized Keyword Spotting Systems}

For some of the above applications, having a certain degree of personalization in the sense that only a specific user is allowed to utilize the KWS system can be a desirable feature. Towards this personalization goal, some research has studied the combination of KWS and speaker verification \cite{Kumar18, Enea19, Jung20, Rajeev21}. While \cite{Enea19, Rajeev21} employ independently trained deep learning models to perform both tasks, \cite{Kumar18, Jung20} address, following a multi-task learning scheme, joint KWS and speaker verification with contradictory conclusions, since KWS performance is negatively and positively affected in \cite{Kumar18} and \cite{Jung20}, respectively, by the integration of speaker verification. A reason for this could be that, unlike in \cite{Jung20}, higher-level features are shared for both tasks in \cite{Kumar18}, so this further preservation of speaker information may contaminate the phonetic information required to carry out KWS.

Personalization can be of particular interest for voice activation of voice assistants \cite{HeySiri} as well as for voice control of hearing assistive devices like hearing aids. These two KWS applications are reviewed in a bit more detail in the next subsections.

\subsection{Voice Activation of Voice Assistants}
\label{ssec:voice_assistants}

\begin{figure}
    \centering
    \includegraphics[width=\linewidth]{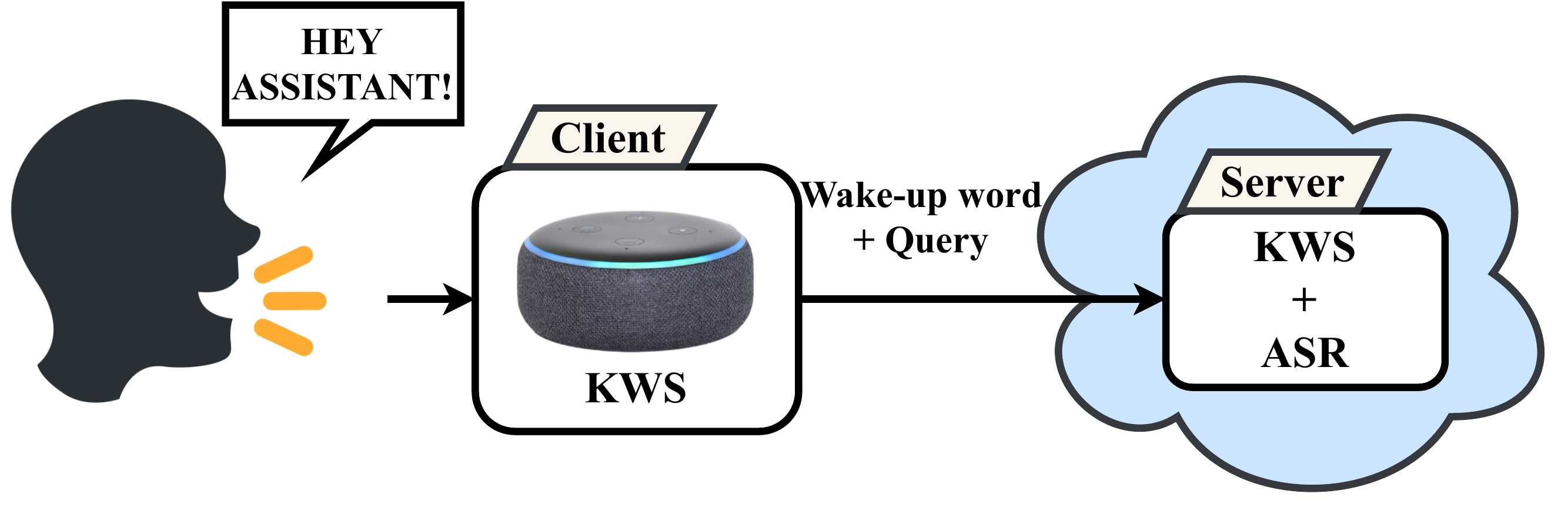}
    \caption{Typical voice assistant client-server framework.}
    \label{fig:voice_assist}
\end{figure}

The flagship application of (deep) KWS is the activation of voice assistants like Amazon's Alexa, Apple's Siri, Google's Assistant and Microsoft's Cortana. Actually, without fear of error, we can say that revitalization of KWS research over the last years is owed to this application \cite{Parada15c}. And there is a compelling reason for this: forecasts suggest that, by 2024, the number of voice assistant units will exceed that of world's population \cite{Stat20}.

Figure \ref{fig:voice_assist} illustrates the typical voice assistant client-server framework. The client consists of an electronic device like a smartwatch or a smart speaker integrating the client-side of a voice assistant and an always-on KWS system to detect when a user wakes up the assistant by uttering a trigger word/phrase, e.g., ``hey assistant!''. To limit the impact on the battery life, the KWS system has to be necessarily light. In this vein, Apple employs a two-pass detection strategy \cite{HeySiri}. By this, a very light, always-on KWS system listens for the corresponding wake-up word. If this is detected, a more complex and accurate KWS system ---also placed on the client device--- is used to double check whether or not the wake-up word has been really uttered.

When the wake-up word is spotted on the client-side, the supposed wake-up word audio and subsequent query audio are sent to a server on which, first, the presence of the wake-up word is checked for a second or third time by using much more powerful and robust LVCSR-based KWS \cite{HeySiri, Cortana21, Assaf17}. If, finally, the LVCSR-based KWS system determines that the wake-up word is not present, the subsequent audio is discarded and the process is ended. Otherwise, ASR is applied to the supposed query audio and the result is further processed ---e.g., using natural language processing techniques--- to provide the client device with a response. In the context of Google's Assistant \cite{Assaf17}, it is shown that this server-side wake-up word check drastically reduces the false alarm rate while marginally increasing the rate of miss detections. Notice that this server-side check could be useful for mitigating privacy issues as a result of pseudo-query audio leakage if it were not for the fact that the supposed wake-up word audio and query audio are inseparably streamed to the server. Interestingly, Garg \emph{et al.} \cite{Vineet21} have recently proposed a streaming Transformer encoder carrying out the double check efficiently on the client-side, which can truly help to mitigate privacy issues.

\subsection{Voice Control of Hearing Assistive Devices}
\label{ssec:hearing_devices}

\begin{figure}
\centering
\subfloat[Legitimate user detected.]{\includegraphics[width=0.48\columnwidth]{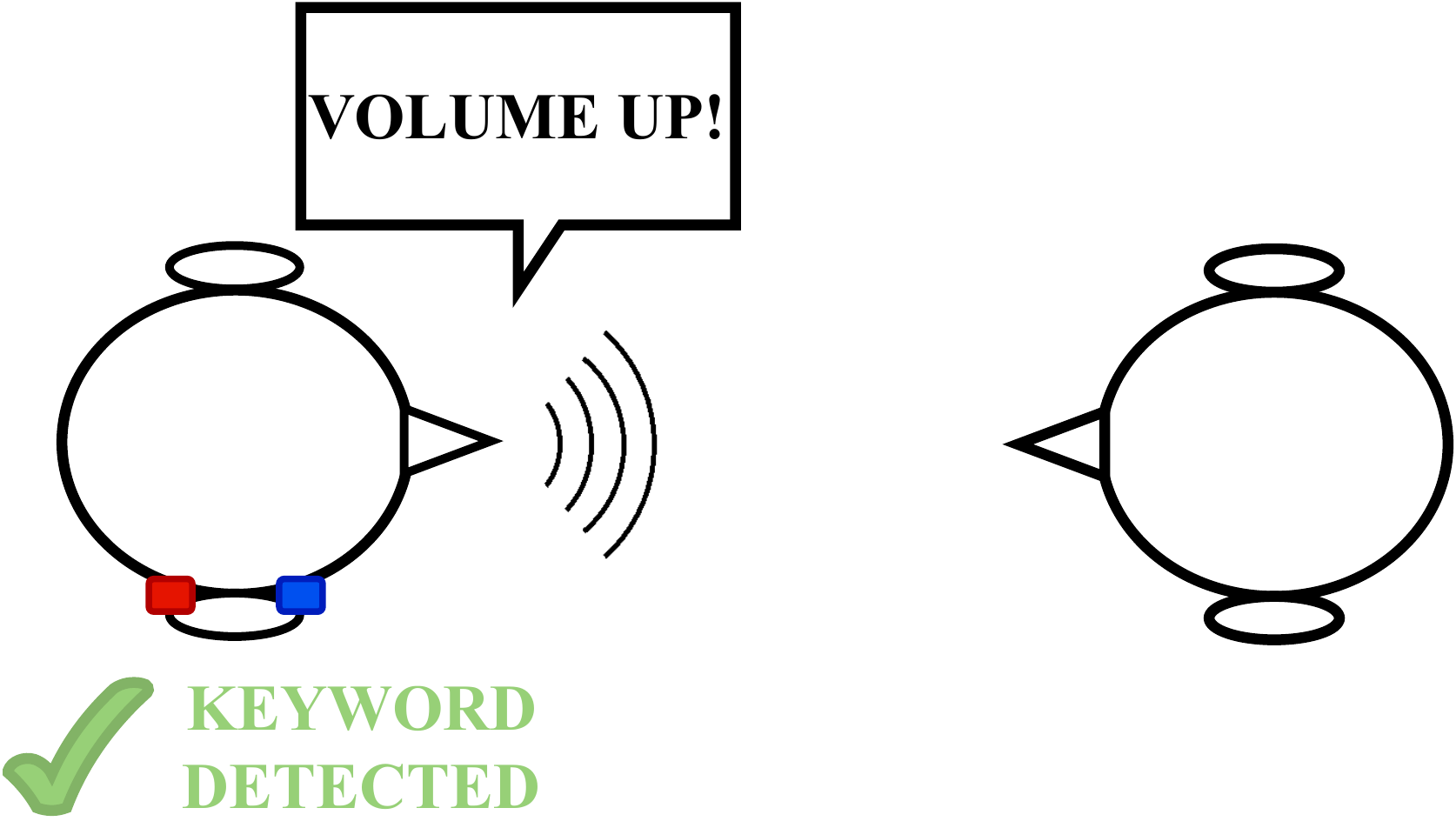}}
\hfil
\rulesep
\subfloat[External speaker detected.]{\includegraphics[width=0.48\columnwidth]{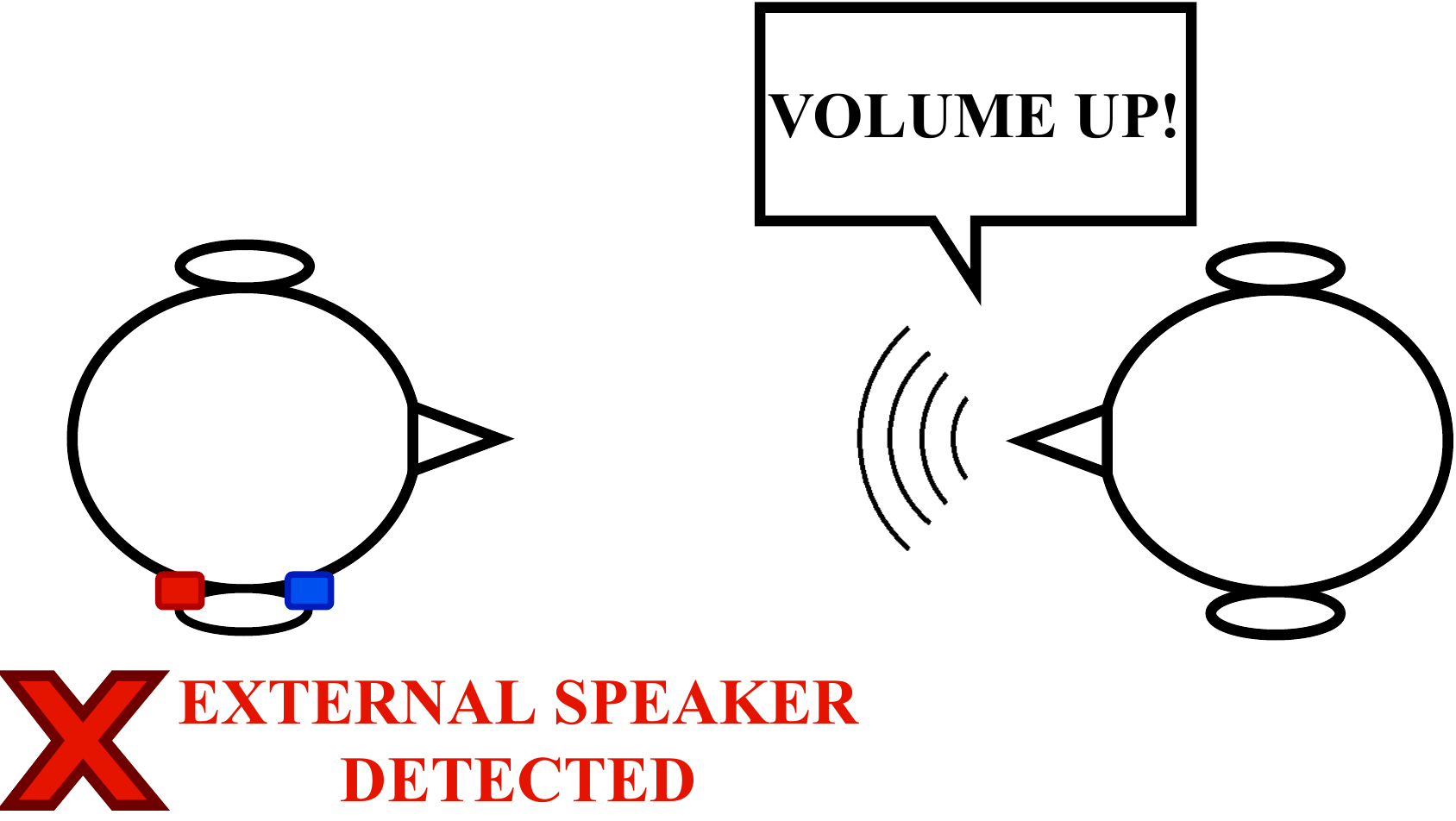}}
\caption{Users' own voice/external speaker detection in the context of voice control of hearing aids. Red and blue dots symbolize the two microphones of a hearing aid sitting behind the ear.}
\label{fig:kws_hads}
\end{figure}

Manually operating small, body-worn devices like hearing aids is not always feasible or can be cumbersome. One reason could be that hands are busy doing other activities like cooking or driving. Another cause could be that the wearer is an elderly person with reduced fine motor skills. Whatever the reason is, KWS can help to deploy voice interfaces to comfortably operate such a kind of devices. Furthermore, these devices are personal devices, so it is desirable that the user is the only person who can handle them.

In the above respect, in \cite{Lopez19, Lopez20} we studied an alternative way to speaker verification to provide robustness against external speakers (i.e., personalization) in KWS for hearing aids as exemplified by Figure \ref{fig:kws_hads}. Particularly, we extended the deep residual learning model proposed by Tang and Lin \cite{Tang18} to jointly perform KWS and users' own voice/external speaker detection following a multi-task learning scheme. A keyword prediction is then taken into account if and only if the multi-task network determines that the spotted keyword was uttered by the legitimate user. Thanks to exploiting GCC-PHAT (Generalized Cross-Correlation with PHAse Transform) \cite{Knapp76} coefficients from dual-microphone hearing aids in the perceptually-motivated constant-Q transform \cite{Brown91} domain, we achieve almost flawless users' own voice/external speaker detection \cite{Lopez20}. This is because phase difference information is extremely useful to characterize the virtually time-invariant position of the user's mouth with respect to that of the hearing aid. It is worth noting that this experimental validation was carried out on a hearing aid speech database created by convolving the Google Speech Commands Dataset \emph{v2} \cite{Warden18} with acoustic transfer functions measured in a hearing aids set-up.

\section{Datasets}
\label{sec:datasets}

\begin{table*}[th]
	\caption{A selection of the most significant speech datasets employed for training and validating deep KWS systems. ``P.A.'' stands for ``publicly available'', while ``Y'' and ``N'' mean ``yes'' and ``no'', respectively. Furthermore, ``\emph{+ sampl.}'' (``\emph{- sampl.}'') refers to the size of the positive/keyword (negative/non-keyword) subset, and ``\emph{Size}'' denotes the magnitude of the whole set. Such sizes are given, depending on the available information, in terms of either the number of samples or time length in hours (h). Unknown information is indicated by hyphens.}
	\label{tab:datasets}
	\centering
	\setlength{\extrarowheight}{3pt}
	\resizebox{\linewidth}{!}{\begin{tabular}{cllclcccccccc}
			\toprule
			\multicolumn{1}{c}{\textbf{Ref.}} & \textbf{Name} & \textbf{Developer} & \textbf{P.A.?} & \textbf{Language} & \textbf{Noisy?} & \textbf{No. of KW} & \multicolumn{3}{c}{\textbf{Training set}} & \multicolumn{3}{c}{\textbf{Test set}} \\ \cline{8-10} \cline{11-13}
			\multicolumn{7}{c}{} & \emph{Size} & \emph{+ sampl.} & \emph{- sampl.} & \emph{Size} & \emph{+ sampl.} & \emph{- sampl.} \\ \midrule
			\multicolumn{1}{c}{\cite{Chen18}} & - & Alibaba & N & Mandarin & Y & 1 & 24k h & - & - & - & 12k & 600 h \\
			\multicolumn{1}{c}{\cite{Arik17}} & - & Baidu & N & English & Y & 1 & 12k & - & - & 2k & - & - \\
			\multicolumn{1}{c}{\cite{Liu19}} & - & \makecell[l]{Chinese\\Academy\\of Sciences} & N & Mandarin & Y & 2 & 47.8k & 8.8k & 39k & - & 1.7k & - \\
			\multicolumn{1}{c}{\cite{Samuel18}} & - & Fluent.ai & N & English & Y & 1 & 50 h & 5.9k & - & 22 h & 1.6k & - \\
			\multicolumn{1}{c}{\cite{Parada14}} & - & Google & N & English & Y & 10 & >3k h & 60.7k & 133k & 81.2k & 11.2k & 70k \\
			\multicolumn{1}{c}{\cite{Parada15c}} & - & Google & N & English & Y & 14 & 326.8k & 10k & 316.8k & 61.3k & 1.9k & 59.4k \\
			\multicolumn{1}{c}{\cite{Xuan19}} & - & \makecell[l]{Harbin\\Institute\\of Technology} & N & Mandarin & - & 1 & 115.2k & 19.2k & 96k & 28.8k & 4.8k & 24k  \\
			\multicolumn{1}{c}{\cite{Sacchi19}} & - & Logitech & N & English & - & 14 & - & - & - & - & - & - \\
			\multicolumn{1}{c}{\cite{Wang19}} & - & Mobvoi & N & Mandarin & Y & 1 & 67 h & 20k & 54k & 7 h & 2k & 5.9k  \\
			\multicolumn{1}{c}{\cite{Bluche20}} & - & Sonos & Y & English & Y & 16 & 0 & 0 & 0 & 1.1k & 1.1k & 0 \\
			\multicolumn{1}{c}{\cite{Meng20}} & - & Tencent & N & Mandarin & Y & 1 & 339 h & 224k & 100k & - & - & - \\
			\multicolumn{1}{c}{\cite{Xiong19}} & - & Tencent & N & Mandarin & Y & 1 & 65.9 h & 6.9 h & 59 h & 8.7 h & 0.9 h & 7.8 h \\
			\multicolumn{1}{c}{\cite{Yuan19}} & - & Tencent & N & Mandarin & Y & 42 & 22.2k & 15.4k & 6.8k & 10.8k & 7.4k & 3.4k  \\
			\multicolumn{1}{c}{\cite{Shan18}} & - & Xiaomi & N & Mandarin & - & 1 & 1.7k h & 188.9k & 1M & 52.2 h & 28.8k & 32.8k  \\
			\multicolumn{1}{c}{\cite{Jiayu18}} & AISHELL-2 (13) & AISHELL & Y & Mandarin & N & 13 & 24.8 h & >24k & - & 16.7 h & >8.4k & - \\
			\multicolumn{1}{c}{\cite{Jiayu18}} & AISHELL-2 (20) & AISHELL & Y & Mandarin & N & 20 & 35 h & >34k & - & 23.9 h & >12k & - \\
			\multicolumn{1}{c}{\cite{Yuriy19}} & ``Alexa'' & Amazon & N & English & Y & 1 & 495 h & - & - & 100 h & - & - \\
			\multicolumn{1}{c}{\cite{GSCDv1}} & \makecell[l]{Google Speech\\Commands Dataset \emph{v1}} & Google & Y & English & Y & 10 & 51.7k & 18.9k & 32.8k & 6.5k & 2.4k & 4.1k \\
			\multicolumn{1}{c}{\cite{Warden18}} & \makecell[l]{Google Speech\\Commands Dataset \emph{v2}} & Google & Y & English & Y & 10 & 84.6k & 30.8k & 53.8k & 10.6k & 3.9k & 6.7k \\
			\multicolumn{1}{c}{\cite{Higuchi20}} & ``Hey Siri'' & Apple & N & English & Y & 1 & 500k & 250k & 250k & - & 6.5k & 2.7k h \\
			\multicolumn{1}{c}{\cite{Kim19}} & \makecell[l]{Hey Snapdragon\\Keyword Dataset} & Qualcomm & Y & English & N & 4 & - & - & - & 4.3k & 4.3k & - \\
			\multicolumn{1}{c}{\cite{Alice19}} & Hey Snips & Snips & Y & English & Y & 1 & 50.5 h & 5.9k & 45.3k & 23.1 h & 2.6k & 20.8k \\
			\multicolumn{1}{c}{\cite{An19}} & ``Narc Ya'' & Netmarble & N & Korean & Y & 1 & 130k & 50k & 80k & 800 & 400 & 400 \\
			\multicolumn{1}{c}{\cite{Alvarez19}} & ``Ok/Hey Google'' & Google & N & English & Y & 2 & - & 1M & - & >3k h & 434k & 213k \\
			\multicolumn{1}{c}{\cite{Yiteng18}} & ``Ok/Hey Google'' & Google & N & English & Y & 2 & - & - & - & 247 h & 4.8k & 7.5k \\
			\multicolumn{1}{c}{\cite{Jingyong19}} & Ticmini2 & Mobvoi & N & Mandarin & Y & 2 & 157.5k & 43.6k & 113.9k & 72.9k & 21.3k & 51.6k \\
			\bottomrule
	\end{tabular}}
\end{table*}

Data are an essential ingredient of any machine learning system for both training the parameters of the algorithm (primarily, in our context, the acoustic model parameters) and validating it. Some well-known speech corpora that have been extensively used over the years in the field of ASR are now also being employed for the development of deep KWS systems. For example, LibriSpeech \cite{Vassil15} has been used by \cite{Chung16, Loren18, Kumar18, Bluche20, Wei21}, TIDIGITS \cite{TIDIGITS}, by \cite{Enea19}, TIMIT \cite{timit}, by \cite{Martin13, Chen16, Ravi18, Bruno18, Erika19}, and the Wall Street Journal (WSJ) corpus \cite{WSJData}, by \cite{Yimeng16, Kumar18, Sacchi19}. The main problem with these speech corpora is that they were not developed for KWS, and, therefore, they do not standardize a way of utilization facilitating KWS technology reproducibility and comparison. By contrast, KWS research work exploiting these corpora employs them in a variety of ways, which is even reflected by, e.g., the set of considered keywords.

In the following we focus on those datasets particularly intended for KWS research and development, which, normally, are comprised of hundreds or thousands of different speakers who do not overlap across sets (i.e., training, development and test sets), e.g., \cite{Hou16, Arik17, Warden18, Wang19, Alice19, HeySnips, Jingyong19, Xuan19, Yuan19, Lee19}. Table \ref{tab:datasets} shows a wide selection of the most significant speech corpora available for training and testing deep KWS systems. From this table, the first inference that we can draw is that the advancement of the KWS technology is led by the private sector of the United States of America (USA) and China. Seven and five out of the seventeen different dataset developers included in Table \ref{tab:datasets} are, respectively, North American and Chinese corporations. Actually, except for the ``Narc Ya'' corpus \cite{An19}, which is in Korean, all the datasets shown in this table are in either English or Mandarin Chinese.

A problem with the above is that the majority of the speech corpora of interest for KWS research and development are not publicly available (P.A.), but they are for (company) internal use only. On many occasions, these datasets are collected by companies to improve their wake-up word detection systems for voice assistants running on smart speakers. For example, this is the case for the speech corpora reported in \cite{Wang19}, \cite{Yiteng18} and \cite{Shan18}, which were collected, respectively, from Mobvoi's TicKasa Fox, Google's Google Home and Xiaomi's AI Speaker smart speakers. Unfortunately, only seven out of twenty six datasets in Table \ref{tab:datasets} are publicly available: one from Sonos \cite{Bluche20}, two different arrangements of AISHELL-2 \cite{Jiayu18} (used in \cite{Yan20}), the Google Speech Commands Dataset \emph{v1} \cite{GSCDv1} and \emph{v2} \cite{Warden18}, the Hey Snapdragon Keyword Dataset \cite{Kim19}, and Hey Snips \cite{Alice19, HeySnips} (also used in, e.g., \cite{Emre20, Kun20}). In case of interest in getting access to any of these speech corpora, the reader is pointed to the corresponding references indicated in Table \ref{tab:datasets}. Among these publicly available datasets, the Google Speech Commands Dataset (\emph{v1} and \emph{v2}) is, by far, the most popular, and has become the \emph{de facto} open reference for KWS development and evaluation. Because of this, the KWS performance comparison presented in Section \ref{sec:performance} is carried out among KWS systems that are evaluated on the Google Speech Commands Dataset. Further information on this corpus is provided in Subsection \ref{ssec:gscd}.

Also from Table \ref{tab:datasets}, we can observe that the great majority of datasets are noisy, which means that speech signals are distorted in different ways, e.g., by natural and realistic background acoustic noise or room acoustics. This is generally a must if we want to minimize the mismatch between the KWS performance at the lab phase and that one observable in the inherently-noisy real-life conditions. In particular, dataset acoustic conditions should be as close as possible as those that we expect to find when deploying KWS systems in real-life \cite{Gao20}. Noisy corpora can be classified as natural and/or simulated noisy speech:
\begin{enumerate}
    \item \emph{Natural noisy speech}: Some of the datasets in Table \ref{tab:datasets} (e.g., \cite{Chen18, Yiteng18, Alvarez19, Jingyong19, Xiong19, Yuan19, An19, Higuchi20}) were partially or totally created from natural noisy speech recorded ---many times in far-field conditions--- by electronic devices such as smart speakers, smartphones and tablets. Often, recording scenarios consist of home environments with background music or TV sound, since this is the target scenario of many KWS systems.
    \item \emph{Simulated noisy speech}: Some other noisy datasets conceived for KWS ---e.g., \cite{Parada14, Parada15, Parada15c, Arik17, Samuel18, Alvarez19, Liu19}--- were partially or totally generated by artificially distorting clean speech signals through a procedure called data augmentation \cite{Malek19}. Typically, given a clean speech signal, noisy copies of it are created by adding different types of background noises (e.g., daily life noises like babble, caf\'e, car, music and street noises) in such a manner that the resulting SNR levels (commonly, within the range $[-5,\;20]$ dB) are under control. Filtering and Noise-adding Tool (FaNT) \cite{FaNT} is a useful software to create such noisy copies. For example, FaNT was employed in \cite{Peter20, Lopez21b} to generate, in a controlled manner, noisier versions of the already noisy Google Speech Commands Dataset. Normally, background noises for data augmentation come from publicly available databases like TUT \cite{Mesaros16}, DEMAND \cite{DEMANDDataset}, MUSAN \cite{Snyder15}, NOISEX-92 \cite{Varga93} and CHiME \cite{CHiME3, CHiME3b}. In addition, alteration of room acoustics, e.g., to simulate far-field conditions from close-talk speech \cite{Arik17}, is another relevant data augmentation strategy.
\end{enumerate}
Collecting a good amount of natural noisy speech data in the desired acoustic conditions is not always feasible. In such cases, simulation of noisy speech is a smart and cheaper alternative allowing us for obtaining similar technology performance \cite{Povey17}.

We can clearly see from Table \ref{tab:datasets} that the number of keywords per dataset is mostly 1 or 2. A reason for this is that datasets mainly fit the application of KWS that, lately, is boosting research on this technology: wake-up word detection for voice assistants.

Finally, the right part of Table \ref{tab:datasets} tells some information about the sizes of the training and test sets\footnote{Many of these corpora also include a development set. However, this part has been omitted for the sake of clarity.} of the different corpora in terms of either the number of samples (i.e., words, normally) or time length in hours (h) ---depending on the available information---. Specifically, ``\emph{+ sampl.}'' (``\emph{- sampl.}'') refers to the size of the positive/keyword (negative/non-keyword) subset, and ``\emph{Size}'' denotes the magnitude of the whole set. Unknown information is indicated by hyphens. From this table, we note that, as a trend, publicly available datasets tend to be smaller than in-house ones. Furthermore, while the ratio between the sizes of the training and test sets is greater than 1 in all the reported cases except \cite{Bluche20}, ratio values tend to differ from one corpus to another. Also, \emph{mainly}, the ratio between the sizes of the corresponding negative/non-keyword and positive/keyword subsets is greater than 1, that is, $\frac{\mbox{\emph{\small - sampl.}}}{\mbox{\emph{\small + sampl.}}}>1$. This is purposely done to accurately reflect potential scenarios of use consisting of always-on KWS applications like wake-up word detection, in which KWS systems, most of the time, will be exposed to other types of words instead of keywords.

\subsection{Google Speech Commands Dataset}
\label{ssec:gscd}

\begin{table}
 \caption{List of the words included in the Google Speech Commands Dataset \emph{v1} (first six rows) and \emph{v2} (all the rows). Words are broken down by the standardized 10 keywords (first two rows) and non-keywords (last five rows).}
 \label{tab:word_list}
 \centering
 \setlength{\extrarowheight}{2pt}
 \resizebox{\linewidth}{!}{\begin{tabular}{ccccccc|c|}
 \hline
 \multicolumn{1}{|c}{\parbox[t]{2mm}{\multirow{6}{*}{\rotatebox[origin=c]{90}{Version 1 (\emph{v1})}}}} & \multicolumn{1}{|c|}{\parbox[t]{2mm}{\multirow{7}{*}{\rotatebox[origin=c]{90}{Version 2 (\emph{v2})}}}} & yes & no & up & down & left & \parbox[t]{2mm}{\multirow{2}{*}{\rotatebox[origin=c]{90}{KW}}}\\
 \multicolumn{1}{|c}{} & \multicolumn{1}{|c|}{} & right & on & off & stop & go & \\ \cline{3-8}
 \multicolumn{1}{|c}{} & \multicolumn{1}{|c|}{} & zero & one & two & three & four & \parbox[t]{2mm}{\multirow{5}{*}{\rotatebox[origin=c]{90}{Non-KW}}}\\
 \multicolumn{1}{|c}{} & \multicolumn{1}{|c|}{} & five & six & seven & eight & nine & \\
 \multicolumn{1}{|c}{} & \multicolumn{1}{|c|}{} & bed & bird & cat & dog & happy & \\
 \multicolumn{1}{|c}{} & \multicolumn{1}{|c|}{} & house & Marvin & Sheila & tree & wow & \\
 \cline{1-1}
 & \multicolumn{1}{|c|}{} & backward & forward & follow & learn & visual & \\
 \cline{2-8}
 \end{tabular}}
\end{table}

The publicly available Google Speech Commands Dataset \cite{GSCDv1, Warden18} has become the \emph{de facto} open benchmark for (deep) KWS development and evaluation. This crowdsourced database was captured at a sampling rate of 16 kHz by means of phone and laptop microphones, being, to some extent, noisy. Its first version, \emph{v1} \cite{GSCDv1}, was released in August 2017 under a Creative Commons BY 4.0 license \cite{cclicense}. Recorded by 1,881 speakers, this first version consists of 64,727 one-second (or less) long speech segments covering one word each out of 30 possible different words. The main difference between the first version and the second version ---which was made publicly available in 2018--- is that the latter incorporates 5 more words (i.e., a total of 35 words), more speech segments, 105,829, and more speakers, 2,618. Table \ref{tab:word_list} lists the words included in the Google Speech Commands Dataset \emph{v1} (first six rows) and \emph{v2} (all the rows). In this table, words are broken down by the standardized 10 keywords (first two rows) and non-keywords (last five rows). To facilitate KWS technology reproducibility and comparison, this benchmark also standardizes the training, development and test sets, as well as other crucial aspects of the experimental framework, including a training data augmentation procedure involving background noises (see, e.g., \cite{Tang18} for further details). Multiple recent deep KWS works have employed either the first version \cite{Raphael18, Tang18, Samuel18, Du18, Bai19, Riviello19, Zeng19, Choi19, Chen19, Peter20, Menglong20, Yang20, Mo20, Oleg20, Simon20, Ximin20, Kim21, Axel21} or the second version \cite{Lopez19, Zeng19, Emad19, Bo19, Lopez20, Chen20, Emre20, Peng20, Lin20, Oleg20, Simon20, Jung20, Lopez21, Lopez21b, Kim21, Axel21} of the Google Speech Commands Dataset.

Despite how valuable this open reference is for KWS research and development, we can raise two relevant points of criticism:
\begin{enumerate}
    \item \emph{Class balancing}: The different keyword and non-keyword classes are rather balanced (i.e., they appear with comparable frequencies) in this benchmark, which, as we know, is generally not realistic. See Subsection \ref{ssec:accuracy} for further comments on this question.
    \item \emph{Non-streaming mode}: Most of the above-referred works using the Google Speech Commands Dataset performs, due to the nature of this corpus, KWS evaluations in non-streaming mode, namely, multi-class classification of independent short input segments. In this mode, a full keyword or non-keyword is surely present within every segment. However, real-life KWS involves the continuous processing of an input audio stream.
\end{enumerate}
A few deep KWS research works \cite{Samuel18, Peter20, Lopez20, Lopez21b} have proposed to overcome the above two limitations by generating more realistic streaming versions of the Google Speech Commands Dataset by concatenation of one-second long utterances in such a manner that the resulting word class distribution is unbalanced. Even though the author of the Google Speech Commands Dataset reports some streaming evaluations in the database description manuscript \cite{Warden18}, still, we think that this point should be standardized for the sake of reproducibility and comparison, thereby enhancing the usefulness of this valuable corpus.

Lastly, we wish to draw attention to the fact that we produced three outcomes revolving around the Google Speech Commands Dataset \emph{v2}: \emph{1)} a variant of it emulating hearing aids as a capturing device (employed, as mentioned in Subsection \ref{ssec:hearing_devices}, for KWS for hearing assistive devices robust to external speakers) \cite{Lopez19, Lopez20}, \emph{2)} another noisier variant with a diversity of noisy conditions\footnote{Tools to create this noisy dataset can be freely downloaded from \url{http://ilopez.es.mialias.net/misc/NoisyGSCD.zip}} (i.e., types of noise and SNR levels) \cite{Lopez21b}, and \emph{3)} manually-annotated speaker gender labels\footnote{These labels are publicly available at \url{https://ilopezes.files.wordpress.com/2019/10/gscd\_spk\_gender.zip}}.

\section{Evaluation Metrics}
\label{sec:metrics}

Obviously, the gold plate test of any speech communication system is a test with relevant end-users. However, such tests tend to be costly and time-consuming. Instead (or in addition to subjective tests), one adheres to objective performance metrics for estimating system performance. It is important to choose a meaningful objective evaluation metric that allows us to determine the goodness of a system and is highly correlated to the subjective user experience. In what follows, we review and provide some criticism of the most common metrics considered in the field of KWS. These metrics are rather intended for binary classification ---e.g., keyword/non-keyword--- tasks. In the event of having multiple keywords, a common approach consists of applying the metric computation for every keyword and, then, the result is averaged, e.g., see \cite{Tang18, Lopez20, Lopez21b}.

\subsection{Accuracy}
\label{ssec:accuracy}

Accuracy can be defined as the ratio between the number of correct predictions/classifications and the total number of them \cite{accuracy}. In the context of binary classification (e.g., keyword/non-keyword), accuracy can also be expressed from the number of true positives (TP), false positives (FP), true negatives (TN) and false negatives (FN) as follows \cite{Mower05}:
\begin{equation}
 \mbox{Accuracy}=\frac{\mbox{TP}+\mbox{TN}}{\mbox{TP}+\mbox{TN}+\mbox{FP}+\mbox{FN}}.
\end{equation}
$\mbox{Accuracy}\in [0,\;1]$, where 0 and 1 indicate, respectively, worst and perfect classification.

\begin{figure}
    \centering
    \includegraphics[width=0.8\linewidth]{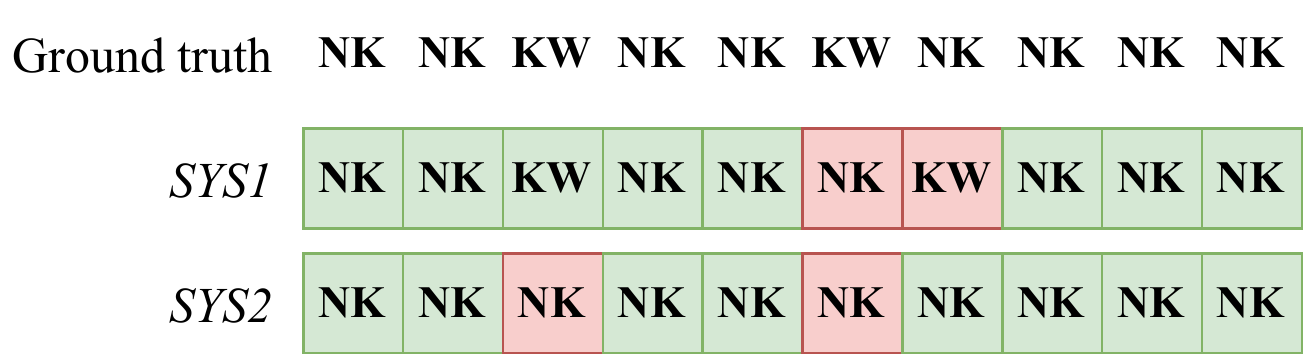}
    \caption{Example of two different KWS systems \emph{SYS1} and \emph{SYS2} recognizing a sequence of keywords (\textbf{KW}) and non-keywords (\textbf{NK}). The ground truth sequence is also shown on top.}
    \label{fig:acc_example}
\end{figure}

It is reasonable to expect that, in real-life applications like wake-up word detection, KWS systems will hear other word types rather than keywords most of the time. In other words, KWS is a task in which, in principle, the keyword and non-keyword classes are quite unbalanced. Under these circumstances, accuracy tends to be an unsuitable evaluation metric yielding potentially misleading conclusions \cite{Chawla2005, Hossin15}. Let us illustrate this statement with the following example. Let us consider two different KWS systems \emph{SYS1} and \emph{SYS2}. While \emph{SYS1} is a relatively decent system, \emph{SYS2} is a totally useless one, since it always outputs ``non-keyword'' regardless of the input. Figure \ref{fig:acc_example} depicts, along with an example ground truth sequence, the sequences of keywords (\textbf{KW}) and non-keywords (\textbf{NK}) predicted by \emph{SYS1} and \emph{SYS2}. In this situation, both KWS systems perform with 80\% accuracy, even though \emph{SYS2} is useless while \emph{SYS1} is not. Thus, particularly in unbalanced situations, more appropriate evaluation metrics than accuracy may be required, and these are discussed in the next subsections.

In spite of its disadvantage in unbalanced situations, accuracy is a widely used evaluation metric for deep KWS, especially when performing evaluations on the popular Google Speech Commands Dataset \cite{GSCDv1, Warden18} in non-streaming mode \cite{Samuel18, Tang18, Bo19, Choi19, Chen19, Zeng19, Riviello19, Bai19, Oleg20, Ximin20, Mo20, Menglong20, Peng20, Chen20, Wang21}. In this latter case, accuracy can still be considered a meaningful metric, since the different word classes are rather balanced in the Google Speech Commands Dataset benchmark. Hence, the main criticism that might be raised here is the lack of realism of the benchmark itself, as discussed in Subsection \ref{ssec:gscd}. Nevertheless, we have experimentally observed for KWS a strong correlation between accuracy on a quite balanced scenario and more suitable metrics like F-score (see Subsection \ref{ssec:fscore}) on a more realistic, unbalanced scenario \cite{Lopez20, Lopez21b}. This might suggest that the employment of accuracy, although not ideal, can still be useful under certain experimental conditions to adequately explain the goodness of KWS systems.

\subsection{Receiver Operating Characteristic and Detection Error Trade-Off Curves}
\label{ssec:roc_det}

\begin{figure}
    \centering
    \includegraphics[width=\linewidth]{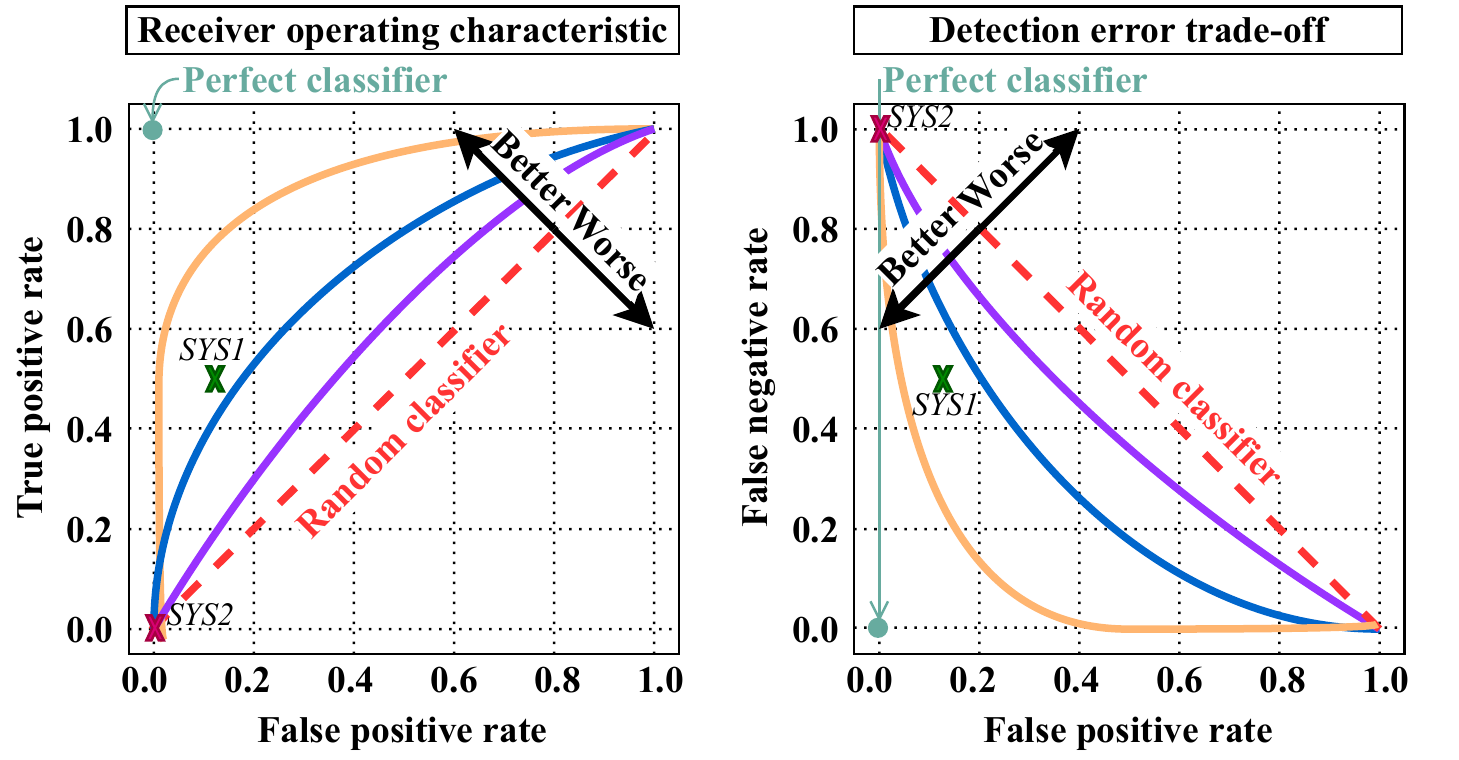}
    \caption{Outlining of the receiver operating characteristic (left) and detection error trade-off (right) curves. The location of \emph{SYS1} and \emph{SYS2} is indicated by green and red crosses, respectively. See the text for further explanation.}
    \label{fig:roc_example}
\end{figure}

Let TPR denote the true positive rate ---also known as recall \cite{Brodersen10}---, which is defined as the ratio
\begin{equation}
    \mbox{TPR}=\mbox{Recall}=\frac{\mbox{TP}}{\mbox{TP}+\mbox{FN}}.
    \label{eq:tpr}
\end{equation}
Notice that Eq. (\ref{eq:tpr}) is the probability that a positive sample (i.e., a keyword in this paper) is correctly detected as such. Similarly, let FPR be the false positive rate ---also known as false alarm rate---, namely, the probability that a negative sample (i.e., a non-keyword in our case) is wrongly classified as a positive one \cite{Pokrywka08}:
\begin{equation}
    \mbox{FPR}=\frac{\mbox{FP}}{\mbox{FP}+\mbox{TN}}.
    \label{eq:fpr}
\end{equation}
Then, a better and prominent way of evaluating the performance of a KWS system is by means of the receiver operating characteristic (ROC) curve, which consists of the plot of pairs of false positive and true positive rate values that are obtained by sweeping the sensitivity (decision) threshold \cite{Tom06}. The left part of Figure \ref{fig:roc_example} outlines example ROC curves. Coordinate $(\mbox{FPR}=0,\;\mbox{TPR}=1)$ in the upper left corner represents a perfect classifier. The closer to this point a ROC curve is, the better a classification system. In addition, a system performing on the ROC space identity line would be randomly guessing. The area under the curve (AUC), which equals the probability that a classifier ranks a randomly-chosen positive sample higher than a randomly-chosen negative one \cite{Tom06}, is also often employed as a ROC summary for KWS evaluation, e.g., \cite{Fuchs17, Kumar18, An19, Menon18, Benisty18, Menon19, Wintrode20, Zeyu20}. The larger the $\mbox{AUC}\in[0,\;1]$, the better a system is \cite{Lee07}.

Let us return for a moment to the example of Figure \ref{fig:acc_example}. It is easy to check that the KWS systems \emph{SYS1} and \emph{SYS2} would be characterized, in the ROC space, by the coordinates $(\mbox{FPR}=0.125,\;\mbox{TPR}=0.5)$ and $(\mbox{FPR}=0,\;\mbox{TPR}=0)$, respectively (see Figure \ref{fig:roc_example}). Unlike what happened when using accuracy, now we can rightly assess that \emph{SYS1} (above the random guessing line) is much better than \emph{SYS2} (on the random  guessing line).

An alternative (with no particular preference) to the ROC curve (e.g., \cite{Sundar15, Sankaran16, Guo18, Kun20}) is the detection error trade-off (DET) curve \cite{Martin97}. From the right part of Figure \ref{fig:roc_example}, it can be seen that a DET curve is like a ROC curve except for the $y$-axis being false negative rate ---also known as miss rate \cite{Broadley18}---, FNR:
\begin{equation}
 \mbox{FNR}=\frac{\mbox{FN}}{\mbox{FN}+\mbox{TP}}.
\end{equation}
This time, coordinate $(\mbox{FPR}=0,\;\mbox{FNR}=0)$ in the bottom left corner represents a perfect classifier. The closer to this point a DET curve is, the better a classification system. Therefore, the smaller the $\mbox{AUC}\in[0,\;1]$ in this case, the better a system is. Notice that, as $\mbox{FNR}=1-\mbox{TPR}$, the DET curve is nothing else but a vertically-flipped version of the ROC curve. From the DET curve we can also straightforwardly obtain the equal error rate (EER) as the intersection point between the identity line and the DET curve (i.e., the point at which $\mbox{FNR}=\mbox{FPR}$) \cite{David07}. Certainly, the lower the EER value, the better. Though the use of EER is much more widespread in the field of speaker verification \cite{Larcher14, Bai20, sarkar2019time}, this DET summary is sometimes considered for KWS evaluation \cite{Yimeng16, Ravi18, Kumar18, Menon18, Leem19, Menon19, Jung20}.

In real-world KWS applications, typically, the cost of a false alarm is significantly greater than that of a miss detection\footnote{Evidently, in these circumstances, EER may not be a good metric candidate for system comparison.} \cite{OpenKWS13}. This is for example the case for voice activation of voice assistants, where privacy is a major concern \cite{tang-etal-2020-howl} since this application involves streaming voice to a cloud server. As a result, a popular variant of the ROC and DET curves is that one replacing false positive rate along the $x$-axis by the number of false alarms per hour \cite{Parada15c, Yuxuan17, He17, Yiyan18, Alvarez19, Yiteng19, Kim19, Higuchi20}. By this, a practitioner can just set a very small number of false alarms per hour (e.g., 1) and identify the system with the highest (lowest) true positive (false negative) rate for deployment. An alternative good selection criterion consists of picking up the system maximizing, at a particular system-dependent sensitivity threshold, the so-called term-weighted value (TWV) \cite{OpenKWS13, Pavel15, Haipeng15, Leung16, Huang16, Nancy16, Trmal17, Pattanayak19}. Given a sensitivity threshold, TWV is a weighted linear combination of the false negative and false positive rates as in
\begin{equation}
    \mbox{TWV}=1-\left(\mbox{FNR}+\beta\mbox{FPR}\right),
\end{equation}
where $\beta\gg 1$ (e.g., $\beta=999.9$ \cite{OpenKWS13}) is a constant expressing the greater cost of a false alarm with respect to that of a miss detection.

\subsection{Precision-Recall and F-Score Curves}
\label{ssec:fscore}

\begin{figure}
    \centering
    \includegraphics[width=\linewidth]{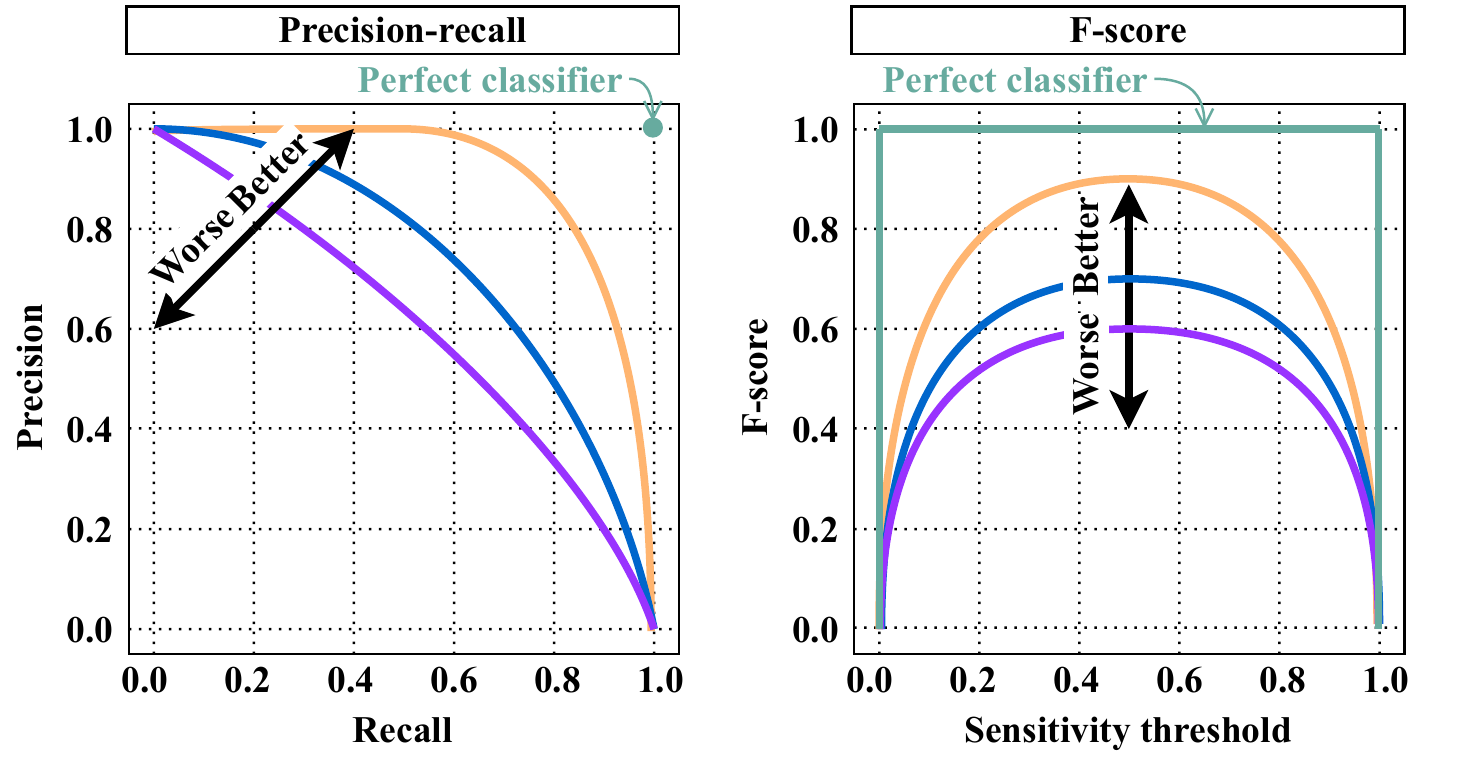}
    \caption{Outlining of the precision-recall (left) and F-score (right) curves. See the text for further explanation.}
    \label{fig:prf_example}
\end{figure}

The precision-recall curve \cite{Davis06} is another important visual performance analysis tool for KWS systems (e.g., \cite{Enea19, Yue19, Chai19, Lopez20}). Let precision, also known as positive predictive value \cite{Saito15}, be the probability that a sample that is classified as positive is actually a positive sample:
\begin{equation}
    \mbox{Precision}=\frac{\mbox{TP}}{\mbox{TP}+\mbox{FP}}.
    \label{eq:precision}
\end{equation}
Then, the precision-recall curve plots pairs of recall (equivalently, TPR, see Eq. (\ref{eq:tpr})) and precision values that, as in the case of the ROC and DET curves, are obtained by sweeping the sensitivity threshold. This definition is schematized by the left part of Figure \ref{fig:prf_example}, where a perfect classifier lies on the coordinate $(\mbox{Recall}=1,\;\mbox{Precision}=1)$. The closer to this point a precision-recall curve is and the larger the $\mbox{AUC}\in[0,\;1]$, the better a classifier. This time, a (precision-recall) random guessing line has not been drawn, since it depends on the proportion of the positive class within both classes \cite{Saito15}. For example, while in a balanced scenario random guessing would be characterized by a horizontal line at a precision of 0.5, we can expect that such a line is closer to 0 precision in the event of the KWS problem due to the highly imbalance nature of it.

The close relationship between the ROC (and DET) and precision-recall curves can be intuited, and, in fact, there exists a one-to-one correspondence between both of them \cite{Davis06}. However, the precision-recall curve is considered to be a more informative visual analysis tool than the ROC one in our context \cite{Saito15}. This is because, thanks to the use of precision, the precision-recall curve allows us to better focus on the \emph{minority} positive (i.e., keyword) class of interest (see Eq. (\ref{eq:precision})). On the precision-recall plane, while \emph{SYS1} lies on the point $(\mbox{Recall}=0.5,\;\mbox{Precision}=0.5)$, precision is undefined (i.e., $\mbox{Precision}=0/0$) for \emph{SYS2}, which should alert us to the existence of a problem with the latter system.

From precision and recall we can formulate the F-score metric \cite{FScore}, $F_1$, which is often used for KWS evaluation, e.g., \cite{Jianbin18, Enea19, Chai19, Lopez20, Lopez21b, Wei21}. F-score is the harmonic mean of precision and recall, that is,
\begin{equation}
    F_1=\frac{2}{\mbox{Recall}^{-1}+\mbox{Precision}^{-1}}=\frac{2\mbox{TP}}{2\mbox{TP}+\mbox{FP}+\mbox{FN}},
    \label{eq:fscore}
\end{equation}
where $0\le F_1\le 1$, and the larger $F_1$, the better. Indeed, as for precision and recall, F-score can be calculated as a function of the sensitivity threshold and plotted as exemplified by the right part of Figure \ref{fig:prf_example}. In this representation, we assume that a KWS system provides confidence scores resulting from posterior probabilities, and this is why the sensitivity threshold ranges from 0 to 1. The larger the $\mbox{AUC}\in[0,\;1]$, the better a system is. A perfect classifier would be characterized by an AUC of 1. As in the case of the precision-recall curve, a random guessing line has not been drawn either on the F-score space, since this similarly depends on the proportion between the positive and negative classes. Finally, let us notice that F-score is 0.5 and 0 for \emph{SYS1} and \emph{SYS2}, respectively, which clearly indicates the superiority of \emph{SYS1} with respect to \emph{SYS2}.

\section{Performance Comparison}
\label{sec:performance}

\begin{table*}[th]
	\caption{Performance comparison among some of the latest deep KWS systems in terms of both accuracy (\%) and computational complexity (i.e., number of parameters and multiplications) of the acoustic model. Accuracy, provided with confidence intervals for some systems, is on the Google Speech Commands Dataset (GSCD) \emph{v1} and \emph{v2}. The reported values are directly taken from the references in the ``Description'' column. Unknown information is indicated by hyphens.}
	\label{tab:results}
	\centering
	\setlength{\extrarowheight}{3pt}
	\resizebox{\linewidth}{!}{\begin{tabular}{clccccc}
			\toprule
			\multicolumn{1}{c}{\textbf{ID}} & \textbf{Description} & \textbf{Year} & \multicolumn{2}{c}{\textbf{Accuracy (\%)}} & \multicolumn{2}{c}{\textbf{Computational complexity}} \\ \cline{4-5} \cline{6-7}
			\multicolumn{3}{c}{} & GSCD \emph{v1} & GSCD \emph{v2} & No. of params. & No. of mults. \\ \midrule
			\multicolumn{1}{c}{1} & Standard FFNN with a pooling layer \cite{Oleg20} & 2020 & 91.2 & 90.6 & 447k & -- \\
			\multicolumn{1}{c}{2} & DenseNet with trainable window function and mixup data augmentation \cite{Du18} & 2018 & 92.8 & -- & -- & -- \\
			\multicolumn{1}{c}{3} & Two-stage TDNN \cite{Samuel18} & 2018 & 94.3 & -- & 251k & 25.1M \\
			\multicolumn{1}{c}{4} & CNN with striding \cite{Oleg20} & 2018 & 95.4 & 95.6 & 529k & -- \\
			\multicolumn{1}{c}{5} & BiLSTM with attention \cite{Coimbra18} & 2018 & 95.6 & 96.9 & 202k & -- \\
			\multicolumn{1}{c}{6} & Residual CNN \texttt{res15} \cite{Tang18} & 2018 & 95.8 $\pm$ 0.484 & -- & 238k & 894M \\
			\multicolumn{1}{c}{7} & TDNN with shared weight self-attention \cite{Bai19} & 2019 & 95.81 $\pm$ 0.191 & -- & 12k & 403k \\
			\multicolumn{1}{c}{8} & DenseNet+BiLSTM with attention \cite{Zeng19} & 2019 & 96.2 & 97.3 & 223k & -- \\
			\multicolumn{1}{c}{9} & Residual CNN with temporal convolutions \texttt{TC-ResNet14} \cite{Choi19} & 2019 & 96.2 & -- & 137k & -- \\
			\multicolumn{1}{c}{10} & SVDF \cite{Oleg20} & 2019 & 96.3 & 96.9 & 354k & -- \\
			\multicolumn{1}{c}{11} & SincConv+(Grouped DS-CNN) \cite{Simon20} & 2020 & 96.4 & 97.3 & 62k & -- \\
			\multicolumn{1}{c}{12} & Graph convolutional network \texttt{CENet-40} \cite{Chen19} & 2019 & 96.4 & -- & 61k & 16.18M \\
			\multicolumn{1}{c}{13} & GRU \cite{Oleg20} & 2020 & 96.6 & 97.2 & 593k & -- \\
			\multicolumn{1}{c}{14} & SincConv+(DS-CNN) \cite{Simon20} & 2020 & 96.6 & 97.4 & 122k & -- \\
			\multicolumn{1}{c}{15} & Temporal CNN with depthwise convolutions \texttt{TENet12} \cite{Ximin20} & 2020 & 96.6 & -- & 100k & 2.90M \\
			\multicolumn{1}{c}{16} & Residual DS-CNN with squeeze-and-excitation \texttt{DS-ResNet18} \cite{Menglong20} & 2020 & 96.71 $\pm$ 0.195 & -- & 72k & 285M \\
			\multicolumn{1}{c}{17} & \texttt{TC-ResNet14} with neural architecture search \texttt{NoisyDARTS-TC14} \cite{Bo21} & 2021 & 96.79 $\pm$ 0.30 & 97.18 $\pm$ 0.26 & 108k & 6.3M \\
			\multicolumn{1}{c}{18} & LSTM \cite{Oleg20} & 2020 & 96.9 & 97.5 & -- & -- \\
			\multicolumn{1}{c}{19} & DS-CNN with striding \cite{Oleg20} & 2018 & 97.0 & 97.1 & 485k & -- \\
			\multicolumn{1}{c}{20} & CRNN \cite{Oleg20} & 2020 & 97.0 & 97.5 & 467k & -- \\
			\multicolumn{1}{c}{21} & BiGRU with multi-head attention \cite{Oleg20} & 2020 & 97.2 & 98.0 & 743k & -- \\
			\multicolumn{1}{c}{22} & CNN with neural architecture search \texttt{NAS2\_6\_36} \cite{Mo20} & 2020 & 97.22 & -- & 886k & -- \\
			\multicolumn{1}{c}{23} & Keyword Transformer \texttt{KWT-3} \cite{Axel21} & 2021 & 97.49 $\pm$ 0.15 & 98.56 $\pm$ 0.07 & 5.3M & -- \\
			\multicolumn{1}{c}{24} & Variant of TC-ResNet with self-attention \texttt{LG-Net6} \cite{Wang21} & 2021 & 97.67 & 96.79 & 313k & -- \\
			\multicolumn{1}{c}{25} & Broadcasted residual CNN \texttt{BC-ResNet-8} \cite{Kim21} & 2021 & 98.0 & 98.7 & 321k & 89.1M \\
			\bottomrule
	\end{tabular}}
\end{table*}

\begin{figure}
    \centering
    \includegraphics[width=\linewidth]{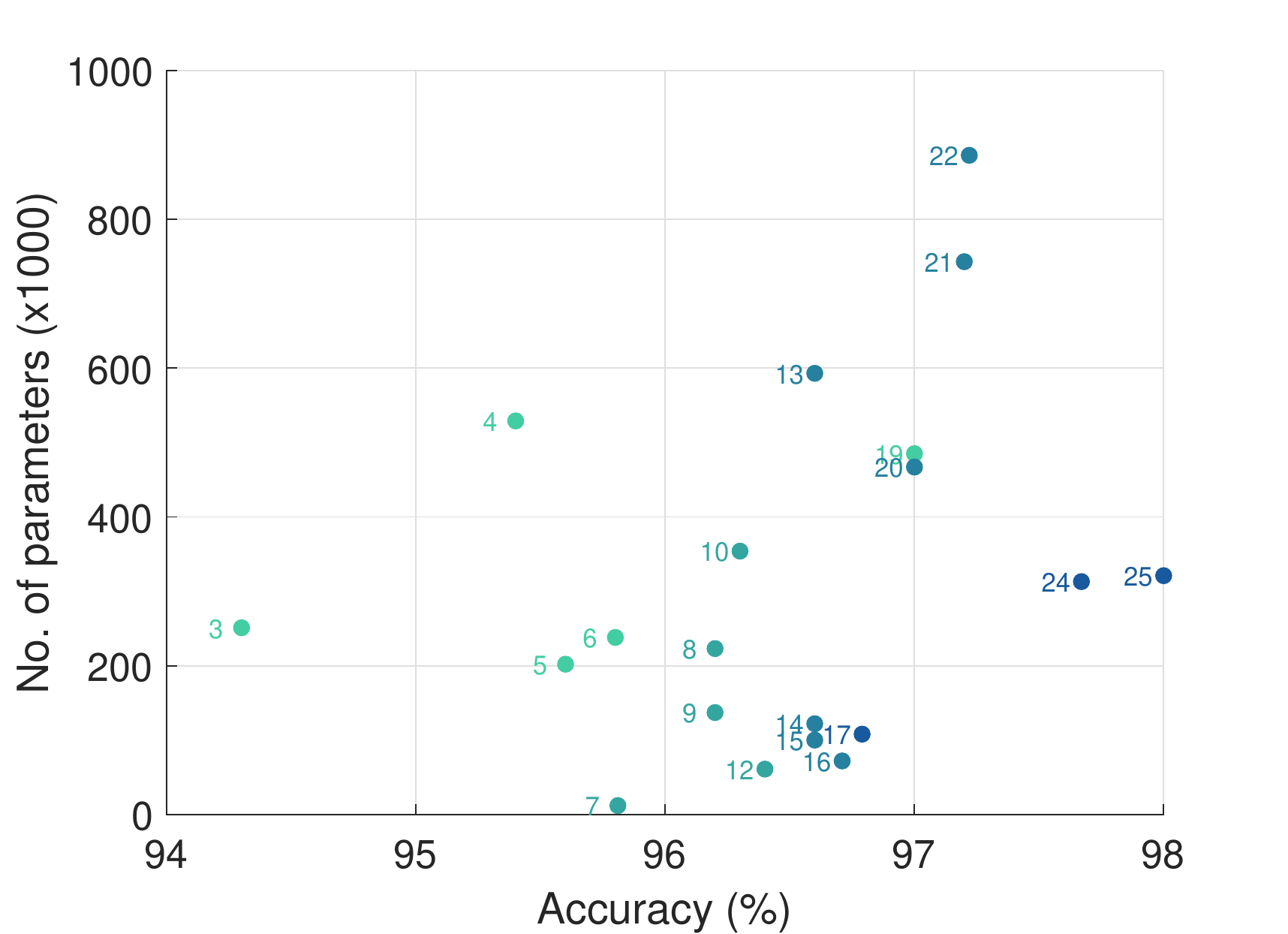}
    \caption{Location of some of the deep KWS systems of Table \ref{tab:results} on the plane defined by the dimensions ``number of parameters'' and ``accuracy'' (on the Google Speech Commands Dataset \emph{v1}). Better systems can be found on the lower right corner of this plane. The systems are identified by the numbers in the ``ID'' column of Table \ref{tab:results}. More recent systems are marked with a darker color.}
    \label{fig:acc_vs_params}
\end{figure}

In this section, we present a performance comparison among some of the latest and most relevant deep KWS systems reviewed throughout this manuscript. This comparison is carried out in terms of both KWS performance and computational complexity of the acoustic model, which is the main distinctive component of every system.

To measure KWS performance, we examine accuracy of systems in non-streaming mode on the Google Speech Commands Dataset (GSCD) \emph{v1} and \emph{v2} (described in Subsection \ref{ssec:gscd}), which standardize 10 keywords (see Table \ref{tab:word_list}). In this way, since the publicly available GSCD has become the \emph{de facto} open benchmark for deep KWS, we can straightforwardly use accuracy values reported in the literature in order to rank the most prominent deep KWS systems. Regarding accuracy as an evaluation metric, recall that this metric, although not ideal, is still meaningful under the GSCD experimental conditions to explain the goodness of KWS systems, as discussed in Subsection \ref{ssec:accuracy}.

On the other hand, the number of parameters and multiplications of the acoustic model is used to evaluate the computational complexity of the systems. Notice that these measures are a good approximation to the complexity of the entire deep KWS system since the acoustic model is, by far, the most demanding component in terms of computation. Actually, in \cite{Raphael18}, Tang \emph{et al.} show that the number of parameters and, especially, the number of multiplications of the acoustic model are solid proxies predicting the power consumption of these systems.

Table \ref{tab:results} shows a performance comparison among some of the latest deep KWS systems in terms of both accuracy on the GSCD \emph{v1} and \emph{v2} (in percentages), and complexity of the acoustic model. The reported values are directly taken from the references in the ``Description'' column, while hyphens indicate non-available information. Notice that some of the accuracy values in Table \ref{tab:results} are shown along with confidence intervals that are calculated across different acoustic models trained with different random model initialization. Deep KWS systems are listed in ascending order in terms of their accuracy on the first version of the GSCD. From Table \ref{tab:results}, it can be observed that KWS performance on GSCD \emph{v2} tends to be slightly better than that on the first version of this dataset. This behavior could be related to the fact that the second version of this dataset has more word samples (see Table \ref{tab:datasets}), which might lead to better trained acoustic models.

Also from Table \ref{tab:results}, we can see the wide variety of architectures (e.g., standard FFNNs, SVDFs, TDNNs, CNNs, RNNs and CRNNs) integrating different elements (e.g., attention, residual connections and/or depthwise separable convolutions) that has been explored for deep KWS. It is not surprising that the worst-performing system is that whose acoustic model is based on a standard and relatively heavy (447k parameters) FFNN \cite{Oleg20} (ID 1 in Table \ref{tab:results}). Besides, the most frequently used acoustic model type is based on CNN. This surely is because CNNs are able to provide a highly competitive performance ---thanks to exploiting local speech time-frequency correlations--- while typically involving lesser computational complexity than other well-performing types of models like RNNs.

Furthermore, it is interesting to note the capability of neural architecture search techniques \cite{Barret17} to automatically produce acoustic models performing better than those manually designed. Thus, the performance of the residual CNN with temporal convolutions \texttt{TC-ResNet14} \cite{Choi19} (ID 9) is improved when \texttt{NoisyDARTS-TC14} \cite{Bo21} (ID 17) automatically searches for kernel sizes, additional skip connections and enabling or not squeeze-and-excitation \cite{Gang18}. Even better, this is achieved by employing fewer parameters, i.e., 137k \emph{versus} 108k. In addition, the CNN with neural architecture search \texttt{NAS2\_6\_36} \cite{Mo20} (ID 22) reaches an outstanding performance (97.22\% accuracy on the GSCD \emph{v1}), though at the expense of using a large number of parameters (886k).

The effectiveness of CRNNs combining CNNs and RNNs (see Subsection \ref{ssec:rnns}) can also be assessed from Table \ref{tab:results}. For instance, the combination of DenseNet\footnote{Recall that DenseNet is an extreme case of residual CNN with a hive of skip connections.} \cite{DenseNet} with a BiLSTM network with attention as in \cite{Zeng19} (ID 8) yields superior KWS accuracy with respect to considering standalone either DenseNet \cite{Du18} (ID 2) or a BiLSTM network with attention \cite{Coimbra18} (ID 5). Moreover, we can see that the performance of a rather basic CRNN incorporating a GRU layer \cite{Oleg20} (ID 20) is quite competitive.

Due to the vast number of disparate factors contributing to the performance of the deep KWS systems of Table \ref{tab:results}, it is extremely difficult to draw strong conclusions and even trends far beyond the ones indicated above. Figure \ref{fig:acc_vs_params} gives another perspective of Table \ref{tab:results} by plotting the location of some of the systems of this table on the plane defined by the dimensions ``number of parameters'' and ``accuracy'' (on the GSCD \emph{v1}). In this figure, the systems are identified by the numbers in the ``ID'' column of Table \ref{tab:results}.

Since more recent deep KWS systems are marked with a darker color in Figure \ref{fig:acc_vs_params}, it can be clearly observed that, primarily, the driving force is the optimization of KWS performance, where the computational complexity, although important, is relegated to a secondary position. A good example of this is the so-called Keyword Transformer \texttt{KWT-3} \cite{Axel21} (ID 23), a fully self-attentional Transformer \cite{Vaswani17} that is an adaptation of Vision Transformer \cite{Alexey21} to the KWS task. \texttt{KWT-3} (not included in Figure \ref{fig:acc_vs_params}), which achieves state-of-the-art performance (97.49\% and 98.56\% accuracy on the GSCD \emph{v1} and \emph{v2}, respectively), has the extraordinary amount of more than 5 million parameters. That being said, generally, we will be more interested in systems exhibiting both high accuracy and a small footprint, i.e., in systems that can be found on the lower right corner of the plane in Figure \ref{fig:acc_vs_params}. In this region of the plane we have the following two groups of systems:
\begin{enumerate}
    \item \emph{Systems with IDs 14, 15, 16 and 17}: These systems are characterized by a good KWS performance along with a particularly reduced number of parameters. All of them are based on CNNs while most of them integrate residual connections and/or depthwise separable convolutions. Furthermore, the three best performing systems (with IDs 15, 16 and 17) integrate either dilated or temporal convolutions to exploit long time-frequency dependencies.
    \item \emph{Systems with IDs 24 and 25}: These two systems are characterized by an outstanding KWS performance along with a relatively small number of parameters. Both of them are based on CNNs and they integrate residual connections and a mechanism to exploit long time-frequency dependencies: dilated convolutions in System 25, and temporal convolutions and self-attention layers in System 24. System 25 also incorporates depthwise separable convolutions.
\end{enumerate}
The analysis of the above two groups of systems very much reinforces our summary reflections concluding Subsection \ref{ssec:cnns}. In other words, a state-of-the-art KWS system comprising a CNN-based acoustic model should cover the following three elements in order to reach a high performance with a small footprint: a mechanism to exploit long time-frequency dependencies, depthwise separable convolutions \cite{Andrew17} and residual connections \cite{He16}.

\section{Audio-Visual Keyword Spotting}
\label{sec:avkws}

\begin{figure}
    \centering
    \includegraphics[width=\linewidth]{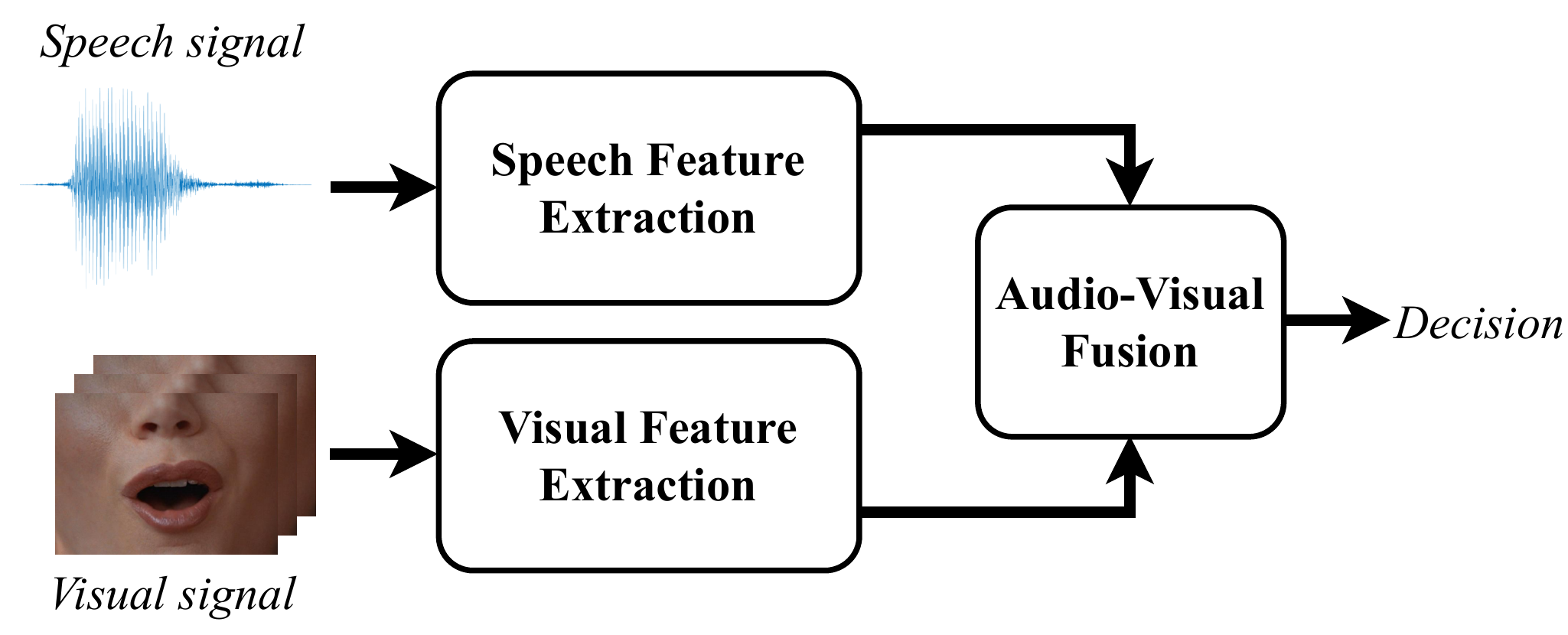}
    \caption{General diagram of a modern audio-visual keyword spotting system.}
    \label{fig:avkws}
\end{figure}

In face-to-face human communication, observable articulators like the lips are an important information source. In other words, human speech perception is bimodal, since it relies on both auditory and visual information. Similarly, speech processing systems such as ASR systems can be benefited from exploiting visual information along with the audio information to enhance their performance \cite{Su19, Makino19, Pan19, Helen20}. This can be particularly fruitful in real-world scenarios where severe acoustic distortions (e.g., strong background noise and reverberation) are present, since the visual information is not affected by acoustic distortions.

While fusion of audio-visual information is a quite active research area in ASR (e.g., see \cite{Su19, Makino19, Pan19, Helen20}), very few works have studied it for (deep) KWS \cite{Wu16, Ding18, Momeni20}. Figure \ref{fig:avkws} illustrates the general diagram of a modern audio-visual KWS system. First, speech and visual features are extracted. In former audio-visual KWS work \cite{Wu16, Ding18}, visual feature extraction consists of a pipeline comprising face detection and lip localization (via landmark estimation), and visual feature extraction itself from the lips crop. Nowadays, the use of a deep learning model fed with raw images containing the uncropped speaker's face seems to be the preferred approach for visual feature extraction \cite{Momeni20}. Finally, the extracted audio-visual information is fused in order to come up with a decision about the presence or not of a keyword. Typically, one of the two following fusion strategies is considered in practice \cite{Lee08}:
\begin{enumerate}
    \item \emph{Feature-level fusion}: Speech and visual features are somehow combined (e.g., concatenated) before their joint classification using a neural network model.
    \item \emph{Decision-level fusion}: The final decision is formed from the combination of the decisions from separate speech and visual neural network-based classifiers. This well-performing approach seems to be preferred \cite{Wu16, Ding18, Momeni20} over the feature-level fusion scheme and is less data-hungry than feature-level fusion \cite{Wu16}.
\end{enumerate}
Notice that thanks to the integration of visual information ---which, as aforementioned, is not affected by acoustic distortions---, audio-visual KWS achieves the greatest relative improvements with respect to audio-only KWS at lower SNRs \cite{Wu16, Ding18, Momeni20}.

For those who are interested in audio-visual KWS research, the following realistic and challenging audio-visual benchmarks can be of interest: Lip Reading in the Wild (LRW) \cite{Joon16}, and Lip Reading Sentences 2 (LRS2) \cite{Chung17} and 3 (LRS3) \cite{LRS3} datasets. While LRW comprises single-word utterances from BBC TV broadcasts, LRS2 and LRS3 consist of thousands of spoken sentences from BBC TV and TED(x) talks, respectively.

\section{Conclusions and Future Directions}
\label{sec:conclusions}

The goal of this article has been to provide a comprehensive overview of state-of-the-art KWS technology, namely, of deep KWS. We have seen that the core of this paradigm is a DNN-based acoustic model whose goal is the generation, from speech features, of posterior probabilities that are subsequently processed to detect the presence of a keyword. Deep spoken KWS has revitalized KWS research by enabling a massive deployment of this technology for real-world applications, especially in the area of voice assistant activation.

We foresee that, as has been happening to date, advances in ASR research will dramatically continue impacting the field of KWS. In particular, we think that the expected progress in end-to-end ASR \cite{Parcollet20} (replacing handcrafted speech features by optimal feature learning integrated in the acoustic model) will also be reflected in KWS.

Immediate future work will keep focusing on advancing acoustic modeling towards two goals simultaneously: \emph{1)} improving KWS performance in real-life acoustic conditions, and \emph{2)} computational complexity reduction. With these two goals in mind, surely, acoustic model research will be mainly focused on the development of novel and efficient convolutional blocks. This is because of the good properties of CNNs allowing us to achieve an outstanding performance with a small footprint, as has been widely discussed throughout this paper. Furthermore, based on its promising initial results for KWS \cite{Mo20, Bo21}, we expect that neural architecture search will play a greater role in acoustic model architecture design.

Specifically within the context of computational complexity reduction, acoustic model compression will be, more than ever, a salient research line \cite{Yao21}. Indeed, this is driven by the numerous applications of KWS that involve embedding KWS technology in small electronic devices characterized by severe memory, computation and power constraints. Acoustic model compression entails three major advantages: \emph{1)} reduced memory footprint, \emph{2)} decreased inference latency, and \emph{3)} less energy consumption. All of this is of utmost importance for, e.g., enabling on-device acoustic model re-training for robustness purposes or personalized keyword inclusion. Acoustic model compression research will undoubtedly encompass model parameter quantization, neural network pruning and knowledge distillation \cite{Hinton14}, among other approaches.

Another line of research that might experience a notable growth in the short term could be semi-supervised learning \cite{Ouali20} for KWS. Especially in an industrial environment, it is simple to daily collect a vast amount of speech data from users of cloud speech services. These data are potentially valuable to strengthen KWS acoustic models. However, the cost of labeling such an enormous amount of data for discriminative model training can easily be prohibitively expensive. To not ``waste'' these unlabeled speech data, semi-supervised learning methodologies can help by allowing hybrid learning based on both small and big volumes of labeled and unlabeled data, respectively.

On the other hand, consumers seem to increasingly demand or, at least, value a certain degree of personalization when it comes to consumer electronics. While some research has already addressed some KWS personalization aspects (as we have discussed in this article), we foresee that KWS personalization will become even more relevant in the immediate future. This means that we can expect new research going deeper into topics like \emph{efficient} open-vocabulary (personalized) KWS and joint KWS and speaker verification \cite{Sigtia20, Jia21}.

Last but not least, recall that KWS technology is many times intended to run on small devices like smart speakers and wearables that typically embed more than one microphone. This type of multi-channel information has been successfully leveraged by ASR in different ways (which includes, e.g., beamforming) to provide robustness against acoustic distortions \cite{Lopez17, Lopez18}. Surprisingly, and as previously outlined in Section \ref{sec:robustness}, multi-channel KWS has only been marginally studied. Therefore, we expect that this rather unexplored area is worthy to be examined, which could lead to contributions further improving KWS performance in real-life (i.e., noisy) conditions.

\bibliographystyle{IEEEtran}
\bibliography{mybibfile}

\begin{IEEEbiography}[{\includegraphics[width=1in,height=1.25in,clip,keepaspectratio]{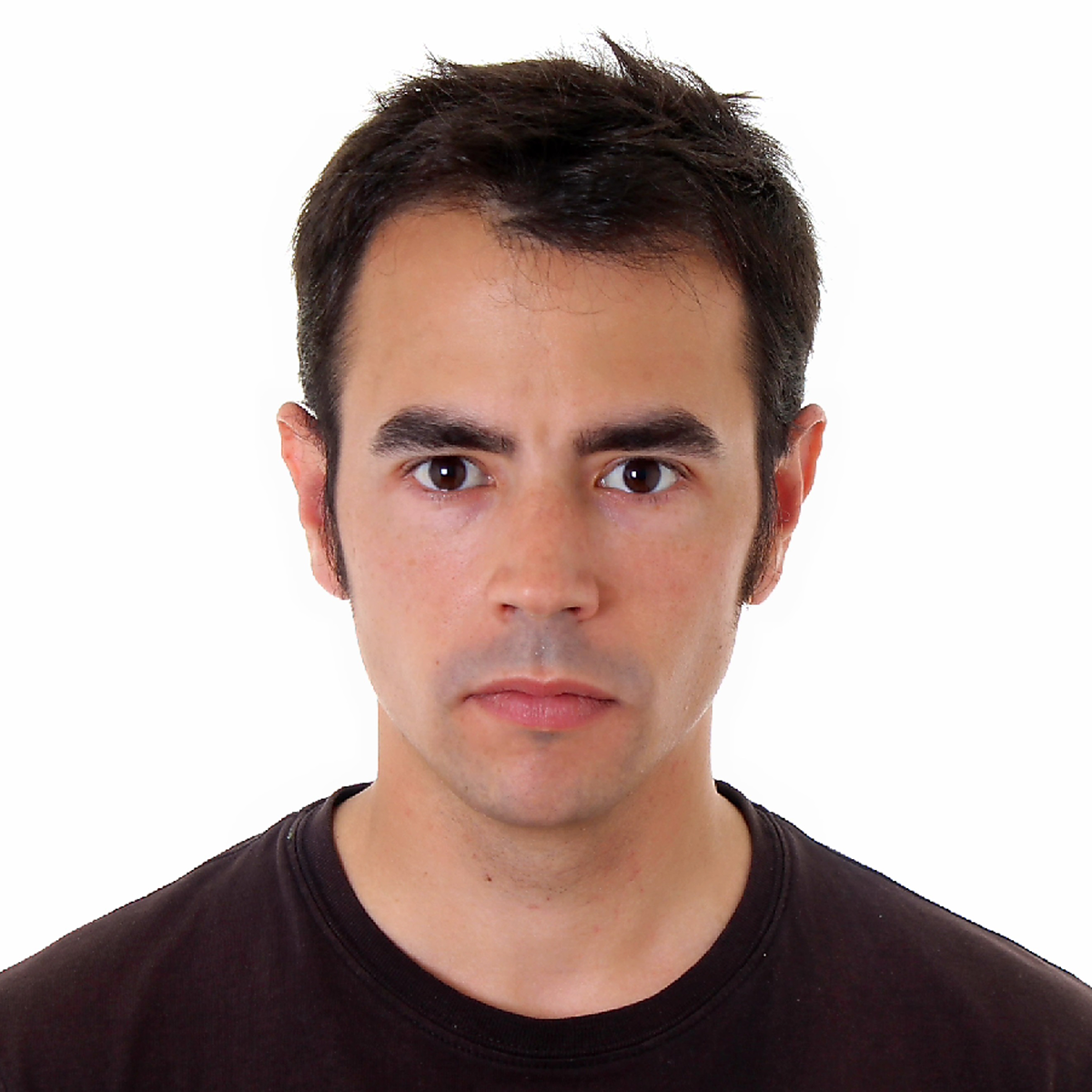}}]{\textbf{Iv\'an L\'opez-Espejo}} received the M.Sc. degree in Telecommunications Engineering, the M.Sc. degree in Electronics Engineering and the Ph.D. degree in Information and Communications Technology, all from the University of Granada, Granada (Spain), in 2011, 2013 and 2017, respectively. In 2018, he was the leader of the speech technology team of Veridas, Pamplona (Spain). Since 2019, he is a post-doctoral researcher at the section for Artificial Intelligence and Sound at the Department of Electronic Systems of Aalborg University, Aalborg (Denmark). His research interests include speech enhancement and robust speech recognition, multi-channel speech processing, and speaker verification.
\end{IEEEbiography}

\begin{IEEEbiography}[{\includegraphics[width=1in,height=1.25in,clip,keepaspectratio]{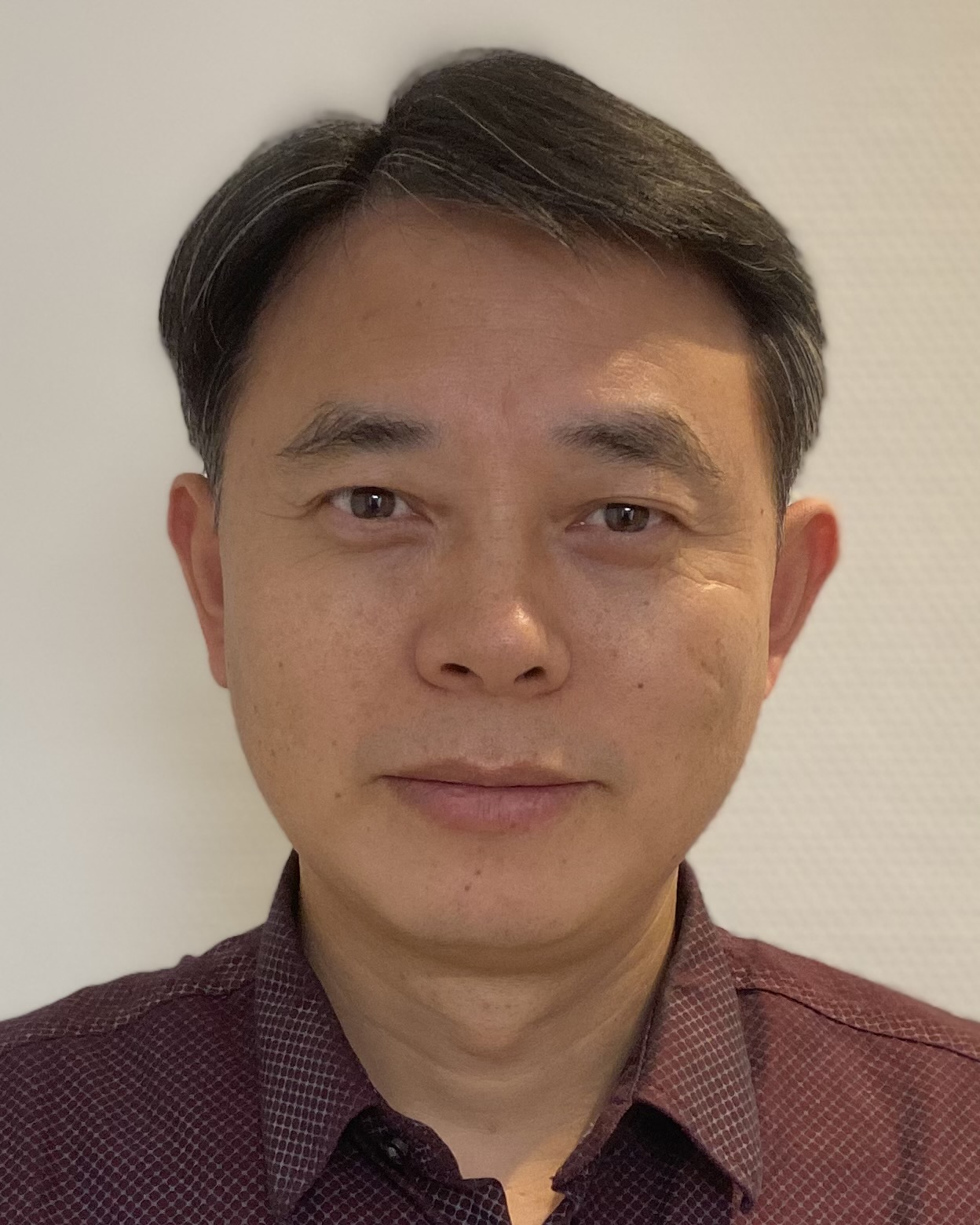}}]{\textbf{Zheng-Hua Tan}} (M'00--SM'06) received the B.Sc. and M.Sc. degrees in electrical engineering from Hunan University, Changsha, China, in 1990 and 1996, respectively, and the Ph.D. degree in electronic engineering from Shanghai Jiao Tong University (SJTU), Shanghai, China, in 1999.

He is a Professor in the Department of Electronic Systems and a Co-Head of the Centre for Acoustic Signal Processing Research at Aalborg University, Aalborg, Denmark. He was a Visiting Scientist at the Computer Science and Artificial Intelligence Laboratory, MIT, Cambridge, USA, an Associate Professor at SJTU, Shanghai, China, and a postdoctoral fellow at KAIST, Daejeon, Korea. His research interests include machine learning, deep learning, pattern recognition, speech and speaker recognition, noise-robust speech processing, multimodal signal processing, and social robotics. He has (co)-authored over 200 refereed publications. He is the Chair of the IEEE Signal Processing Society Machine Learning for Signal Processing Technical Committee (MLSP TC). He is an Associate Editor for the IEEE/ACM TRANSACTIONS ON AUDIO, SPEECH AND LANGUAGE PROCESSING. He has served as an Editorial Board Member for Computer Speech and Language and was a Guest Editor for the IEEE JOURNAL OF SELECTED TOPICS IN SIGNAL PROCESSING and Neurocomputing. He was the General Chair for IEEE MLSP 2018 and a TPC Co-Chair for IEEE SLT 2016.
\end{IEEEbiography}

\begin{IEEEbiography}[{\includegraphics[width=1in,height=1.25in,clip,keepaspectratio]{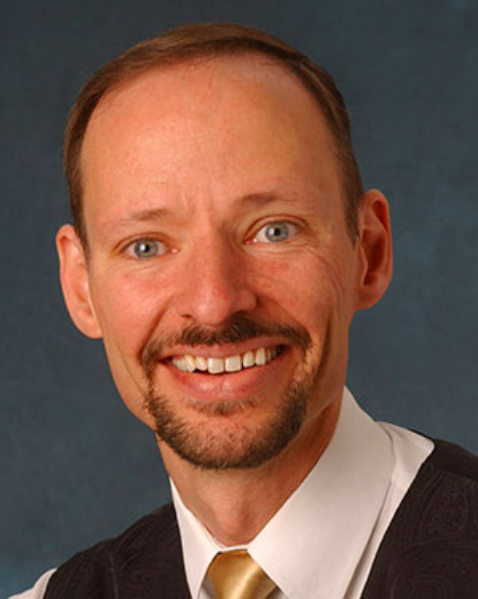}}]{\textbf{John H. L. Hansen}} (Fellow, IEEE) received the B.S.E.E. degree from the College of Engineering, Rutgers University, New Brunswick, NJ, USA, and the M.S. and Ph.D. degrees in electrical engineering from the Georgia Institute of Technology, Atlanta, GA, USA. In 2005, he joined the Erik Jonsson School of Engineering and Computer Science, the University of Texas at Dallas, Richardson, TX, USA, where he is currently an Associate Dean for research and a Professor of electrical and computer engineering. He also holds the Distinguished University Chair in telecommunications engineering and a joint appointment as a Professor of speech and hearing with the School of Behavioral and Brain Sciences. From 2005 to 2012, he was the Head of the Department of Electrical Engineering, the University of Texas at Dallas. At UT Dallas, he established the Center for Robust Speech Systems. From 1998 to 2005, he was the Department Chair and a Professor of speech, language, and hearing sciences, and a Professor of electrical and computer engineering with the University of Colorado Boulder, Boulder, CO, USA, where he co-founded and was an Associate Director of the Center for Spoken Language Research. In 1988, he established the Robust Speech Processing Laboratory. He has supervised 92 Ph.D. or M.S. thesis students, which include 51 Ph.D. and 41 M.S. or M.A. He has authored or coauthored 765 journal and conference papers including 13 textbooks in the field of speech processing and language technology, signal processing for vehicle systems, co-author of the textbook \emph{Discrete-Time Processing of Speech Signals} (IEEE Press, 2000), \emph{Vehicles, Drivers and Safety: Intelligent Vehicles and Transportation} (vol. 2 DeGruyter, 2020), \emph{Digital Signal Processing for In-Vehicle Systems and Safety} (Springer, 2012), and the lead author of The Impact of Speech Under `Stress' on Military Speech Technology (NATO RTO-TR-10, 2000). His research interests include machine learning for speech and language processing, speech processing, analysis, and modeling of speech and speaker traits, speech enhancement, signal processing for hearing impaired or cochlear implants, machine learning-based knowledge estimation and extraction of naturalistic audio, and in-vehicle driver modeling and distraction assessment for human–machine interaction. He is an IEEE Fellow for contributions to robust speech recognition in stress and noise, and ISCA Fellow for contributions to research for speech processing of signals under adverse conditions. He was the recipient of Acoustical Society of America's 25 Year Award in 2010, and is currently serving as ISCA President (2017–2022). He is also a Member and the past Vice-Chair on U.S. Office of Scientific Advisory Committees (OSAC) for OSAC-Speaker in the voice forensics domain from 2015 to 2021. He was the IEEE Technical Committee (TC) Chair and a Member of the IEEE Signal Processing Society: Speech-Language Processing Technical Committee (SLTC) from 2005 to 2008 and from 2010 to 2014, elected the IEEE SLTC Chairman from 2011 to 2013, and elected an ISCA Distinguished Lecturer from 2011 to 2012. He was a Member of the IEEE Signal Processing Society Educational Technical Committee from 2005 to 2010, a Technical Advisor to the U.S. Delegate for NATO (IST/TG-01), an IEEE Signal Processing Society Distinguished Lecturer from 2005 to 2006, an Associate Editor for the IEEE TRANSACTIONS ON AUDIO, SPEECH, AND LANGUAGE PROCESSING from 1992 to 1999 and the IEEE SIGNAL PROCESSING LETTERS from 1998 to 2000, Editorial Board Member for the IEEE \emph{Signal Processing Magazine} from 2001 to 2003, and the Guest Editor in October 1994 for Special Issue on Robust Speech Recognition for the IEEE TRANSACTIONS ON AUDIO, SPEECH, AND LANGUAGE PROCESSING. He is currently an Associate Editor for the JASA, and was on the Speech Communications Technical Committee for Acoustical Society of America from 2000 to 2003. In 2016, he was awarded the honorary degree Doctor Technices Honoris Causa from Aalborg University, Aalborg, Denmark in recognition of his contributions to the field of speech signal processing and speech or language or hearing sciences. He was the recipient of the 2020 Provost's Award for Excellence in Graduate Student Supervision from the University of Texas at Dallas and the 2005 University of Colorado Teacher Recognition Award. He organized and was General Chair for ISCA Interspeech-2002, Co-Organizer and Technical Program Chair for the IEEE ICASSP-2010, Dallas, TX, and Co-Chair and Organizer for IEEE SLT-2014, Lake Tahoe, NV. He will be the Tech. Program Chair for the IEEE ICASSP-2024, and Co-Chair and Organizer for ISCA INTERSPEECH-2022.
\end{IEEEbiography}

\begin{IEEEbiography}[{\includegraphics[width=1in,height=1.25in,clip,keepaspectratio]{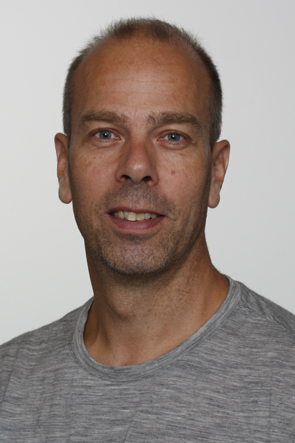}}]{\textbf{Jesper Jensen}} received the M.Sc. degree in electrical engineering and the Ph.D. degree in signal processing from Aalborg University, Aalborg, Denmark, in 1996 and 2000, respectively. From 1996 to 2000, he was with the Center for Person Kommunikation (CPK), Aalborg University, as a Ph.D. student and Assistant Research Professor. From 2000 to 2007, he was a Post-Doctoral Researcher and Assistant Professor with Delft University of Technology, Delft, The Netherlands, and an External Associate Professor with Aalborg University. Currently, he is a Senior Principal Scientist with Oticon A/S, Copenhagen, Denmark, where his main responsibility is scouting and development of new signal processing concepts for hearing aid applications. He is a Professor with the Section for Artificial Intelligence and Sound (AIS), Department of Electronic Systems, at Aalborg University. He is also a co-founder of the Centre for Acoustic Signal Processing Research (CASPR) at Aalborg University. His main interests are in the area of acoustic signal processing, including signal retrieval from noisy observations, coding, speech and audio modification and synthesis, intelligibility enhancement of speech signals, signal processing for hearing aid applications, and perceptual aspects of signal processing.
\end{IEEEbiography}

\EOD

\end{document}